\documentclass{aastex6}
\usepackage{setspace}
\usepackage{cellspace}
\usepackage{graphicx}
\usepackage{booktabs}
\usepackage{latexsym,color}
\usepackage{hyperref}
\usepackage{bookmark}
\def\be{\begin{equation}}
\def\ee{\end{equation}}
\def\bea{\begin{eqnarray}}
\def\eea{\end{eqnarray}}

\newcommand{\pp}{{\footnotesize\textbf{()}}}
\newcommand{\non}[1]{{\LARGE{\not}}{#1}}
\newcommand{\il}{~}
\newcommand{\jj}{\mathrm{J}}
\newcommand{\mso}{\mathrm{mso}}
\newcommand{\mbo}{\mathrm{mbo}}

  \newcommand{\cc}{\mathrm{C}}
    \newcommand{\oo}{\mathrm{O}}

\fullcollaborationName{The Friends of AASTeX Collaboration}

\begin{document}


\title{Ringed accretion disks: evolution of double toroidal configurations}

%% Use \author, \affil, plus the \and command to format author and affiliation
%% information.  If done correctly the peer review system will be able to
%% automatically put the author and affiliation information from the manuscript
%% and save the corresponding author the trouble of entering it by hand.
%%
%% The \affil should be used to document primary affiliations and the
%% \altaffil should be used for secondary affiliations, titles, or email.

%% Authors with the same affiliation can be grouped in a single
\author{D. Pugliese\altaffilmark{1} and Z. Stuchl\'{\i}k\altaffilmark{2}}
\affil{Institute of Physics and Research Centre of Theoretical Physics and Astrophysics,\\  Faculty of Philosophy \& Science,\\
  Silesian University in Opava,\\
 Bezru\v{c}ovo n\'{a}m\v{e}st\'{i} 13, CZ-74601 Opava, Czech Republic
}

%% Use the \and command so offset the last author.


\altaffiltext{1}{daniela.pugliese@fpf.slu.cz}
\altaffiltext{2}{zdenek.stuchlik@physics.cz}

\begin{abstract}
We investigate  ringed  accretion disks constituted by two   tori (rings)  orbiting  on the equatorial plane of  a  central super-massive Kerr black hole. We discuss the emergence of  the instability phases of each ring of the macro-configuration (ringed disk) according to Paczynski  violation of mechanical equilibrium.
  In the  full general relativistic treatment, we  consider   the   effects of the geometry of the Kerr spacetimes relevant in the characterization of the evolution of these configurations. The discussion of the  rings stability in different  spacetimes  enables us to  identify particular classes  of  central  Kerr attractors   in dependence of   their dimensionless spin. As a result of this analysis we set constraints of the evolutionary  schemes  of the ringed disks  related to the  tori morphology and   their  rotation relative to  the  central black hole and to each other.
{{ The
dynamics of the unstable phases of this system is significant   for the high energy phenomena    related  to accretion onto super-massive black holes in active galactic nuclei (AGNs), and the extremely energetic phenomena in quasars which could be observable in their  X-ray  emission.}}
\end{abstract}



%%%%%%%%%%%%%%%%%%%%%%%%%%%%%%%%%%%%%%\input{D-Sys-Introduction}
\section{Introduction}
%r
The physics of accretion disks around super-massive attractors  is characterized  by  several
   processes of very diversified  nature and it is  ground  of many  phenomena  of the  high energy astrophysics.  However, the existence of a unique  satisfactory framework for a  complete  theoretical interpretation of  such observations   remains still to be proved, as the  phenomenology associated with these systems includes  very different events  all supposedly related to the physics of strong attractors and their environment: from the issue of jet generation and collimation, to the Gamma-ray bursts (GRBs) and  the  accretion  process itself.
Particularly, accretion of matter from  disks orbiting
 super-massive black holes  (\textbf{SMBHs}), hosted in the centre of
active galactic nuclei (AGN) or quasars,   is a subject which seems to require consistent  additional  investigations. The location
and definition of  the inner edge of an accretion disk  for example is  one of  the topics that are continuously debated within several argumentations  leading  often to different conclusions \citep{Krolik:2002ae,BMP98,2010A&A...521A..15A,Agol:1999dn,Paczynski:2000tz}.
The interaction between the \textbf{SMBHs} and   orbiting matter is  certainly  a complicated subject of investigation  which   entangles  the central attractor and the  embedded material in one  dynamical picture. Interaction between the attractor and environment     could  give rise potentially to a mutation of the  geometrical characteristics of spacetime which,  initially considered as  ``frozen background'', may change following  a spin-down or eventually  spin-up of the non isolated black hole (\textbf{BH})\citep{Abra83,Abramowicz:1997sg,Rez-Zan-Fon:2003:ASTRA:,Font:2002bi,Hamersky:2013cza,Lot2013}.
 Jets emission is then  a further key element  of such systems: the relation between the rotational energy of the central attractor,  disk inner edge  location  and jet emission (jet-accretion correlation) is still substantially obscure\citep{Romanova,McKinneyScience,Allen:2006mh,Stuchlik-Kolos2015,NatureMa,Maraschi:2002pp,Chen:2015cga,Yu:2015lqj,Zhang:2015eka,Sbarrato:2014uxa,Coughlin:2013lva,MSBNNW2009,Fender:2015kta,AbramowiczSharp(1983),Sadowski:2015ena,Okuda:2004zv,Ferreira:2003yy,Lyutikov(2009),Ghisellini:2014pwa,FragileW2012}.

However, considering the huge variety of approaches and  studies  on each of these individual issues, it appears necessary, in addressing these topics,  to formulate the  investigation  in  a  more global perspective reformulating the problem in terms of structures and macro-structures, the  first  understood as   isolated objects the second  as isolated clusters of  individual  interacting objects,    to consider the relation between the different components  in these systems, and  involving the knowledge acquired on certain specific processes such as  the accretion mechanism, jet properties and black hole physics  in a more inclusive picture.

{{Accretion disk models may be distinguished by at
   least three important aspects: the geometry  (the vertical thickness defines  geometrically thin or thick disks), the matter accretion rate (sub or super-Eddington luminosity), and the optical depth  (transparent or opaque disks)--see \citep{AbraFra}.
Geometrically thick disks are  well  modelled as the Polish Doughnuts (P-D) \citep{KJA78,Abr-Jar-Sik:1978:ASTRA:,Jaroszynski(1980),Stu-Sla-Hle:2000:ASTRA:,
Rez-Zan-Fon:2003:ASTRA:,Sla-Stu:2005:CLAQG:,
Stu:2005:MPLA,PuMonBe12,PuMon13}, or also the ion tori \citep{Rees1982}. The P-D  tori (with very high, super-Eddington, accretion rates) and slim disks have high optical depth while the ion tori and the ADAF (Advection-Dominated Accretion Flow) disks have low optical depth and relatively low accretion rates (sub-Eddington).
Geometrically thin disks are generally  modelled as the standard Shakura-Sunayev (Keplerian) disks \citep{[S73],[SS73],Nov-Thorne73,Page-Thorne74}, the ADAF  disks \citep{Ab-Ac-Schl14,Narayan:1998ft}, and the slim disks \citep{AbraFra}.
In these disks, dissipative viscosity processes are relevant for accretion, being usually attributed to the magnetorotational instability of the local magnetic fields \citep{Hawley1984,Hawley1987,Hawley1991,DeVilliers}.  On the other hand in the toroidal disks, pressure gradients are crucial \citep{Abr-Jar-Sik:1978:ASTRA:}.
 As proposed in \citep{Pac-Wii},  accretion disks can been  modelled by using an appropriately defined Pseudo-Newtonian potential, see also  \citep{Nov-Thorne73,Abr-Jar-Sik:1978:ASTRA:,Stu:2005:MPLA,Stuchlik:2009jv,AbraFra}.
}}

In \cite{ringed}  we considered the possibility that  during   several accretion regimes  occurred in the lifetime of non isolated super-massive Kerr black hole several  toroidal fluid configurations   might be   formed  from  the  interaction of the  central attractor    with the  environment in {AGNs}, where   corotating  and counterrotating accretion stages are
mixed \citep{Dyda:2014pia,Aligetal(2013),Carmona-Loaiza:2015fqa,Lovelace:1996kx,Gafton:2015jja}.
 These systems can be  then
   reanimated  in some subsequent stages of the \textbf{BH}-accretion disks life, for example  in colliding galaxies, or in galactic center, in some kinds of binary systems,  where    some additional matter  could be  supplied into the vicinity of the
central black hole due to  tidal distortion of a star, or if some cloud of interstellar matter is captured by the strong gravity.

We formulated   an analytic model of a  macro-structure, the ringed accretion disk,  made by  several  toroidal   axis-symmetric sub-configurations (rings)  of  corotating and counterrotating  fluid structures (tori) orbiting  one center super-massive Kerr black hole,  with symmetry plane coinciding with   the equatorial plane of the central Kerr   \textbf{BH}. The emergence of instabilities for each ring and  the entire macro-structure was then addressed in \cite{open,opentobe}.   Similar studies on analogue problems  are in \cite{Cremaschini:2013jia}  where off-equatorial tori around compact objects were considered
 and also in \cite{Pict-to}.
In \cite{letter} we drew some conclusions
for the case of only two toroidal  disks  orbiting  a   central   Kerr attractor.   We demonstrated that only  under    specific conditions  a double accretion   system may be formed.  Rings of the macro-structure  can then  interact colliding. The center-of-mass energy  during ring collision was evaluated within the  test particle approximation  demonstrating  that  energy  efficiency of the collisions increases with increasing  dimensionless black hole spin,  being very high for near-extreme black holes.
The collisional energy efficiency  could be even higher  in near-extreme  Kerr naked singularity spacetimes \cite{cos1,cos2,Stuchlik(1980)}.

{{ Using  numerical methods, multi-disks have been  also analyzed in more complex, non-symmetric situations.
Formation  of several accretion disks in the geometries of the  \textbf{SMBH}   in AGNs or in binary systems, have  been  considered in relation to various factors, where the rupture of symmetries  has been addressed, for example for titled, warped, not coplanar disks. Attention  has been paid to the investigations of the relevance of the disk geometry in  the attractor-disk interaction.
  Initial stages of the formation of such systems has been addressed in   \cite{bib:decoha-sewi}.
Concerning  counter-aligned accretion disks in  AGN, we point out
\cite{King:2006uu}, where   the Bardeen-Petterson effect is proposed as a possible  cause of  the counter-alignment of \textbf{BH} and
disk spins:  it is shown that  \textbf{BH} can grow rapidly  if
they acquire most of accreting mass  it in a sequence of randomly oriented accretion episodes.
In \cite{Lodato:2006kv}, the evolution of misaligned accretion disks and spinning
\textbf{BH}  are considered especially in
AGN, where  the
\textbf{BH} spin  changes  under the action of the disk torques, as the disk, being
subjected to Lense-Thirring precession,  becomes  twisted
and warped.  It is shown that
accretion from  misaligned disk in  galactic nuclei  would be  significantly more luminous than  accreting from a flat disk. Aligning  of Kerr \textbf{BH}s and accretion disks are studied in \cite{King:2005mv}.
In \cite{Nealon:2015jya} the
effects of \textbf{BH} spin on warped or misaligned accretion disks are studied in connections to  the role of the  inner edge of the disk in the alignment of the angular momentum  with the \textbf{BH} spin.
Stable counter-alignment of a circumbinary disk is focused in  \cite{bib:spso-sa}.
\cite{King:2008au} argue  that there is a generic
tendency of AGN accretion disks to become self-gravitating at a certain radius from the attractor.
The study of  particular accretion processes including merging of the AGN accretion  disk, demonstrates that the disk has  generally a lower angular momentum than the \textbf{\textbf{BH}}, for an analogue limit in  \cite{pugtot}.
The chaotical
accretion  in AGNs could produce counter–rotating accretion
disks or  strongly misaligned disks with respect to the central \textbf{SMBH} spin.
Rapid AGN accretion from counter–rotating disks  is particularly addressed in \cite{No-Ni-Yiz}.
 Authors studied the
 angular momentum cancelation  in accretion disks characterized by a significant  tilt between inner and outer disk parts. }
}%%%%%%%%%%%%%%%%%Tearing up%%%%%%%%%%%%%%%%%%%%%%%%%%%%%%%%
{{These  studies show that  evolution of a misaligned disks around a Kerr
\textbf{BH} might lead to a tearing up of the  disk into several   planes with different inclinations.
 Tearing up the disk in misaligned accretion onto a binary system is considered for example in
\cite{Nixon:2013qfa}.}}

{{Tearing up process  has been also considered as  possible mechanism behind the almost  periodic  emission in X-ray emission  band known as QPOs.
Tearing up a misaligned accretion disk with a binary companion is addressed in \cite{Dogan:2015ida}. disk formation by tidal disruptions of stars on eccentric
orbits by a spherically symmetric black hole is considered in
\cite{Bonnerot:2015ara}. For misaligned gas disks around eccentric \textbf{BHs} binaries see
\cite{Aly:2015vqa}.}}

{{As explained in \cite{Pict-to}, in realistic cases of AGN accretion,  or also in  stellar-mass X-ray binaries,
 there is a  break in the  central part of a  tilted accretion disks  orbiting    Kerr {\textbf{BH}s} due to the  Lense-Thirring effect. The disk is thus  splitted into several,  essentially  separated,  planes.
It is observed that  also for small  tilt angles the disk may still break and this must  be connected with  some  observable phenomena as for example   QPOs.
For a brief review of the
\textbf{{SMBH}} accretion  mergers  and  accretion flows on to \textbf{SMBH}  see \cite{King:2013vva}.}
}

{{
  The existence  of ringed disks in general  may lead  us to reinterpret  action of the phenomena so far analyzed in a  single disk  framework in terms of  orbiting multi-toroidal  structures. Especially, this shift could be reinforced in modelling the spectral  features  of multi-disk structures.
It is generally assumed  that the  X-ray emission
from AGNs is   related to accretion disks and surrounding corona.  Assuming to be related to the  accretion disk instabilities, the
spectra interpretation of  X-ray emission is taken to  constrain the main  \textbf{BH}-disk  model parameters.
We argue that this spectra profile should provide also a fingerprint of the ringed disk structure,
 possibly showing as  a radially stratified emission profile.
 In fact, the simplest structures of  this kind are thin radiating rings. Signature of alternative gravity, as exotic objects, given by spectral lines from the radiating  rings is investigated in  \cite{Schee:2008fc,Schee:2013asiposs,Bambi:2016sac,Ni:2016uik}.
In \cite{S11etal}  the authors propose  that  the \textbf{BH} accretion rings models may be  revealed by future X-ray spectroscopy, from the
 study of  relatively indistinct excesses
on top of the relativistically broadened spectral line  profile,
 unlike
the main body of the broad line of the spectral line  profile, connected to an   extended (continuous) region of the accretion disk. They predicted relatively indistinct excesses
 of the relativistically broadened  emission-line components, arising in a well-confined
radial distance in the accretion disk, envisaging  therefore a sort of rings model which may be adapted as a special case  of the  model discussed in \cite{ringed,open}.
Specifically, in \cite{KS10}
extremal energy shifts of radiation from a ring near a rotating black hole were particularly  studied:
radiation from a narrow circular ring shows a  double-horn profile with photons having
energy around the maximum or minimum of the  range (see also \cite{Schee:2008fc}).
This energy span of  spectral lines is a function of the observer's viewing angle, the black hole spin and the  ring radius. The authors
describe a method to calculate the extremal energy shift in the regime of strong gravity. The  accretion disk  is modelled by a rings located in a Kerr \textbf{BH} equatorial plane,  originating by a series of  episodic accretion events. It is argued that the proposed  geometric and emission ringed structure should be evident
from  the extremal energy shifts of the each  rings. Accordingly,
the ringed disks  may be revealed thought
detailed spectroscopy of the spectral line wings.
Although the method has been  specifically   adapted  to the case of  geometrically thin disks, an extension to   thick rings should be possible. Furthermore,  as detailed in  \cite{open,ringed}, some of the general geometric  characteristics of the ringed disk  structure  are  well applicable to the thin  disk case.
}}

Here we  extend the study in \cite{ringed,open} considering an orbiting pair of axi-symmetric tori  governed by   the   relativistic
   hydrodynamic Boyer  condition  of equilibrium configurations  of rotating perfect fluids \cite{Boy:1965:PCPS:}.
Our primary
result is the  characterization the rings-attractor systems in terms of equilibrium or unstable (critical) topology, constraining    the  formation of such a system on the basis of   the (frozen) dimensionless spin-mass ratio of the attractor and the relative rotation of the fluids. We  investigate the possible dynamical evolution of the tori, generally  considered  as transition from the  topological state of equilibrium to a topology  of instability,  and the  evolution for   the entire macro-configuration  when  accretion onto the central black hole  and collision among the tori may occur. We enlighten the situation where tori collisions lead to the destruction of the macro-configuration.
We summarized this  analysis developing  some  evolutionary schemes which  provide   indications of the topology transition and  the  situations  where  these systems could potentially be found and then observed due to the associated phenomena. These schemes are constrained by  spin of the attractor and the relative rotation of the rings with respect to attractor or each other.
From the methodological viewpoint we represented  evolutionary   schemes with graph models, which we consider here also   as reference in our discussion.

In our model we primarily evaluate the  general relativistic effects  on the orbiting matter  in those situations where  curvature effects  and the fluid rotation are considerable in   determination  of the toroidal topology and morphology. We focus on toroidal disk model orbiting the super-massive Kerr attractors using  the  geometrically thick disk as Polish Doughnuts (P-D),  opaque and with  very high (super-Eddington) accretion rates    where  pressure gradients  are crucial \citep{KJA78,Abr-Jar-Sik:1978:ASTRA:,Jaroszynski(1980),Stu-Sla-Hle:2000:ASTRA:,Rez-Zan-Fon:2003:ASTRA:,Sla-Stu:2005:CLAQG:,Stu:2005:MPLA,PuMonBe12}.
These configurations are  often adopted as the initial conditions in the set up for simulations of the MHD (magnetohydrodynamic) accretion structures\citep{Igumenshchev,Fragile:2007dk,DeVilliers}.
In fact, the majority of the current analytical and numerical models of accretion configurations assumes the  axial symmetry of the extended accreting matter.

{For the  geometrically thick configurations it is generally assumed that
 the time scale of the dynamical processes $\tau_{dyn}$  (regulated by the gravitational and inertial forces, the timescale for  pressure to balance the  gravitational and centrifugal force) is much lower than the time scale of the thermal ones $\tau_{the}$  (i.e. heating and cooling processes, timescale of  radiation entropy redistribution) that is lower than the time scale of the viscous processes $\tau_{vis}$, and the effects of strong gravitational fields are dominant with respect to the  dissipative ones and predominant to determine  the unstable phases of the systems \citep{F-D-02,Igumenshchev,AbraFra}, i.e. $\tau_{dyn}\ll\tau_{the}\ll\tau_{vis}$
see also \citet{Fragile:2007dk,DeVilliers,Hawley1987,Hawley1991,Hawley1984}.
This in turn grounded the assumption of  perfect fluid energy-momentum tensor. Thus  the effects of strong gravitational fields dominate  the  dissipative ones \citep{F-D-02,AbraFra,Pac-Wii}.
Consequently  during the evolution of dynamical processes, the functional form of the angular
momentum and entropy distribution depends on the initial conditions of the system and on
the details of the dissipative processes.
Paczy\'nski realized that it is physically
reasonable to assume an ad hoc distributions \cite{Abramowicz:2008bk}. This feature  constitutes a great advantage of these models  and  render their  adoption   extremely useful and predictive (the angular momentum transport in the
fluid is perhaps one of the most controversial aspects in thin accretion disk).
Moreover, we should note that the Paczy\'nski accretion mechanics from  a  Roche lobe overflow  induces   the mass loss  from tori being an important
 local stabilizing mechanism  against thermal and viscous instabilities, and globally   against the Papaloizou-Pringle instability (for a review we refer to \cite{AbraFra}).}

In this models the entropy is constant along the flow. According to the von Zeipel condition, the surfaces of constant angular velocity $\Omega$ and of constant specific angular momentum $\ell$ coincide \citep{M.A.Abramowicz,Chakrabarti0,Chakrabarti,Zanotti:2014haa} and  the rotation law $\ell=\ell(\Omega)$ is independent of the equation of state \citep{Lei:2008ui,Abramowicz:2008bk}.

\medskip

\textbf{Article layout}

\medskip

{In details the plan of this article is as follows:
The introduction of the
thick accretion disks model  in a  Kerr spacetime is  summarized in Section\il(\ref{Sec:Taa-DISK})  where the main notation considered through this work is   presented. This section constitutes   first introductive  part of this work  and also  the   disclosure of the  methodological tools used throughout.
We provide main definitions of   the  major morphological features of the ringed disks. Then, we specialize the  concepts for the case of system composed by only  two tori.
After writing the Euler equations  for the orbiting fluids we  cast the set  of hydrodynamic  equations for the tori by introducing  an effective potential function for the macro-configuration.
We   then investigate the parameter space for this model; one set of  couple parameters  provides   the boundary conditions for the description of two tori in the macro-configuration.
We proceed by dividing the discussion  for  the $\ell$corotating  and $\ell$counterrotating  tori-if the tori are  both corotating   or counterrotating   with respect  to the central Kerr \textbf{BH}  they are $\ell$corotating, if   one torus is corotating and the other counterrotating they are $\ell$counterrotating.
 However, even in the case of    one couple of tori orbiting around a single central
Kerr \textbf{BH},  a remarkably large number of possible configurations is possible.
Therefore,  in order to  simplify and illustrate the discussion, we  made use  of special graphs for the representation of  a couple of accretion tori and their
evolution, within the constraints they are subjected to.
The use of these  graphic schemes has been reveled to be
crucial for the study and  representation of these evolutionary  cases.
Although the following analysis may be   followed quite independently from
the graph formalism,  they can be used also to quickly collect the different constraints on the existence and evolution
of the tori  and for reference in our discussion.
 Therefore we include here  also a brief description of  the
graph construction
and basic concepts related to these structures.
 Appendix\il(\ref{Sec:graph-app}) discloses details on the construction and interpretation
of graphs. Main graph blocks are listed in  Fig.\il(\ref{Table:Graphs-models}).
The main analysis of the present work  is in Section\il(\ref{Sec:basic-intro}), where
 the  double tori disk system is discussed in details. We specialize the investigation detailing  the double system on the basis  of relative rotation of fluids in the disks and with respect to the central \textbf{BH} attractor; therefore in  Sec.\il(\ref{Sec:lc-or})  the $\ell$corotating couple of tori is addressed while  in Sec.\il(\ref{Sec:lcounterrsec}) we focus  attention on the $\ell$counterrotating case. We shall see that the results of Section\il(\ref{Sec:lc-or}) also apply to the description of $ \ell$counterrotating tori in a Schwarzschild (static) spacetime.
The  double tori disk system is characterized by the   existence and  stability  conditions.
We consider first   all the possible  states for the  couple  of accretion disks   with  fixed  topology,  and then we concentrate   on their evolutions.
We will prove that some configurations  are  prohibited.  Then we narrow the space  of the  system parameters to  specific regions according to the dimensionless \textbf{BH} spin.
The case of  $\ell$counterrotating couples  around a rotating attractor is in fact  much more articulated in  comparison to $\ell$corotating (or $\ell$counterrotating torii orbiting a Schwarzschild black hole). This  case  is hugely diversified for classes
of attractors, and for the disk spin orientation with respect to the central attractor. Therefore it requested a different approach adapted to the diversification of the cases.
In order to better analyze the situation we have split the analysis in the two sub-sections (\ref{Sec:coun-co}) and (\ref{Sec:tex-dual}); in the first we consider the case in which the inner  torus of the couple is counterrotating with respect to the attractor, then we address the inner corotating torus.
The accurate analysis in the space of parameter also allows us to discuss the possible and forbidden lines of evolution  for a  fixed couple.
 We close  Section\il(\ref{Sec:basic-intro})   in the
 Section\il(\ref{Sec:non-rigid}) where  the possibility of collision between tori  and the possibility of tori merging is considered.
 We investigate the  conditions for collision occurrence,  drowning a  description of the  associated unstable macro-configurations. Both $\ell$corotaing and $\ell$counterrotating cases are addressed.
We discuss mechanisms  which  may lead to tori collision according to  our model prescription.
This section also refers to the Appendix\il(\ref{App:deep-loop}), where further details are provided.
Indications on possible    observational evidence of doubled tori disks and their evolution are provided in  brief Section\il(\ref{Sec:a-ew-one}). We close this article in  Sec.\il(\ref{Sec:conclusion}) with a summary and brief discussion of future  prospectives.  Appendix\il(\ref{Sec:graph-app}) and Appendix\il(\ref{App:notes-tables}) follow.}

\section{Thick accretion disks in a Kerr spacetime}\label{Sec:Taa-DISK}
The Kerr  metric line element  in the Boyer-Lindquist (BL)  coordinates
\( \{t,r,\theta ,\phi \}\) reads
%r
%
\bea \label{alai}&& ds^2=-dt^2+\frac{\rho^2}{\Delta}dr^2+\rho^2
d\theta^2+(r^2+a^2)\sin^2\theta
d\phi^2+\frac{2M}{\rho^2}r(dt-a\sin^2\theta d\phi)^2\ ,
\\
\nonumber
&&
 \mbox{where}\quad\rho^2\equiv r^2+a^2\cos\theta^2\quad \mbox{and } \quad \Delta\equiv r^2-2 M r+a^2,
\eea
and    $a=J/M\in]0,M]$  is the specific angular momentum, $J$ is the
total angular momentum of the gravitational source and $M$ is the  gravitational mass parameter.  The horizons $r_-<r_+$ and the outer static limit $r_{\epsilon}^+$ are respectively given by\footnote{We adopt the
geometrical  units $c=1=G$ and  the $(-,+,+,+)$ signature, Greek indices run in $\{0,1,2,3\}$.  The   four-velocity  satisfy $u^{\alpha} u_{\alpha}=-1$. The radius $r$ has unit of
mass $[M]$, and the angular momentum  units of $[M]^2$, the velocities  $[u^t]=[u^r]=1$
and $[u^{\varphi}]=[u^{\theta}]=[M]^{-1}$ with $[u^{\varphi}/u^{t}]=[M]^{-1}$ and
$[u_{\varphi}/u_{t}]=[M]$. For the seek of convenience, we always consider the
dimensionless  energy and effective potential $[V_{eff}]=1$ and an angular momentum per
unit of mass $[L]/[M]=[M]$.}:
\bea
r_{\pm}\equiv M\pm\sqrt{M^2-a^2};\quad r_{\epsilon}^{+}\equiv M+\sqrt{M^2- a^2 \cos\theta^2};
%&&\nonumber
\eea
where $r_+<r_{\epsilon}^+$ on   $\theta\neq0$  and  $r_{\epsilon}^+=2M$  in the equatorial plane $\theta=\pi/2$.
The non-rotating  limiting case $a=0$ is the   Schwarzschild metric while the extreme Kerr black hole  has dimensionless spin $a/M=1$.  In the Kerr  geometry the quantities
\be\label{Eq:after}
{E} \equiv -g_{\alpha \beta}\xi_{t}^{\alpha} p^{\beta}=-p_t,\quad L \equiv
g_{\alpha \beta}\xi_{\phi}^{\alpha}p^{\beta}=p_{\phi}\ ,
\ee
are  constants of motion, where $\xi_{\phi}=\partial_{\phi} $ is the rotational Killing field,
      $\xi_{t}=\partial_{t} $   is
the Killing field representing the stationarity of the  spacetime, %, where the covariant
 and $p^{\alpha}$  is the particle four--momentum. The constant $L$ in Eq.\il(\ref{Eq:after}) may be interpreted       as the axial component of the angular momentum  of a test    particle following
timelike geodesics and $E$ is  representing the total energy of the test particle
 coming from radial infinity, as measured  by  a static observer at infinity. %are
%conserved along the   geodesics.
Due to the symmetries of the   metric tensor (\ref{alai}),  the     test particle dynamics is invariant under the mutual transformation of the parameters
$(a,L)\rightarrow(-a,-L)$,
and we could   restrict  the  analysis of the test particle circular motion to the case of  positive values of $a$
for corotating  $(L>0)$ and counterrotating   $(L<0)$ orbits.

In this work we specialize our analysis to    toroidal  configurations of  perfect fluid orbiting a    Kerr black hole \textbf{(BH)} attractor.
The  energy momentum tensor  for   one-species particle perfect  fluid system  is   described by
\be\label{E:Tm}
T_{\alpha \beta}=(\varrho +p) u_{\alpha} u_{\beta}+\  p g_{\alpha \beta},
\ee
where $u^{\alpha}$  is
a timelike flow vector field and  $\varrho$ and $p$ are  the total energy density and
pressure  respectively, as measured by an observer comoving with the fluid with velocity $u^{\alpha}$.
For the
symmetries of the problem, we  assume $\partial_t \mathbf{Q}=0$ and
$\partial_{\varphi} \mathbf{Q}=0$,  with $\mathbf{Q}$ being a generic spacetime tensor.
According to these assumptions  the  continuity equation
is  identically satisfied and the  fluid dynamics  is  governed by the \emph{Euler equation}:
\bea\label{E:1a0}
%u^\alpha\nabla_\alpha\varrho+(p+\varrho)\nabla^\alpha u_\alpha=0\, ,\quad
%\label{Eulerif0}
(p+\varrho)u^\alpha \nabla_\alpha u^\gamma+ \ h^{\beta\gamma}\nabla_\beta p=0\, ,
\eea
where  $\nabla_\alpha g_{\beta\gamma}=0$,  $h_{\alpha \beta}=g_{\alpha \beta}+ u_\alpha u_\beta$ is  the projection tensor  \citep{Pugliese:2011aa,pugtot}.
Assuming  a barotropic equation of state $p=p(\varrho)$, and orbital motion with  $u^{\theta}=0$ and
$u^r=0$, Eq.\il(\ref{E:1a0}) implies
\be\label{Eq:scond-d}
\frac{\partial_{\mu}p}{\varrho+p}=-{\partial_{\mu }W}+\frac{\Omega \partial_{\mu}\ell}{1-\Omega \ell},\quad \ell\equiv \frac{L}{E},\quad W\equiv\ln V_{eff}(\ell),\quad V_{eff}(\ell)=u_t= \pm\sqrt{\frac{g_{\phi t}^2-g_{tt} g_{\phi \phi}}{g_{\phi \phi}+2 \ell g_{\phi t} +\ell^2g_{tt}}},
\ee
where  $\Omega=u^{\phi}/u^{t}$ is the relativistic angular frequency of the fluid relative to the distant observer, and
the Pacz y {\'n}ski-Wiita  (P-W) potential  $W(r;\ell,a)$ and the \emph{effective potential} for the fluid  $V_{eff}(r;\ell,a)$ were introduced. These functions of position reflect the background  Kerr geometry through  the parameter $a$, and the centrifugal effects through the fluid specific angular momenta $\ell$,  here assumed  constant  and conserved  (see also \citep{Lei:2008ui,Abramowicz:2008bk}).  A natural extremal limit on the extension of both corotating and counterrotating tori occurs due to the cosmic repulsion at the so called static radius that is independent of the black hole spin
\citep{Stuchlik:2009jv,Sla-Stu:2005:CLAQG:,Stuchlik:2006ht,Stu-Sla-Hle:2000:ASTRA:,StuchlikHledik1999,Stuchlik(1983),Stu:2005:MPLA}.

The effective potential    in Eq.\il(\ref{Eq:scond-d})  is invariant under the mutual transformation of  the parameters
$(a,\ell)\rightarrow(-a,-\ell)$. Therefore
analogously to the analysis of   test particle dynamics, we  can assume $a>0$ and consider $\ell>0$
for \emph{corotating}   and $\ell<0$  for  \emph{counterrotating} fluids, within  the notation $(\mp)$   respectively.

\begin{figure}[h!]
\centering
\begin{tabular}{cc}
\includegraphics[width=.4\textwidth]{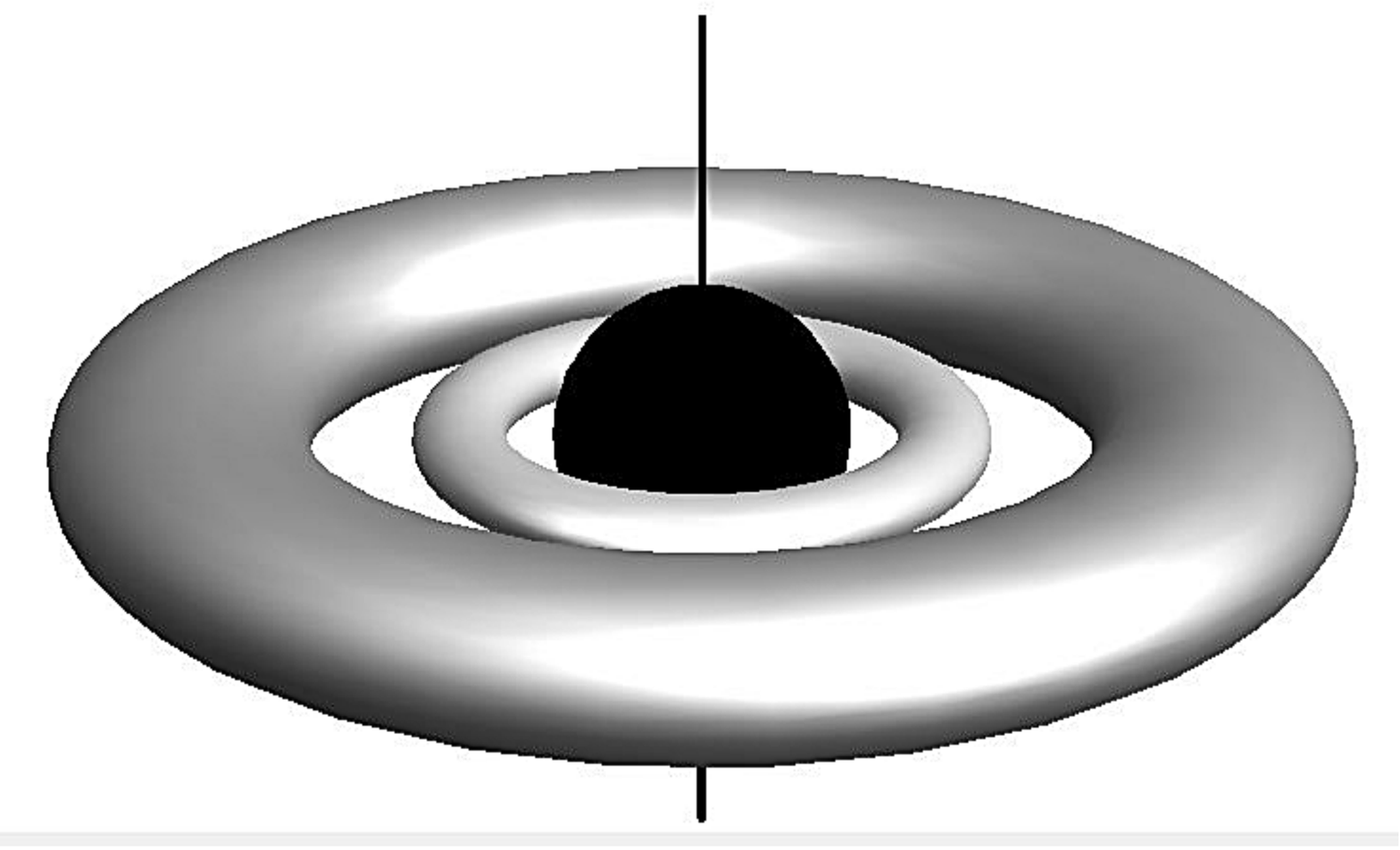}
\\
\includegraphics[width=.48\textwidth]{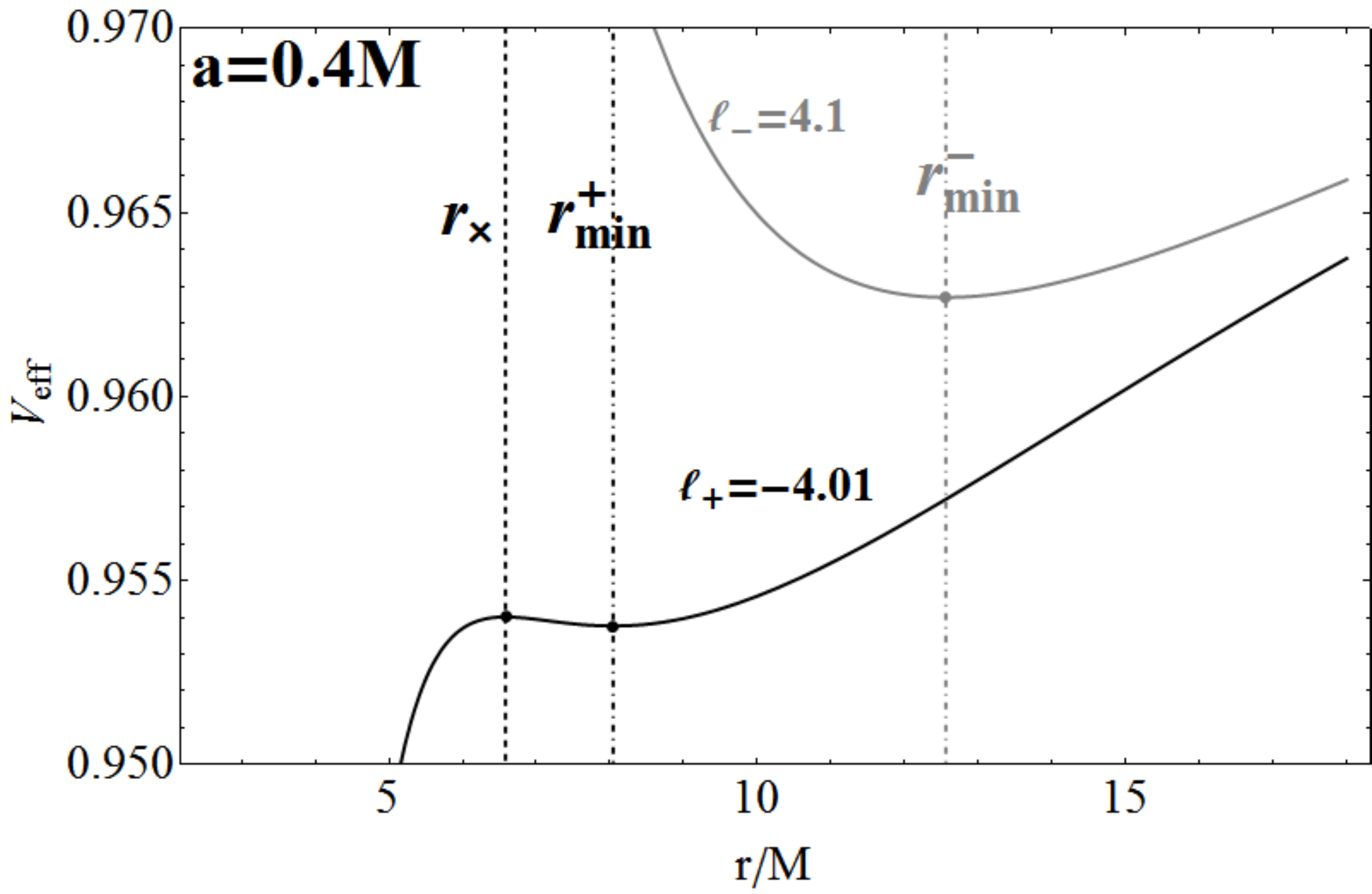}
\includegraphics[width=.48\textwidth]{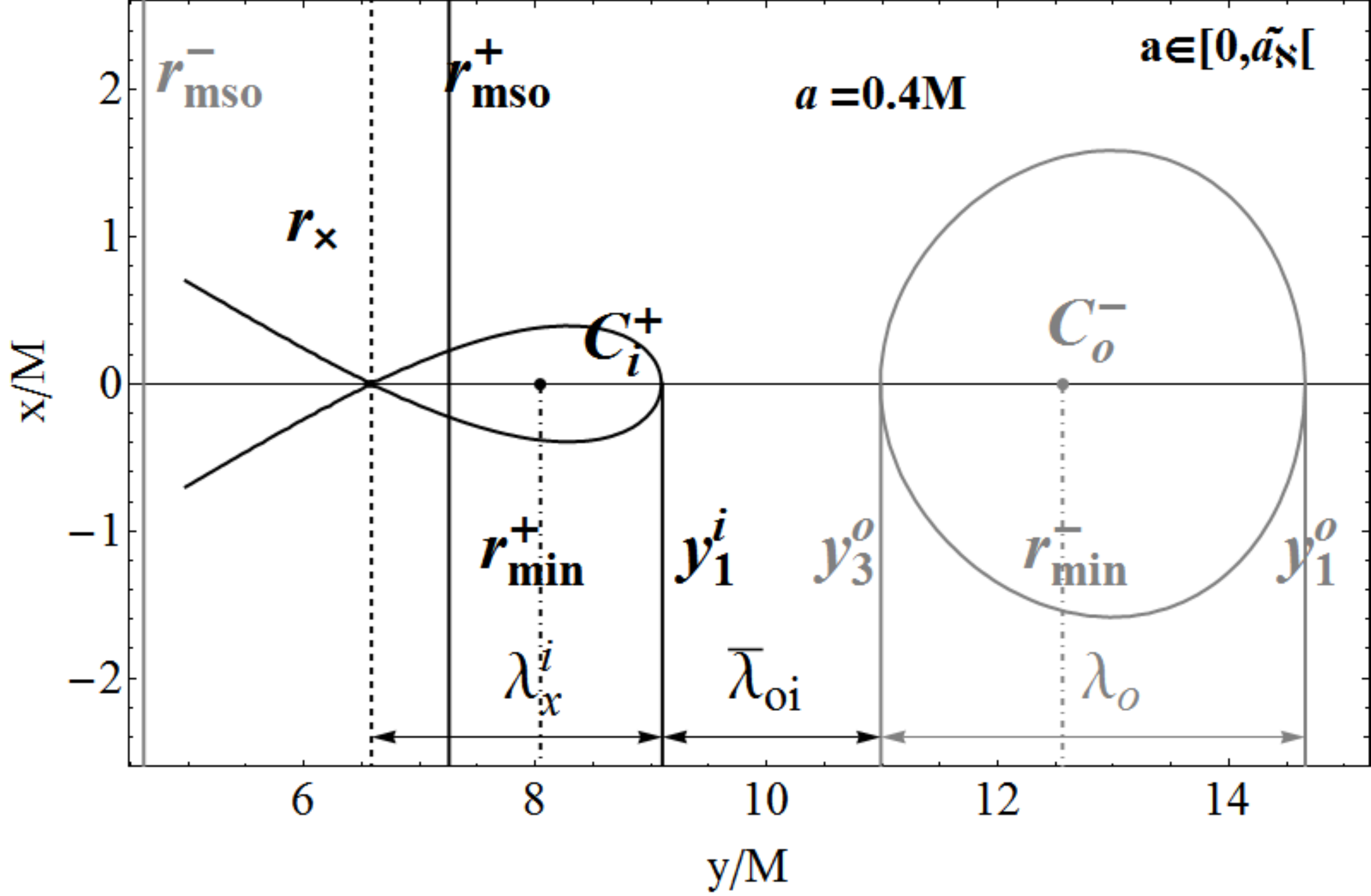}
\end{tabular}
\caption{\emph{Upper}: Pictorial representation of a doubled accretion toroidal systems orbiting a central Kerr black hole (black region)--see also in Fig.\il(\ref{Table:Torc}). \emph{Below}: $\ell$counterrotating  couple of accretion disks orbiting a central Kerr black hole   attractor with spin $a=0.4 M$. Effective potentials (\emph{left-panel}), and  cross sections on the equatorial plane of the outer  Roche lobes  (\emph{right-panel}) for a couple $(\cc_{i}^+,\cc_{o}^-)$ made by an  inner counterrotating disk and outer corotating disk  corresponding   to  \textsl{scheme III} of Fig.\il(\ref{Table:Torc}).  $(x, y)$ are Cartesian coordinates and  $r_{\mso}^{\pm}$ are the marginally stable circular orbits for counterrotating and corotating matter respectively, $r_{\min}^{\pm}$ are the center of the outer Roche lobe (point of minimum of the fluid effective potentials): $(y_3, y_o)$ is in general the disk inner  and the outer  torus edge  respectively, $\lambda_o=y_1^o-y_3^o$ is the elongation of the outer disk on the equatorial plane, $\lambda^i_{\times}$ is the elongation of the inner accreting disk, $\bar{\lambda}_{oi}$ is the spacing between the disks. Accretion for this couple (from the $r_{\times}$ point) may emerge only from the inner disk--Sec.\il(\ref{Sec:coun-co}).  }\label{Figs:ApproxPlo}
%%\end{tabular}
\end{figure}

\begin{figure}[h!]
\begin{center}
\begin{tabular}{c}
\includegraphics[width=.7\columnwidth]{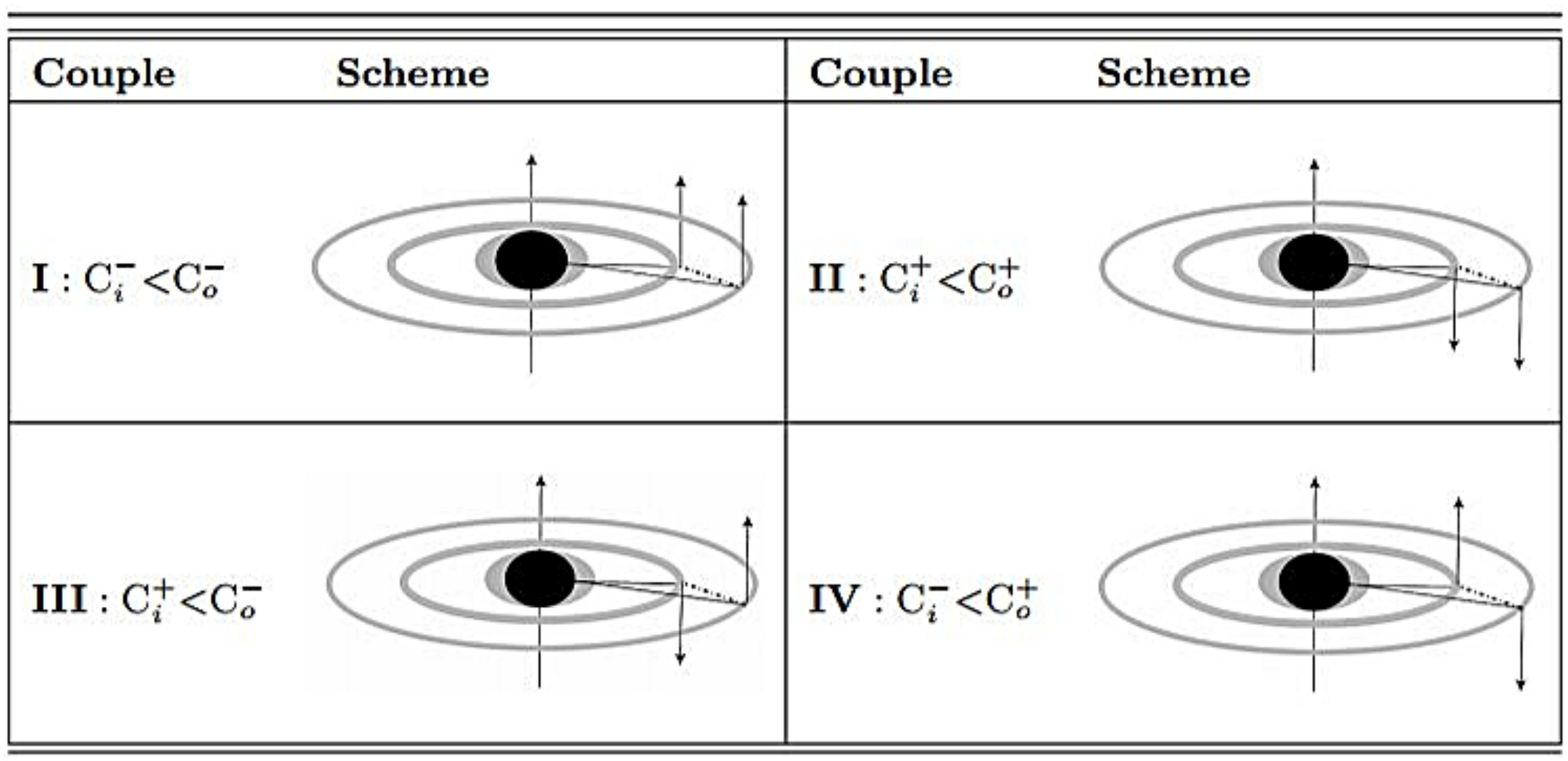}
\end{tabular}
\caption{Pictorial schemes of a double system of accretion disks  (gray thin rings) orbiting a Kerr black hole  attractor. A ringed accretion disk of the  order $n=2$ in the four principal  states: black region is the black hole, gray region is the ergosphere.  The distances between the disk and attractor are not in scale. The rings, $\cc_i$ for the inner and $\cc_o$ for the outer,  are schematically represented  as two-dimensional objects correspondent to the equilibrium topology.
The arrows represent the rotation: the dimensionless spin of the attractor $a/M\geq0$ is considered always positive, ``spin-up'' in the  picture,  or vanishing for the limiting case of the  static Schwarzschild solution. The fluid specific angular momentum of  an accretion disk $\ell$ can be positive, $\ell a>0$, for corotating $(-)$ (``spin-up'') or negative, $\ell a<0$, for counterrotating $(+)$ (``spin-down'')   with respect to the central black hole.
Rings are  $\ell$corotating if $\ell_i\ell_o>0$ ({\textsl{scheme I} } and \textsl{{II}}-see also Figs\il(\ref{Fig:Goatms})) or $\ell$counterrotating if $\ell_i\ell_o<0$ (\emph{scheme III} and \emph{IV}-see also Figs\il(\ref{Figs:ApproxPlo})  and (\ref{Figs:Kind-End}) respectively). A pictorial representation of this system can be found in Fig.\il(\ref{Figs:ApproxPlo})}\label{Table:Torc}
\end{center}
\end{figure}

The ringed accretion disks, introduced in \cite{pugtot,ringed,open}, represent  a fully general relativistic model  of toroidal disk configurations $\mathbf{C}^n=\bigcup^n\cc_i$, consisting of a collection of  $n$ sub-configurations  (configuration \emph{order}  $n$) of     corotating and counterrotating toroidal rings orbiting a supermassive Kerr attractor--Figs\il(\ref{Figs:ApproxPlo}).
Since tori  can be corotating or counterrotating with respect to the black hole,  assuming    first a couple $(\cc_a, \cc_b)$,  orbiting  in   the equatorial plane of a given Kerr \textbf{\textbf{BH}}  with specific angular momentum $(\ell_a, \ell_b)$,   we need to introduce   the concept  of
 \emph{$\ell$corotating} disks,  defined by  the condition $\ell_{a}\ell_{b}>0$, and \emph{$\ell$counterrotating} disks defined  by the relations   $\ell_{a}\ell_{b}<0$.  The two $\ell$corotating tori  can be both corotating, $\ell a>0$, or counterrotating,  $\ell a<0$, with respect to the central attractor--see Fig.\il(\ref{Table:Torc}).

\begin{figure}[h!]
\begin{center}
\begin{tabular}{cc}
\includegraphics[width=0.5\columnwidth]{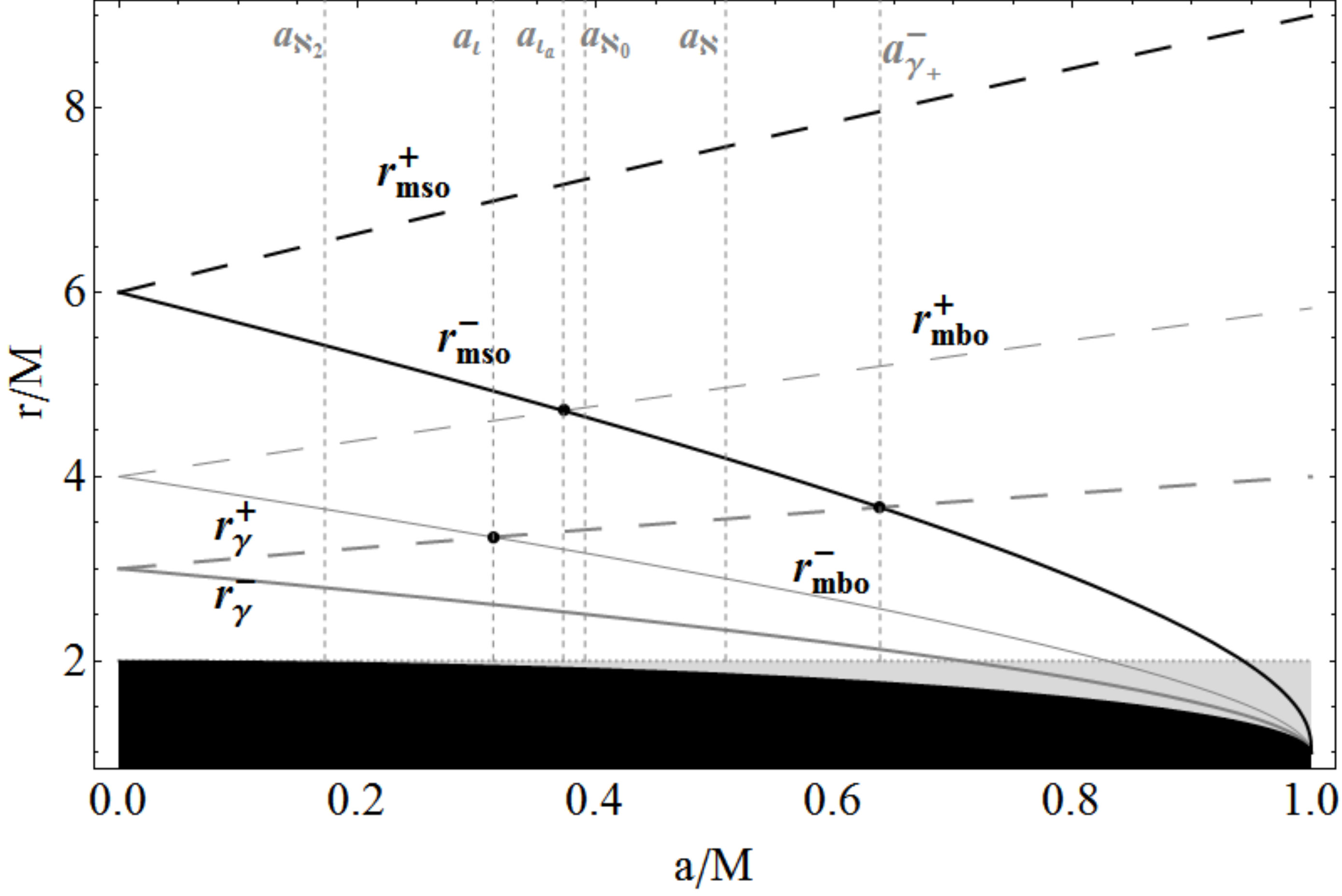}
\includegraphics[width=0.5\columnwidth]{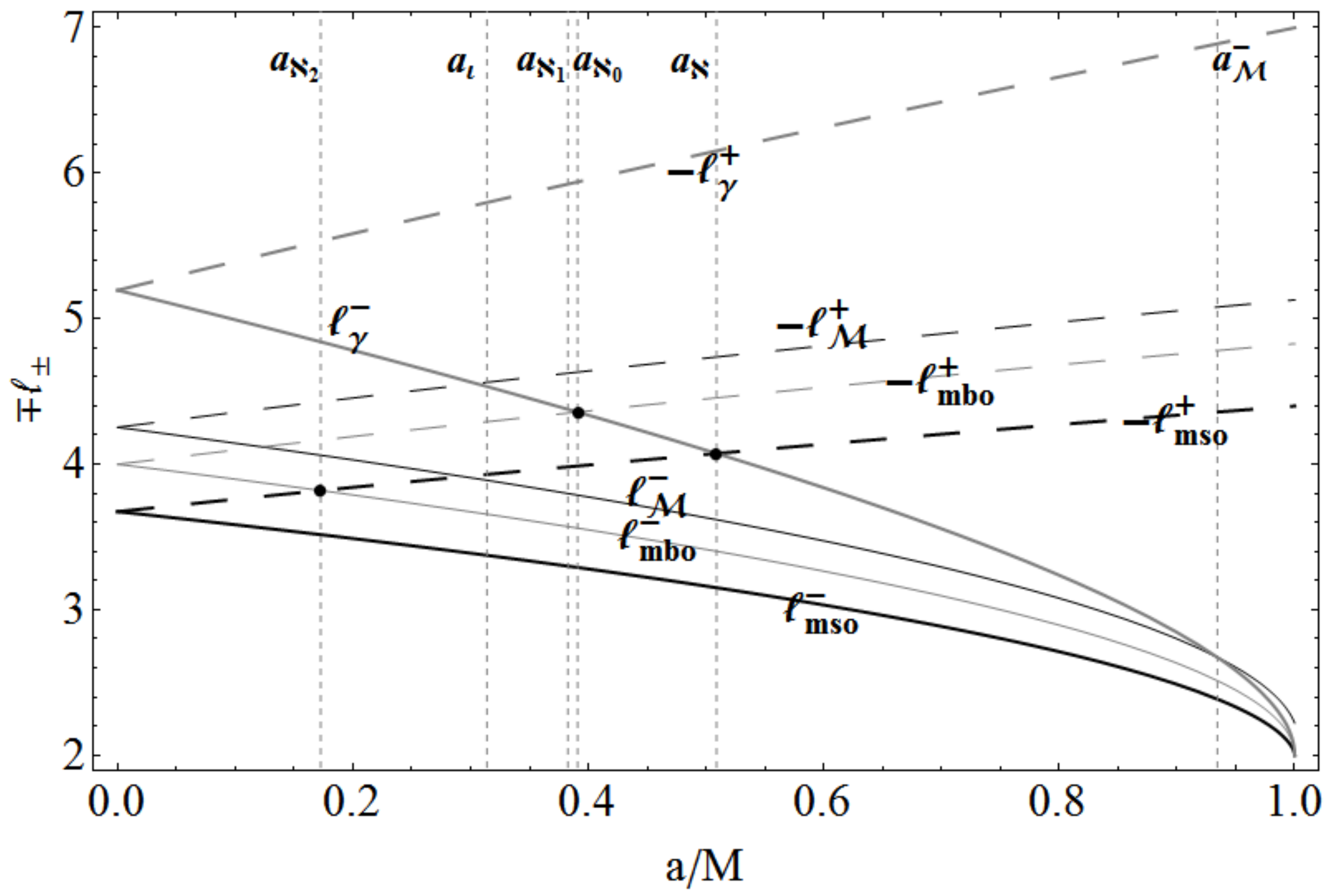}%\\
\end{tabular}
\caption{Geodesic structure of the Kerr geometry: notable radii  $\mathbf{R}_{\mathrm{N}}\equiv \{r_{\gamma}^{\pm}, r_{\mbo}^{\pm},r_{\mso}^{\pm}\}$ (left panel), and  the respective  fluid specific angular momentum $\ell^{\pm}_i=\ell^{\pm}(r^{\pm}_i)$   where $r^{\pm}_i\in \{\mathbf{R}_{\mathrm{N}}^{\pm},r_{\mathcal{M}}^{\pm}\}$,  $r_{\mathcal{M}}^{\pm}$ is the maximum point    of
derivative $\partial_r(\mp \ell^{\pm})$ for $a/M$ respectively. Some notable spacetime spin-mass ratios are also plotted,  a list can found in Table\il(\ref{Table:nature-Att}). Black region is $r<r_+$, $r_+$ being the outer horizon of the Kerr geometry, gray region is $r<r_{\epsilon}^+$, $r_{\epsilon}^+$ is the outer ergosurface.}\label{Figs:Ptherepdiverg}
\end{center}
\end{figure}

\begin{figure}[h!]
\centering
\begin{tabular}{cc}
\includegraphics[width=0.5\columnwidth]{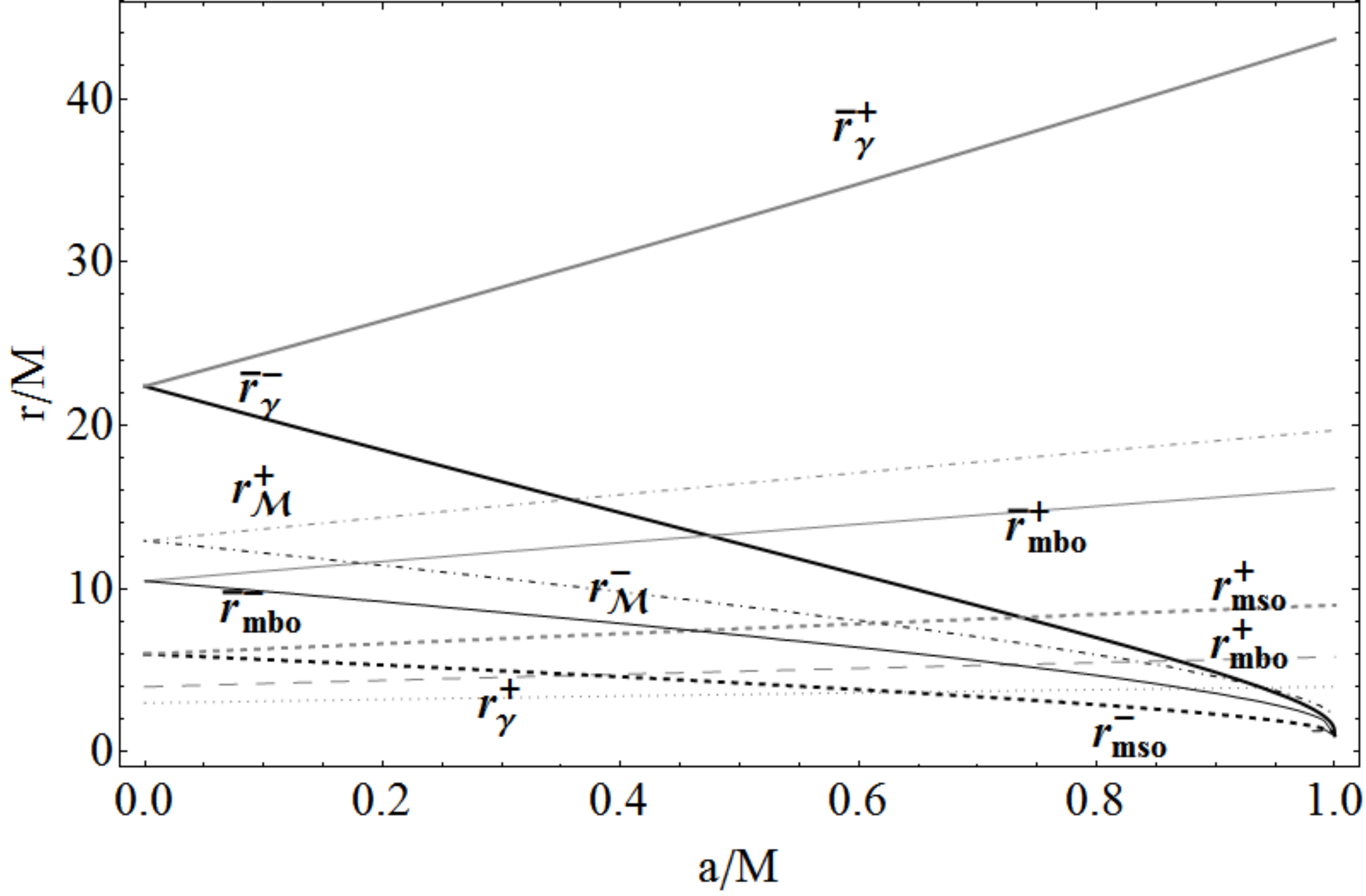}
\includegraphics[width=0.5\columnwidth]{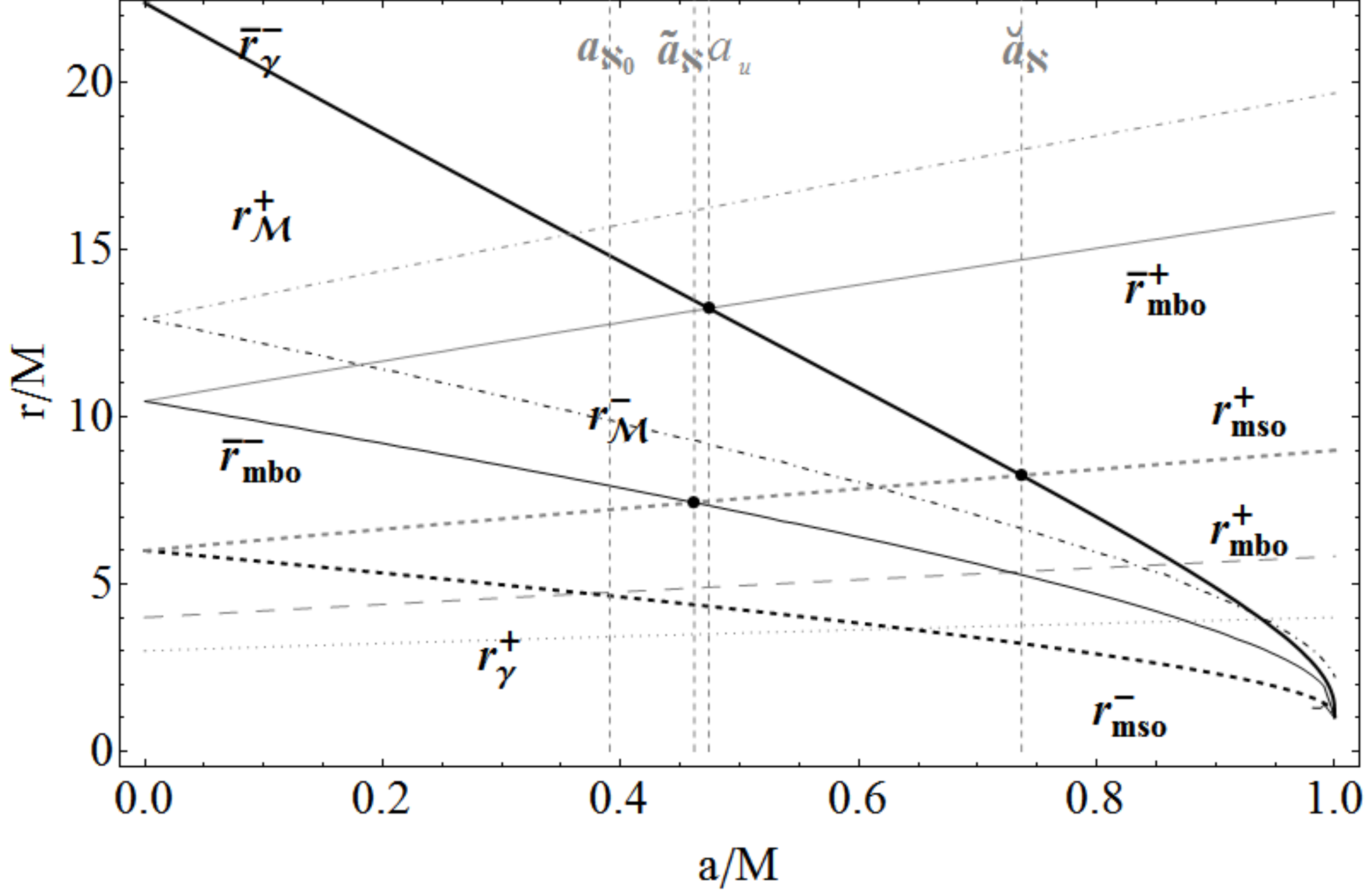}
\end{tabular}
\caption{\footnotesize{
Geodesic structure of the Kerr geometry: notable radii  $\mathbf{R}_{\mathrm{N}}\equiv \{r_{\gamma}^{\pm}, r_{\mbo}^{\pm},r_{\mso}^{\pm}\}$  and complementary geodesic structure  $\overline{\mathbf{R}}_{\mathrm{N}}\equiv \{\bar{\mathfrak{r}}_{\gamma}^{\pm}, \bar{\mathfrak{r}}_{\mbo}^{\pm},\bar{\mathfrak{r}}_{\mso}^{\pm}\}$. Some notable spacetime spin-mass ratios are also plotted,  a list can found in Table\il(\ref{Table:nature-Att}).
Orbits $\bar{r}_{\mathrm{N}}>{r_{{\mathrm{N}}}}:\; \ell(\bar{\mathfrak{r}}_{\mathrm{N}})=\ell({r_{\mathrm{N}}})\equiv\ell_{\mathrm{N}}$  where  $r_{\mathrm{N}}\in \mathbf{R}_{\mathrm{N}}$. The only solution at $a\neq0$ of  $\bar{\mathfrak{r}}_{\mathrm{N}}={r_{{\mathrm{N}}}}$ is  the    marginally stable orbit $r_{\mso}^{\pm}$ respectively. The radii $r_{\mathcal{M}}^{\pm}$ is the maximum point    of
derivative $\partial_r(\mp \ell^{\pm})$ for $a/M$ respectively.
 Right panel show a limited orbital range at $r\leq\bar{\mathfrak{r}}_{\gamma}^+$.}}\label{Fig:OWayveShowno}
%\end{tabular}
\end{figure}

The construction of the ringed configurations  is actually independent of the adopted model  for the single accretion disk (sub-configuration or ring).
However, to simplify discussion  we consider here each toroid of the ringed  disk  governed by the   General Relativity
   hydrodynamic Boyer  condition  of equilibrium configurations  of rotating perfect fluids.
  We will see that  in situations where the curvature  effects of the Kerr geometry are significant,  results are largely independent of the specific characteristics of the model for the single disk configuration,  being primarily  based on the characteristics of the   geodesic structure  the Kerr spacetime  related to  the matter distribution. This is a geometric property   consisting of the   union of the   orbital regions with boundaries  at the notable radii  $\mathbf{R}_{\mathrm{N}}^{\pm}\equiv \{r_{\gamma}^{\pm}, r_{\mbo}^{\pm},r_{\mso}^{\pm}\}$.
It can be decomposed, for $a\neq0$, into   $\mathbf{R}_{\mathrm{N}}^-$ for the corotating and   $\mathbf{R}_{\mathrm{N}}^+$ for  counterrotating matter.  %{according to the convention adopted here and in \citep{ringed}}.
Specifically, for timelike particle circular geodetical orbits, $r_{\gamma}^{\pm}$ is  the \emph{marginally circular orbit}  or  the photon circular orbit, timelike  circular orbits  can fill  the spacetime region $r>r_{\gamma}^{\pm}$. The \emph{marginally stable circular orbit}  $r_{\mso}^{\pm}$: stable orbits are in $r>r_{\mso}^{\pm}$ for counterrotating and corotating particles respectively.  The \emph{marginally  bounded circular  orbit}  is $r_{\mbo}^{\pm}$, where
 $E_{\pm}(r_{\mbo}^{\pm})=1$  \citep{Pugliese:2011xn,Pugliese:2013zma,Pugliese:2010ps,ergon,Stuchlik(1981a),Stuchlik(1981b),Stuchlik:2008fy,Stuchlik:2003dt}
 --see Fig.\il(\ref{Figs:Ptherepdiverg})  and Fig.\il(\ref{Fig:OWayveShowno}). Given $r_i\in \mathcal{R}$,  we adopt  the  following notation for any function $\mathbf{Q}(r):\;\mathbf{Q}_i\equiv\mathbf{Q}(r_i)$,  for example $\ell_{\mso}^+\equiv\ell_+(r_{\mso}^+)$ and, more generally, given the radius  $r_{\bullet}$ and the function  $\mathbf{Q}(r)$,  there is $\mathbf{Q}_{\bullet}\equiv\mathbf{Q}(r_{\bullet})$.
Since the intersection set  of $r_{\mathrm{N}}^{\pm}$ is not empty, the character  of the geodesic structure  will be particularly  relevant in  the characterization of the   $\ell$counterrotating sequences\citep{ringed}.

According to the  Boyer theory on the equipressure surfaces applied to a  P-D  torus,
the toroidal surfaces  are the equipotential surfaces of the effective potential  $V_{eff}(\ell,r)$, being solutions of  $V_{eff}=K=$constant  or $ \ln(V_{eff})=\rm{c}=\rm{constant}$   \citep{Boy:1965:PCPS:,KJA78}.  These  correspond also to the surfaces of constant density, specific angular momentum $\ell$, and constant  relativistic angular frequency  $\Omega$, where $\Omega=\Omega(\ell)$  as a consequence of the von Zeipel theorem \citep{M.A.Abramowicz,Zanotti:2014haa,KJA78}.
Then, each Boyer surface is  uniquely identified by the couple of parameters $\mathbf{p}\equiv (\ell,K)$.
We focus on the solution of Eq.\il(\ref{Eq:scond-d}), $W=$constant, associated to the critical points of the effective potential, assuming  constant specific angular momentum and parameter $K$.
Considering $\Delta_{crit}\equiv[r_{\max}, r_{\min}]$, whose boundaries correspond to the  maximum and minimum points of the effective potential respectively, we have  that
the centers $r_{cent}$  of the closed configurations $\cc_{\pm}$ are located at the minimum points  $r_{\min}>r_{\mso}^{\pm}$  of the effective potential, where the hydrostatic pressure reaches a  maximum. The toroidal surfaces are characterized by $K_{\pm}\in [K^{\pm}_{\min}, K^{\pm}_{\max}[ \subset]K_{\mso}^{\pm},1[\equiv \mathbf{K0}$ and  momentum $\ell_{\pm}\lessgtr\ell_{\mso}^{\pm}\lessgtr0$  respectively.
The inner edge  of the Boyer surface  is at $r_{in}\in\Delta_{crit}$, or $r_{in}\equiv y_3$ on the equatorial plane,
the outer edge  is at $r_{out}>r_{\min}$,  or $r_{out}\equiv y_1$ on the equatorial plane  as in Fig.\il(\ref{Figs:ApproxPlo}).   A further  matter configuration   closest to the black hole is  at $r_{in}<r_{\max}$. %In the closed cusped  configuration $\cc_{\times}$, there  is $y_3=y_2$
 The limiting case of $K_{\pm}=K_{\min}^{\pm}$ corresponds to a one-dimensional ring of matter  located in  $r_{\min}^{\pm}$.
 Equilibrium configurations, with topology $\cc$,  exist for $\pm\ell_{\mp}>\pm\ell_{\mso}^{\mp}$ centered in  $r>r_{\mso}^{\mp}$, respectively.
In general, we   denote by the  label $(i)$, with $i\in\{1,2,3\}$ respectively, any  quantity $\mathbf{Q}$ related to the range  of specific angular momentum $\mathbf{Li}$ respectively;  for example,
$\cc_1^+$ indicates a closed  regular counterrotating configuration with specific angular momentum  $\ell_1^+\in\mathbf{L1}^+$.

The local  maxima of the effective potential $r_{\max}$  correspond to minimum points of the hydrostatic pressure and the P-W  points  of  gravitational and hydrostatic instability.
No  maxima of the effective potential exist for $\pm\ell_{\mp}>\ell_{\gamma}^{\pm}$ ($\mathbf{L3}^{\mp}$) therefore, only equilibrium configurations $\cc_3$ are possible.
An  accretion  overflow of matter from the  closed, cusped  configurations in   $\cc^{\pm}_{\times}$ (see Fig.\il(\ref{Figs:ApproxPlo})) towards the attractor  can occur from the instability point  $r^{\pm}_{\times}\equiv r_{\max}\in]r_{\mbo}^{\pm},r_{\mso}^{\pm}[$, if $K_{\max}\in \mathbf{K0}^{\pm}$  with specific angular momentum $\ell\in]\ell_{\mbo}^+,\ell_{\mso}^+[\equiv\mathbf{L1}^+$  or $\ell\in]\ell_{\mso}^-,\ell_{\mbo}^-[\equiv \mathbf{L1}^-$. Otherwise,  there can be  funnels of  material along an open configuration   $\oo^{\pm}_{\times}$,  proto-jets or for brevity jets, which represent limiting topologies for the  closed  surfaces \citep{KJA78,Sadowski:2015jaa,Lasota:2015bii,Lyutikov(2009),Madau(1988),Sikora(1981)}
 with   $K^{\pm}_{\max}\geq1$ ($\mathbf{K1}^{\pm}$), ``launched'' from the point $r^{\pm}_{\jj}\equiv r_{\max}\in]r_{\gamma}^{\pm},r_{{\mbo}}^{\pm}]$ with specific angular momentum $\ell\in ]\ell_{\gamma}^+,\ell_{\mbo}^+[\equiv \mathbf{L2}^+ $ or $]\ell_{\mbo}^-,
  \ell_{\gamma}^-[\equiv\mathbf{L2}^-$.
However, we can  locate the points of maximum pressure, which correspond to the center of each torus, at $r^{\pm}_{\min}>r_{\mso}^{\pm}$
  more precisely,   by
 introducing  the   ``\emph{complementary}'' geodesic structure,  associated  to  the geodesic  structure constituted by the notable radii $\mathbf{R}_{\mathrm{N}}$, by  defining the radii  $\bar{\mathfrak{r}}_{\mathrm{N}}\in \bar{\mathbf{{R}}}_{\mathrm{N}}:$ $\bar{\mathfrak{r}}_{\mathrm{N}}>{r}_{\mathrm{N}}$ solutions of $\bar{\ell}_{\mathrm{N}}\equiv\ell(\bar{\mathfrak{r}}_{\mathrm{N}})=\ell(r_{\mathrm{N}})\equiv\ell_{\mathrm{N}}$--see Fig.\il(\ref{Figs:Ptherepdiverg}) and Fig.\il(\ref{Fig:OWayveShowno}).
 These radii satisfy the same equation as are  the  notable radii $r_{\mathrm{N}}\in \mathbf{R}_{\mathrm{N}}$ for corotating and counterrotating configurations,
 analogously  to the couples  $r_{\mathcal{M}}^{\pm}$  and  $\bar{\mathfrak{r}}_{\mathcal{M}}^{\pm}$ where $r_{\mathcal{M}}^{\pm}>r_{\mso}^{\pm}$, where associated $\ell_{\mathcal{M}}^{\pm}$,  is a maximum of $\partial_r |\ell(r)|$--\cite{open}.
The geodesic structure of spacetime and the complementary geodesic structure are both significant in the analysis,  especially in the case of $\ell$counterrotating  couples.
There is  $r_{\gamma}^{\pm}<r_{\mbo}^{\pm}<r_{\mso}^{\pm}<\bar{\mathfrak{r}}_{\mbo}^{\pm}<\bar{\mathfrak{r}}_{\gamma}^{\pm}$,
the location of the radii $r_{\mathcal{M}}$ and $\bar{\mathfrak{r}}_{\mathcal{M}}$ depends   on the rotation with respect to the Kerr attractor.
 Clearly the marginally stable orbit $r_{\mso}$ is the  only  solution of   ${r}_{\mathrm{N}}=\bar{\mathfrak{r}}_{\mathrm{N}}$. Thus the configurations $\pp_1$ are centered in $]r_{\mso},\bar{\mathfrak{r}}_{\mbo}[$ (with accretion point in $r_{\times}\in]r_{\mbo},r_{\mso}[$), the  $\pp_2$ rings have centers in the range  $[\bar{\mathfrak{r}}_{\mbo},\bar{\mathfrak{r}}_{\gamma}[$  (with $r_{\jj}\in]r_{\gamma}, r_{\mbo}[$), finally the  $\cc_3$ disks are centered at $r\geq \bar{\mathfrak{r}}_{\gamma}$.

However, a global instability of the entire macro-configuration  may be associated to
 two distinct models  of unstable ringed torus with degenerate topology.   Related to these there are two  types of instabilities emerging in an orbiting macro-structure.  First, the emergence of a P-W instability in one of its ring  and the collision among the sub-configurations.
 The P-W  local  instability affects  one or more    rings of the ringed disk  decomposition, and  then it can  destabilize  the macro-configuration when the rings are   no more separated and a feeding (overlapping)   of material  occurs. Second,
   a contact (or \emph{geometrical correlation}) in this model causes  collision and  penetration of matter, eventually with the feeding of one sub-configuration with material and supply of specific angular momentum of  another consecutive  ring of the decomposition. This mechanism  could possibly end  in a change of   the ringed disk  morphology and topology.  Accordingly,  there is the macro-structure
$\mathbf{C_{\odot}^n}$, with the number $\mathfrak{r}\in[0,n-1]$  of contact points between the boundaries of two consecutive rings (\emph{rank}  of the
$\mathbf{C_{\odot}^n}$),   and the macro-structure $\mathbf{\cc^n_{\times}}$, with    $\mathfrak{r_{\times}}\in[0,n]$  instability P-W points. The number  $\mathfrak{r_{\times}}$ is called \emph{rank} of the ringed disk  $\mathbf{\cc^n_{\times}}$. Finally, we have  the macro-structure $\mathbf{\cc_{\odot}^x}^n$, characterized   at lest by one  contact  point that is also an instability point.

If    $\mathfrak{r}_{\times}=1$  and     the inner ring $\mathbf{C}_{\times}^1$   of its decomposition is in accretion, then  the whole  ringed disk could  be globally stable \citet{ringed}.
We  shall  describe the system made up by two tori  in a Kerr geometry as a ringed accretion disks  $\mathbf{C}^2$ of the order $n=2$ (\emph{state})-Fig.\il(\ref{Table:Torc}).
We can introduce
 \emph{elongation}  $\Lambda_{\mathbf{C}^2}$ of $\mathbf{C}^2:\; \cc_a<\cc_b$   and the  \emph{spacing} $\bar{\Lambda}_{2,1}\equiv [y_1^a,y_3^b]$ by the relations
 \be\label{Eq:chan-L-corby}
  \Lambda_{\mathbf{C}^2}\equiv[y_3^a, y_1^b]=
\left(\bigcup_{i=1}^{2}{\Lambda}_{i}\right)+\bar{\Lambda}_{2,1},\quad\lambda_{\mathbf{C}^2}\equiv y_1^b-y_3^a=\sum_i^2 \lambda_i+\bar{\lambda}_{2,1}\geq \lambda_{\mathbf{C}^2}^{inf}\equiv\left.\sum_i^2 \lambda_i\right|_{\sum_i\lambda_i}, %=
%{\Lambda}_{2}\cup\left(\bar{\Lambda}_{2,1}\cup{\Lambda}_{1}\right),
 \ee
where  $\bar{\Lambda}_{i}$  and  ${\Lambda}_{i}$ are the spacing and elongation   of each ring and
$\lambda_{\mathbf{C}^2}$ is the measure of   the  elongation of the  (separated) configuration  $\mathbf{C}^2$--see Fig.\il(\ref{Figs:ApproxPlo}).
  Equation (\ref{Eq:chan-L-corby}) shows that the minimum value $\lambda_{\mathbf{C}^2}^{inf}$ of the elongation $\lambda_{\mathbf{C}^2}$ is achieved, \emph{at fixed $\sum_i^n \lambda_i$}, when $\bar{\lambda}_{2,1}=0$ that is for a $\mathbf{C}^{2}_{\odot}$ configuration of rank $\mathfrak{r}=\mathfrak{r}_{\max}$.
%
%
%
%\medskip
As demostrated  in \cite{ringed}, we can introduce
 the effective potential $\left.V_{eff}^{\mathbf{C}^2}\right|_{K_i}$ of the  \emph{decomposed} $\mathbf{C}^n$ macro-structure  and {the effective potential  $V_{eff}^{\mathbf{C}^2}$ of the configuration}:
\bea\label{Eq:def-partialeK}
\left.V_{eff}^{\mathbf{C}^2}\right|_{K_i}\equiv V_{eff}^{1}\Theta(-K_1)\bigcup V_{eff}^{2}\Theta(-K_2)\quad\mbox{and}\quad
%\\
%\label{Eq:Vcomplessibo}
V_{eff}^{\mathbf{C}^2}\equiv V_{eff}^{i}(\ell_i)\Theta(r_{\min}^{o}-r)\Theta(r-r_+)\bigcup V_{eff}^{o}(\ell_o)\Theta(r-r_{\min}^{i}),
\eea
where $\Theta(-K_i)$ is the Heaviside (step) function  such that $\Theta(-K_i)=1$  for $V_{eff}^{i}<K_i$ and $\Theta(-K_i)=0$ for $V_{eff}^{i}>K_i$, so that the curve $V_{eff}^{\mathbf{C}}(r)$ is the union of all curves $V_{eff}^{i}(r)<K_i$ of its decomposition.
 Potential  $\left.V_{eff}^{\mathbf{C}^2}\right|_{K_i}$ regulates  behavior of each ring, taking into  account the  gravitational effects induced by the background, and  the centrifugal effect induced by the motion of the fluid, while the potential  $V_{eff}^{\mathbf{C}^2}$  governs the individual configurations considered as part of the macro-configuration--Fig.\il(\ref{Figs:ApproxPlo}). Details on the effective potential,
   definition of   differential rotation of the decomposition, specific angular momentum of the  ringed  disk and  also for the thickness on the ringed disk can be found in \cite{ringed}, where
 these configurations were first introduced, and then detailed in  \cite{open} for a configuration order $n\geq2$.
Here, we specialize the  introduced concepts  to the case of only two rings.
In Sec.\il(\ref{Sec:basic-intro}) we characterize   the  double accretion disk system, focusing in Sec.\il(\ref{Sec:lc-or}) on the $\ell$corotating couples, while in Sec.\il(\ref{Sec:lcounterrsec}) we discuss the case of  $\ell$counterrotating couples.

To simplify and illustrate the  discussion,   we use special graphs  representing a couple of accretion disks and their evolution within  the constraints   they are subjected to. The case of a  couple of tori orbiting around a single central   Kerr black hole involves  in general  a remarkably  large number of possible configurations:  for a couple with fixed and equal  critical topology, there could be  $n=8$ different states according to their rotation and   relative position of the centers. The couple $({\cc_{\times},\oo_{\times}})$, with different but fixed topology,  could  be in $n=16$ different states,
 while for the state  $\cc_i-\pp_{\times}$,  with one equilibrium topology,   we need to address  $n=48$ different cases--{see Figs.\il(\ref{Figs:new-sun-graph}) for a sample of cases.
\begin{figure}[h!]
\begin{center}
\begin{tabular}{cc}
\includegraphics[width=0.5\columnwidth]{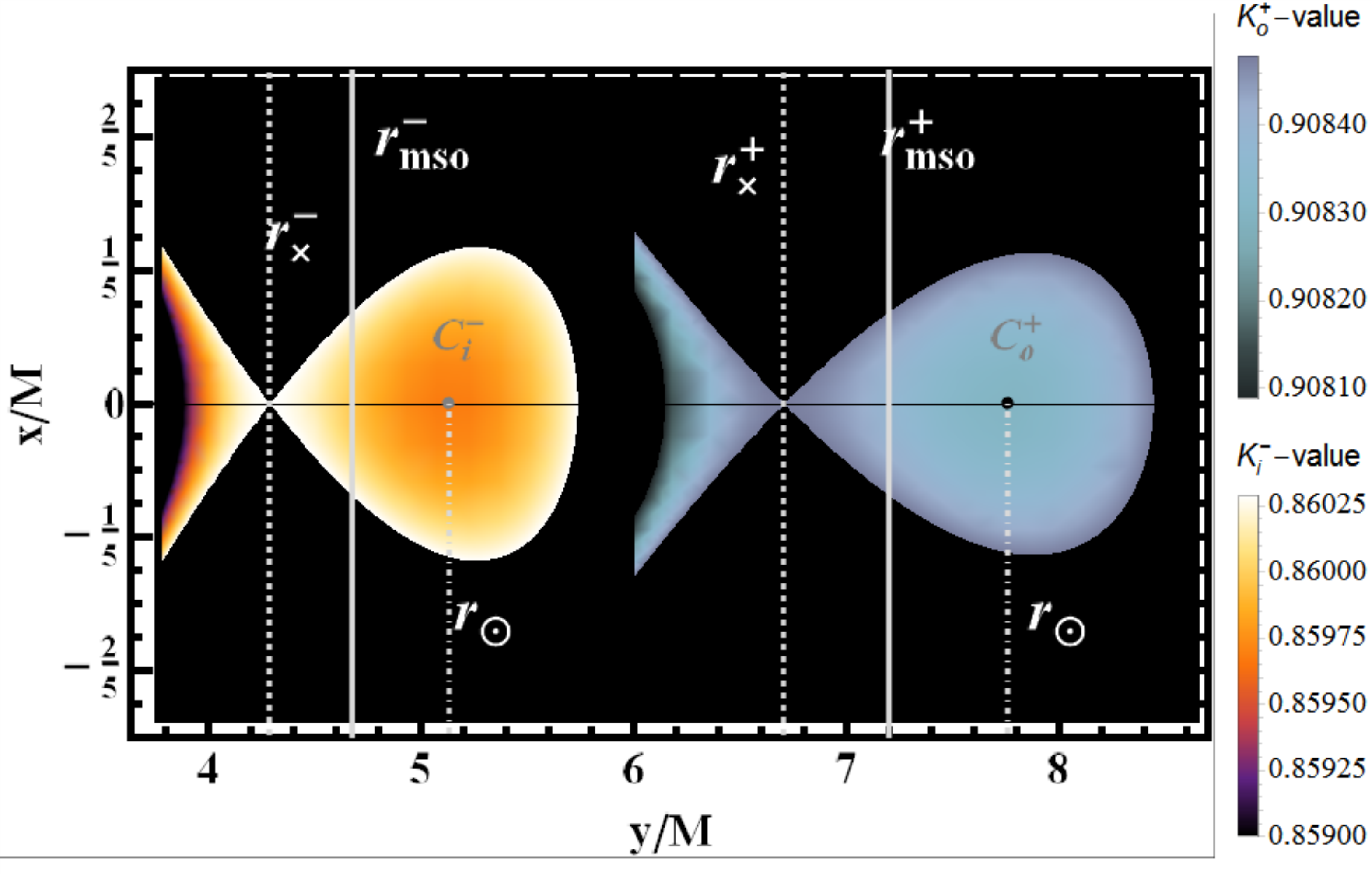}
\includegraphics[width=0.5\columnwidth]{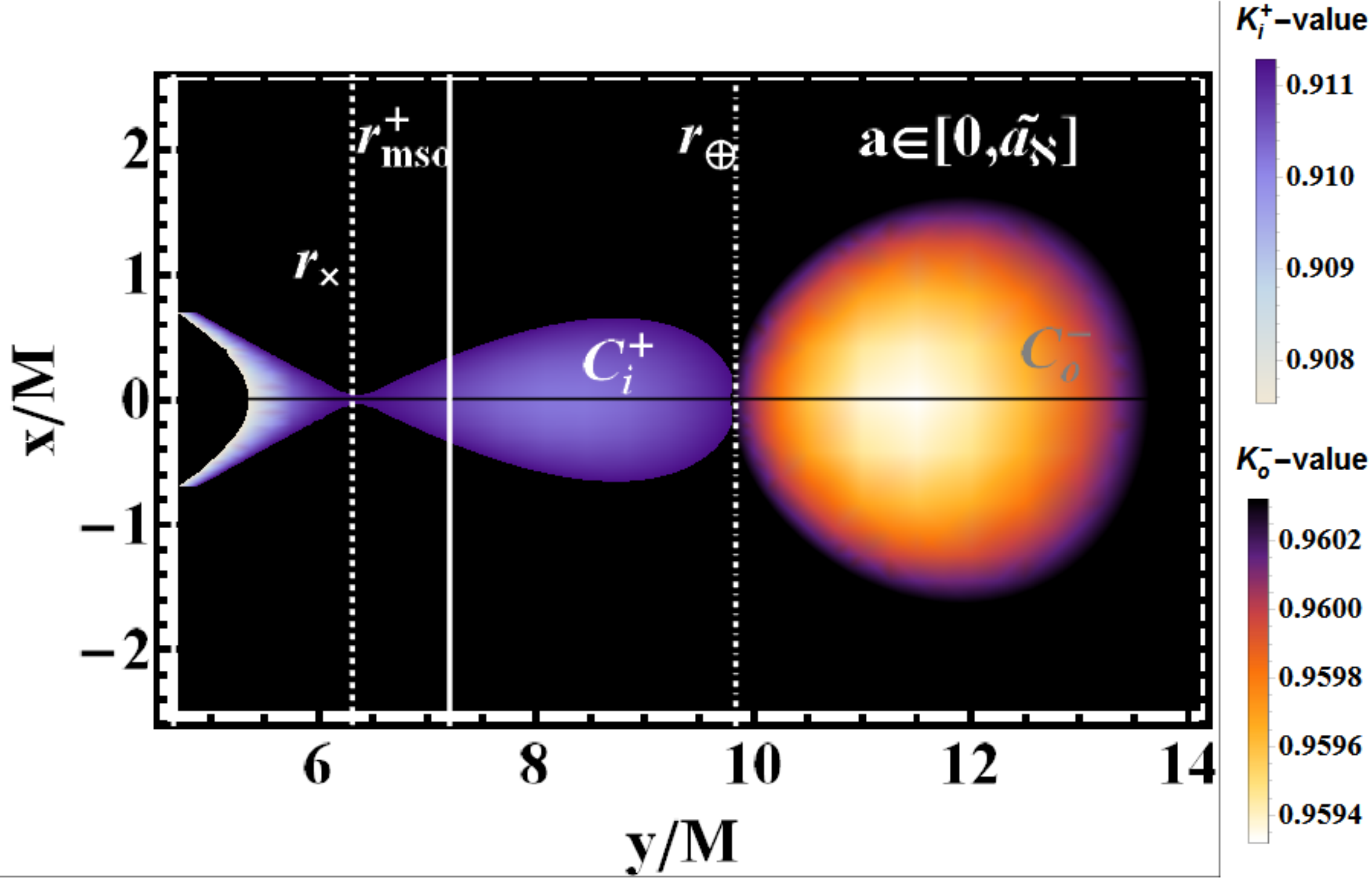}\\
\includegraphics[width=0.5\columnwidth]{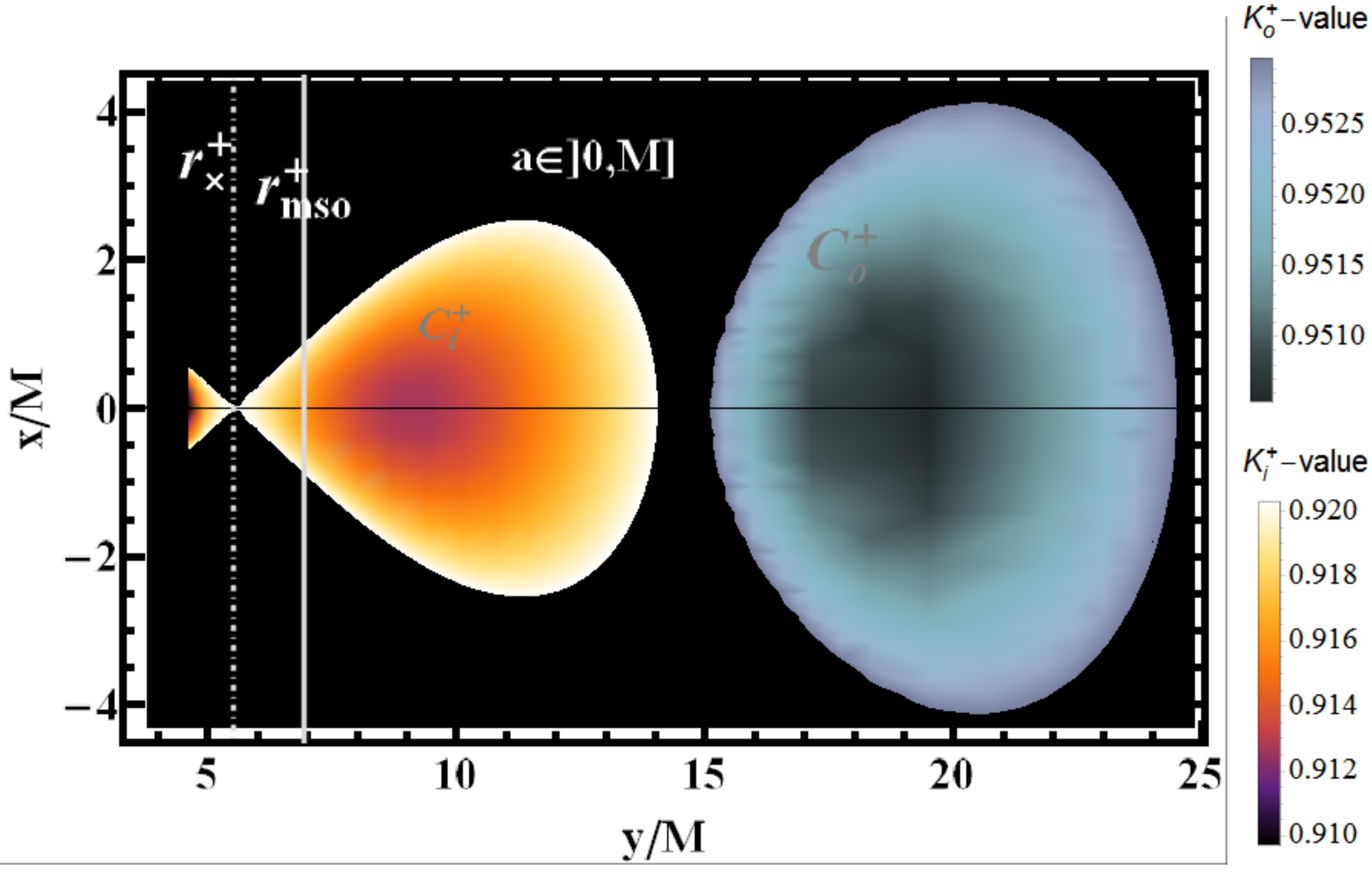}
\includegraphics[width=0.5\columnwidth]{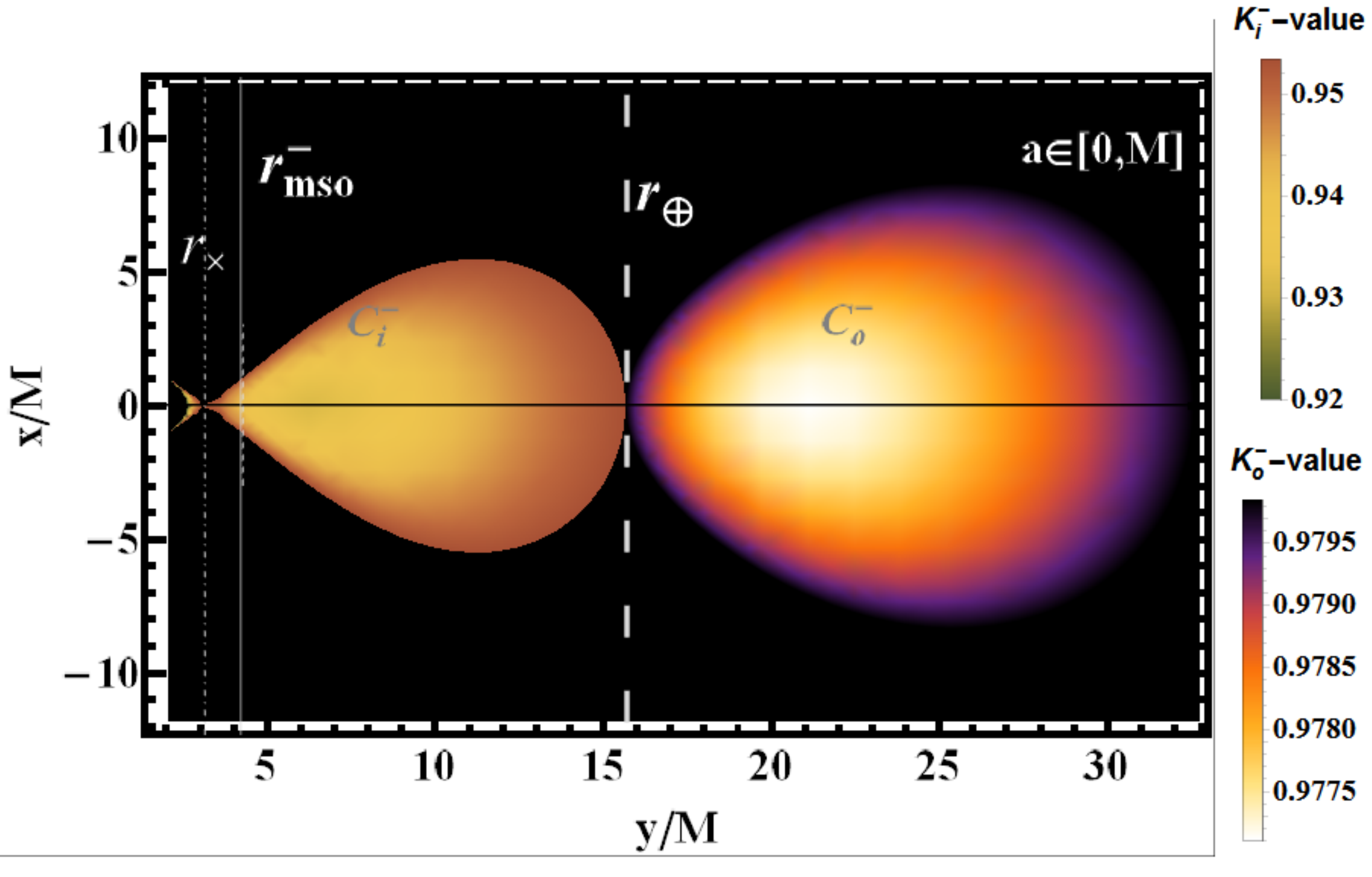}
\end{tabular}
\caption{Density plot. \emph{Upper left}: The $\ell$counterrotating colliding couple $\cc_{\times}^-<
\cc_{\times}^+$, where
$\ell_+ = -3.99$, $\ell_- = 3.31$
and $a = 0.3825M$. Integration has been truncated prior collision. \emph{Upper right}: $\ell$counterrotating
couple $\cc_{\times}^+<\cc^-$, the contact point $r_{\otimes}$ is also signed where $a = 0.385M$, $\ell_- = 4.1$ and $\ell_+ =- 4.01$. \emph{Bottom
left} : $\ell$counterrotating couple $\cc_{\times}^+<\cc^+$, where $a = 0.3M$, $\ell_i = -4.01$, $\ell_o = -4.9$. \emph{Bottom right} : Colliding
$\ell$corotating tori, $\cc_{\times}^-<\cc^-$, where $a = 0.5M$, $\ell_o= 5$,  $\ell_i= 3.3$. $(x, y)$ are Cartesian coordinates.}\label{Figs:new-sun-graph}
\end{center}
\end{figure}}

\begin{figure}[h!]
\centering
\begin{tabular}{c}
\includegraphics[width=.71\textwidth]{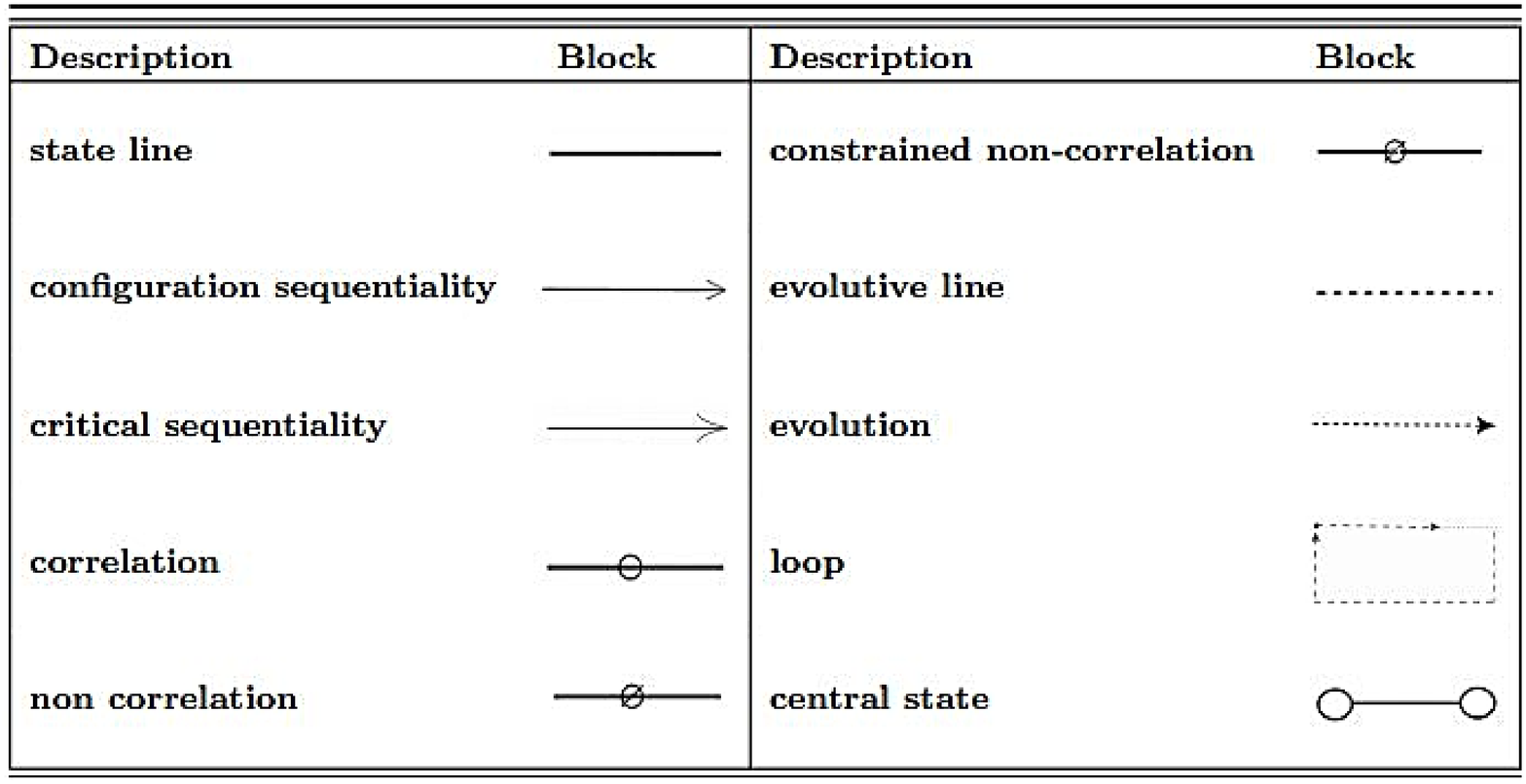}
\end{tabular}
\caption{Graphs construction. Main blocks used in graphs of Figs\il(\ref{Fig:doub-grap-ll-cor}) and  (\ref{Fig:CC-CONT})--see Sec.\il(\ref{Sec:Taa-DISK}). Further details can be found in Sec.\il(\ref{Sec:graph-app}).
}\label{Table:Graphs-models}
%\end{tabular}
\end{figure}

  The use of {graphic} schemes is crucial for the representation of these cases, to quickly collect the different constraints  on the existence and evolution of the states and  for reference in our discussion.  Therefore, although  the following analysis  is quite independent  from the graph formalism, for easy  reference,  we include  here a brief description of this formalism and  discussion on the graph  construction, introducing   the
essential blocks composing  the graphs used in this work, and the list of notations and basic concepts related to these structures.
We refer to  Sec.\il(\ref{Sec:graph-app}) for details on the construction and interpretation of graphs  associated with these systems,
 while
in Fig.\il(\ref{Table:Graphs-models}) we present the main blocks  the  graphs are made of, with a brief description which   provides also a list of the  main notation and definitions  used throughout this work.
\medskip

\textbf{List of principal notation in the graph construction with reference to Fig.\il(\ref{Table:Graphs-models}).}

\medskip

A graph \textbf{vertex}  represents  one configuration of the couple of  tori as defined  by the ringed  disk   topology  and  fluid rotation with respect to the central Kerr black hole attractor; then a vertex stands for  one  configuration of the  set   $\pp^{\pm}=\{\cc^{\pm}, \cc^{\pm}_{\times}, \oo^{\pm}_{\times}\}$. The  \textbf{State lines } connect two vertexes of the graph  and represent a fixed couple of accretion tori disks.
A \textbf{monochromatic graph}  has one monochromatic state i.e. a state line connecting two $\ell$corotating configurations.
A \textbf{bichromatic graph } has one bichromatic states i.e a  state line connecting two $\ell$counterrotating  configurations.
For  \textbf{configuration sequentiality}, signed on a state line and
associated with  the notation $\mathbf{<}$ or $\mathbf{>}$,  we intend  the ordered sequence of  maximum points of the pressure, or $r_{\min}$, minimum of the effective potential which corresponds to  the configuration centers. Therefore, in relation to a couple  of rings,   the terms ``internal'' (inner-$i$) or ``external'' (outer-$o$), will always refer, unless otherwise specified,  to the   sequence ordered according to the center location.
For  \textbf{critical sequentiality}, attached to a state line and
associated with
symbols $\mathbf{\succ}$ and $\mathbf{\prec}$,  we refer  to the sequentiality according to the  location of the \emph{minimum} points of the pressure,   or $r_{\max}$, maximum point  of the effective potential (in $\mathbf{L1}$ or $\mathbf{L2}$).
 A state line is completely oriented if both the configuration   and  critical sequentiality are specified, when the last  one may be defined.
 Two configurations are \textbf{correlated} if they can be in contact, which implies  collision in accordance with the  constraints. In some cases  there are  particularly restrictive conditions to be satisfied for a correlation
to occur (constrained non correlation).
The addition of specific information on the lines and vertices of the graph, for example, the color the correlation and  sequentiality is  called graph  \emph{decoration}.
 An \textbf{ evolutive line} connects two vertexes of two different state lines of the graph, and it represents the evolution of one configuration from one (starting) topology   (vertex)  to another topology (a  vertex of a different state), for example from a $\cc$ configuration to a $\cc_{\times}$ in accretion. Evolutive lines    may be composed  to be closed on an initial  vertex of the initial state lines creating  a  \textbf{loop}--Sec.\il(\ref{App:deep-loop}).
  A \textbf{central} state  of the graph is the couple,   the graph configurations describes the evolution towards different states (every evolutive lines starts, ends or passes trough the central state). In this work the central state is  the initial state line according to the evolution signed by the evolutive lines.
Further details can be found in Sec.\il(\ref{Sec:graph-app}).   State lines for  $\ell$corotating  couple are listed in  Fig.\il(\ref{Table:statesdcc}) and state lines for  $\ell$counterrotating couples are in Fig.\il(\ref{Table:statescc}).  Fig.\il(\ref{Fig:doub-grap-ll-cor}) describes monochromatic graphs while   Fig.\il(\ref{Fig:CC-CONT}) exploit the  bichromatic graphs.
%
%%
%\begin{table}[h!]
%\caption{Graphs construction. Main blocks used in graphs of Figs\il(\ref{Fig:doub-grap-ll-cor}) and  (\ref{Fig:CC-CONT})--see Sec.\il(\ref{Sec:Taa-DISK}). Further details can be found in Sec.\il(\ref{Sec:graph-app}).
%}\label{Table:Graphs-models}
%\centering
%\resizebox{.71\textwidth}{!}{%
%\begin{tabular}{|ll|ll|}
%\toprule\hline \hline
%\textbf{{Description}} &$\;$ \textbf{Block} & \textbf{Description} & $\;$\textbf{Block}\\
%\hline
%\textbf{state line} & \raisebox{-4ex}{\includegraphics[width=.1\textwidth]{state-line1}}&
%\textbf{constrained non-correlation } &\raisebox{-4ex}{\includegraphics[width=.1\textwidth]{non-corr-rest}}
%\\
%\textbf{configuration sequentiality} &\raisebox{-4ex}{\includegraphics[width=.1\textwidth]{order-conf}}&\textbf{evolutive line} & \raisebox{-4ex}{\includegraphics[width=.1\textwidth]{Ev-line}}
%\\
%\textbf{critical sequentiality} & \raisebox{-4ex}{\includegraphics[width=.1\textwidth]{critical-seq}}&
%\textbf{evolution} &\raisebox{-4ex}{\includegraphics[width=.1\textwidth]{evol-order}}
%\\
%\textbf{correlation} &\raisebox{-4ex}{\includegraphics[width=.1\textwidth]{corr-state}}&\textbf{loop}& \raisebox{-4ex}{\includegraphics[width=.1\textwidth]{loop}}
%\\\textbf{non correlation}& \raisebox{-4ex}{\includegraphics[width=.1\textwidth]{not-corrstate}}&\textbf{central state}& \raisebox{-4ex}{\includegraphics[width=.1\textwidth]{central-state}}\\
%\hline\hline
%\end{tabular}}%,width=71mm,scale=.51angle =-90
%\end{table}
%
%%%%%%%%%%%%%%%%%%%%%%%%%%%%%%%%%%%%%%\input{D-Sys-Analysis}
\section{Characterization  of the  double tori disk system}\label{Sec:basic-intro}
In this Section we discuss the existence and the stability of  the ringed disk $\mathbf{C}^2$  of the   order $n=2$, made up by   two toroidal configurations  orbiting a spinning black hole attractor.

We first consider all the possible  states for the  couple  of accretion disks   with  fixed  topology.
In the graph formalism  their  analysis is representing  research  of all  the   possible  state and evolutive lines  and their  decoration (see Fig.\il(\ref{Table:Graphs-models}) and end of Sec.\il(\ref{Sec:Taa-DISK})) according to the separation constraint\footnote{Separated  tori  are defined,   for a $n$-order macro-structure $\mathbf{C}^n=\bigcup_1^n \cc_{i}$, according to the conditions
$
\cc_i\bigcap \cc_j=0$ and $\partial \cc_i\bigcap \partial \cc_j=\{\emptyset,y_1^i=y_3^j\}$ where $i<j$.
Particularly for $n=2$, a double configuration, ${C}_{i}\cap {C}_{o}=\emptyset$ or those with  $y_1^{i}=y_3^{o} $ where  the outer edge of the inner rings ($i$) coincides with the inner  edge of the outer ring ($o$).
 In other words for macro-configurations made by  separated tori, the penetration  of  a ring within another ring  is  not possible. However, as the condition $y_1^{i}=y_3^{o} $  can hold,   in a limit situation  the  collision of matter between the two surfaces at contact point $y_1^i=y_3^j$ could be possible\cite{ringed}.
}.
We refer to  Sec.\il(\ref{Sec:graph-app}) for details on the construction and interpretation of graphs associated with these systems.

We shall prove that some states, or some decorations for a state are  prohibited by several conditions,  determined   mainly by the dimensionless spin of the attractor  and   by the separation constraint. Specifically, we discuss the evolution of the configurations towards the phase of accretion onto the attractor  which could lead to violate   the separation condition. We  study the collisions between the rings of the couple setting  the
 emergence of the $\mathbf{C}_{\odot}^{2}$ (critical) macro-configuration,  causing eventually
the  rings merging.

The states  could be further constrained by the  maximum possible extension of the closed configurations for  fixed angular  momentum,  defined by  the supremum of parameter $K$, $\sup{K}$. It is clear that    for the $\cc_1$ configurations we should consider  the maximum of the  elongation at the accretion  $\lambda_{\times}$ and for the $\cc_2$ disks the superior for  $K_{\max}=1$. On the other hand, there is no similar constraint for the  $\cc_3$ configurations since there are  no minimum points of the hydrostatic pressure.
 However,  we can infer
  the presence of the constraints in terms of the  location of the inner and outer edge of the torus with  respect to the  notable radii by considering  the results of  \cite{open}.

In Sec.\il(\ref{Sec:lc-or}), we will show how a  monochromatic graph, generally describing the situation for a $\ell$corotating couple in any Kerr spacetime with  $a\in[0,M] $ also  describes the states and the evolution of  a  $\ell$counterrotating couple orbiting a  Schwarzschild attractor ($a=0$), due to the particular geodesic structure of this  static  spacetime.
Fig.\il(\ref{Table:statesdcc}) shows the  possible state lines  for the
$\ell$corotating couples,
while the possible state lines for the   $\ell$counterrotating couples  {in a Kerr spacetime} are listed in Fig.\il(\ref{Table:statescc}).
 Table\il(\ref{Table:REDUCtIN}) also  provides guidance on the sequentiality of the $\ell$counterrotating  couples according to criticality and the configuration order.
The decorations of state lines  show generally  the emergence of possible collisions in accordance with the criteria used in the construction of the table, the location of the tori and the possible relation between the critical points.

Restricting our study to   $\mathbf{C}^2$ configurations, we concentrate our attention onto the classification of the configurations with specific angular momenta  $\ell\in \mathbf{Li}$ with $i\in\{1,2\}$--\cite{ringed}. Some of these ringed disks are  constrained to a configuration order $n_{\max}=2$.
\begin{enumerate}
\item
The configuration:
\bea\label{C0}
\bar{\mathfrak{C}}_0:\;\Delta_{cri}^{i}\cap\Delta_{cri}^{o}=\emptyset,\quad\mbox{there is}\quad a\neq0,\; \ell_i\ell_o<0,\;
\pp_i^-<\pp_o^+\quad\mbox{with}\quad   \pp_i^-\prec \pp_o^+,\quad n_{\max}(\bar{\mathfrak{C}}_0)=2,
\eea
 and we have:
 \bea\label{Eq:(15)}
&& -\ell^o_+\in]-\ell_{\mso}^+,-\ell_{+}(r_{\mso}^-)[
\quad\mbox{and}\quad\ell^i_-\in]\ell_{\mso}^-,\ell_-(r_{\mso}^+)[.
 \\
&& \mbox{For }a\in]0,a_{\aleph_1}[:\quad \pp_o^+\neq \oo_{\times}^+\quad\mbox{for}\quad r_{\mbo}^+<r_{\mso}^-
 \quad
\mbox{where}\quad a_{\aleph_1}:\; r_{\mbo}^+=r_{\mso}^-.
 \eea
Table\il(\ref{Table:nature-Att}) lists and  summarizes the main features of the spin values  singled out  by analysis.

 Eq.\il(\ref{Eq:(15)}) is fulfilled  for the following topologies:
at $a<\tilde{a}_{\aleph}$  for $\pp^-=\pp_1^-$
 and at  $a <a_{\iota_a}<\tilde{a}_{\aleph}$
there could be only $\pp_1^-<\pp_1^+$--Fig.\il(\ref{Fig:CC-CONT}).
Then, in  $[a_{\iota_a},\tilde{a}_{\aleph}]$,
there is $\pp_1^-<\pp_2^+$ and   $\pp_1^-<\pp_1^+$.
Whereas at
$a\in[\tilde{a}_{\aleph},a_{\gamma_+}^-]$ there is  also
$\pp_2^-< \pp_2^+$, and
in
$[a_{\gamma_+}^-,\breve{a}_{\aleph}]$
there is also
$\pp_2^-<\pp_3^+$. Finally  for $a>\breve{a}_{\aleph}$  also the couple $\pp^-_3<\pp_3^+$ is possible.
These constraints, however, are not sufficient to fully characterize the couples $\pp^-<\pp^+$ as discussed in Sec.\il(\ref{Sec:tex-dual}), in fact  not all the couples $\pp^-<\pp^+$ belong to the $\bar{\mathfrak{C}}_0$ class.
%%\textbf{Elongation}
%
%
\begin{table*}[ht!]
\caption{Classes of attractors.
In general given a spin value $a_{\bullet}$ the classes $\mathbf{A}_{\bullet}^{\lessgtr}$ stands for  the ranges $a\in[0, a_{\bullet}[$ and $a\in] a_{\bullet},M]$ respectively. Some of these classes are given alongside the spins.
}\label{Table:nature-Att}
\centering
%\resizebox{1\textwidth}{!}{%
\begin{tabular}{|l|l|l|}
\textbf{{{Spins and classes of attractors}}}
& & \\
\hline
${a_{\aleph_2}}\equiv 0.172564M:-\ell_{\mso}^+=\ell_{\mbo}^-$
&
${a_{\iota}}\equiv0.3137M:r_{\mbo}^-=r_{\gamma}^+$-$\left({\mathbf{A}}_{\iota}^{\lessgtr}\right)$
&
${a_{\iota_a}}\equiv0.372583M:r_{\mso}^-= r_{\mbo}^+ $ -$\left(\mathbf{A}_{\iota_a}^{\lessgtr}\right)$
\\
$a_{\aleph_1}=0.382542M:\ell_{\gamma}^-=-\ell_{+}\left(r_{\mso}^-\right)$
&
${a_{\aleph_0}}\equiv0.390781M:\ell_{\gamma}^-=-\ell_{\mbo}^+$
&
 $ {\tilde{a}_{\aleph}}\approx 0.461854M:\ell_-(r_{\mso}^+)=\ell_{\mbo}^-$
%${\breve{a}_*}\equiv 0.401642 M:\breve{\ell}_*=\ell_{\gamma}^-$
%  $(\mathbf{\breve{A}^{\lessgtr}_*})$
  \\
$a_{{}_u}=0.474033M:\; \bar{\mathfrak{r}}_{\mbo}^+=\bar{\mathfrak{r}}_{\gamma}^-$
 &
 ${a_{\aleph}}\approx0.5089M:-\ell_{\mso}^+=\ell_{\gamma}^-$
 &
 ${a_{\gamma_+}^-}\equiv 0.638285 M:r_{\gamma}^+=r_{\mso}^-$
%     ${a_{\iota}^*} =0.618034 M:\breve{\ell}_{+}=\breve{\ell}_{2_+}^-$
     \\
 %    ${a_{\gamma_-}^{\beta}}\equiv0.628201 M:\ell_{\beta}^-=\ell_{\gamma}^-$
${{a}_1}\approx0.707107M:r_{\gamma}^-=r_{\epsilon}^+$
&
 ${\breve{a}_{\aleph}}=0.73688M:\ell_-(r_{\mso}^+)=\ell_{\gamma}^-$
&
%  ${a_{\gamma_-}^{\Gamma}}\equiv0.777271M: \ell_{\Gamma}^-=\ell_{\gamma}^-$
   ${{a}_b}\approx0.828427M:r_{\mbo}^-=r_{\epsilon}^+$
   \\
    ${a_{\mathcal{M}}^-}\equiv 0.934313M:\ell_{\gamma}^-=\ell_{\mathcal{M}}^-$
  &
    ${a_2}\approx0.942809M: r_{\mso}^-=r_{\epsilon}^+$
    &
${\breve{a}}\equiv 0.969174M:\breve{\ell}^-=r_{\gamma}^-$ $\left(\mathbf{\breve{A}_{\lessgtr}}\right)$
\\
\hline
\end{tabular}%}
\end{table*}
 \item
  The configuration:
  \bea\label{C1a}
  \bar{\mathfrak{C}}_{1a}:&&\;r_{\max}^{o}\in\Delta_{crit}^{i},\quad\mbox{thus}\quad \ell_i\ell_o<0\quad
  \pp_i^-\lessgtr \pp_o^+ \quad \pp_o=!\cc_o\quad n_{\max}(\bar{\mathfrak{C}}_{1a})=2
  \\\label{Eq:so1}
 && \mbox{if}\quad
\pp_o^-<\pp_o^+:\quad\mbox{then}\quad
%r_{\max}^-< r_{\max}^+<r_{\min}^-<r_{\min}^+,\quad
 \pp_i^-\prec \pp_o^+\quad \pp_o^+=!\cc^+\quad   n_{\max}(\bar{\mathfrak{C}}_{1a})=2
  \\\label{Eq:so2}
  &&\mbox{if}\quad
\pp_i^+<\pp_o^-:\quad\mbox{then}%\quad r_{\max}^+< r_{\max}^-<r_{\mso}^-<r_{\mso}^+<r_{\min}^+<r_{\min}^-
\quad \pp_o^-=!\cc^-\quad \pp_i^+\prec \pp_o^-:\quad(a\gtrapprox0).
\eea
Here, for any relation $\mathbf{\Large{\bowtie}}$  among two quantities, in  $\bowtie !$ the intensifier $(!)$  a reinforcement of the relation, indicates that this is  a necessary relation which is  \emph{always} satisfied.
 \item
The configuration:
\bea \label{C1b}
\bar{\mathfrak{C}}_{1b}:\quad \Delta_{crit}^{i}\subset\Delta_{crit}^{o}\quad\mbox{then there is } \ell_i\ell_o\lessgtr0\quad{{\textbf{\pp}}}_i<\pp_o\quad\pp_i\succ \pp_o \quad\pp_o=!\cc_o\quad   n_{\max}(\bar{\mathfrak{C}}_{1b})
=\infty.
\eea
A special  case of this class of ringed disks are the couples $\ell_{(i+1)/i}=-1$,  which can have
$\ell_-/-\ell_+\gtrless1$.
 \end{enumerate}

We finally note that,
in a Kerr spacetime $(a\neq0)$,  the  chromaticity of the graphs is determined by the  relative rotation of the disks  together with the rotation  with respect to  the attractor, however  the situation is different in the case of static limit for the attractor with $a=0$, where monochromatic graphs describe also  $\ell$counterrotating (i.e. $\ell_i\ell_o<0$)  (and $\ell$corotating $\ell_i\ell_o>0$) couples.
 In  fact, in the case of  a Schwarzschild attractor (static spacetime),   it is still possible to consider a {bichromatic} graph with an arbitrary choice of  tori  relative rotation $\ell_i\ell_j<0$, but  the  spacetime geodesic structure    is singled out by the properties of the Schwarzschild geometry, independently of  the sign of the fluid angular momentum. Therefore  at all the effects this bichromatic graph must undergo the analysis on the  monochromatic graph.
However, a major difference  between a bichromatic graph where $a=0$ and the monochromatic one occurs  in the static spacetime for the $\ell$counterrotating  case due to the  possible evolutive {loops} of the bichromatic vertices, where   collision between tori with $\ell$counterrotating angular momentum may occur.
In  the following Sec.\il(\ref{Sec:lc-or}) we specialize the discussion for the Schwarzschild geometry and the  $\ell$corotating couples  in the Kerr spacetimes,  while the $\ell$counterrotating couples orbiting a   Kerr attractor are analysed in Sec.\il(\ref{Sec:lcounterrsec}).
\subsection{The $\ell$corotating couples in the Kerr spacetime and the case of the  Schwarzschild geometry}\label{Sec:lc-or}
 Two $\ell$corotating tori must have different specific angular momentum, i.e.,  $\ell_o/\ell_i\equiv \ell_{o/i}>1$.
 They  have to   be both corotating   or counterrotating   with respect  to the central black hole,   as in  \textsl{{scheme I}} and \textsl{{II}} of Fig.\il(\ref{Table:Torc}).
\begin{figure}[h!]
\centering
\begin{tabular}{cc}
\includegraphics[width=.42\textwidth]{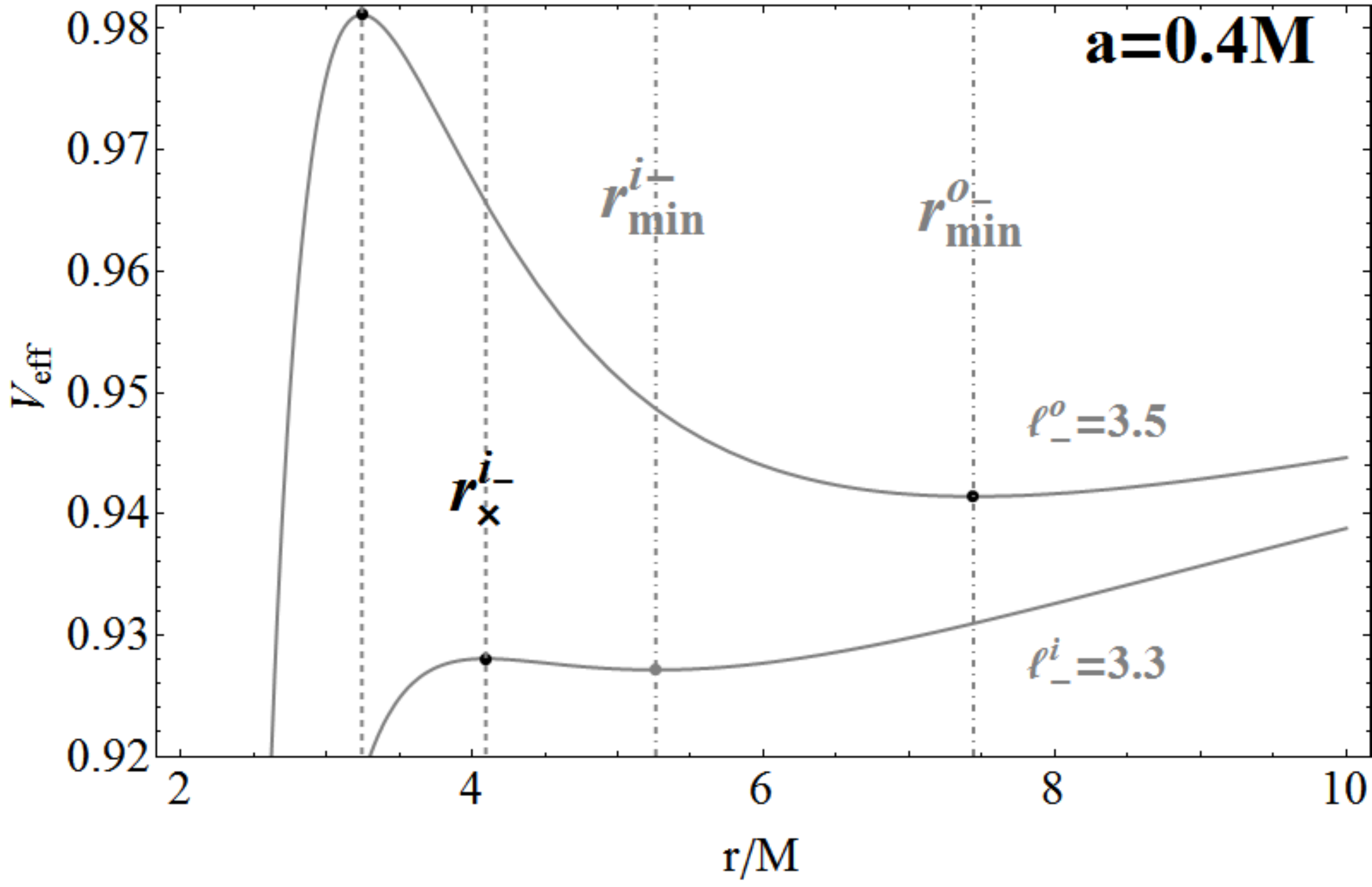}
\includegraphics[width=.42\textwidth]{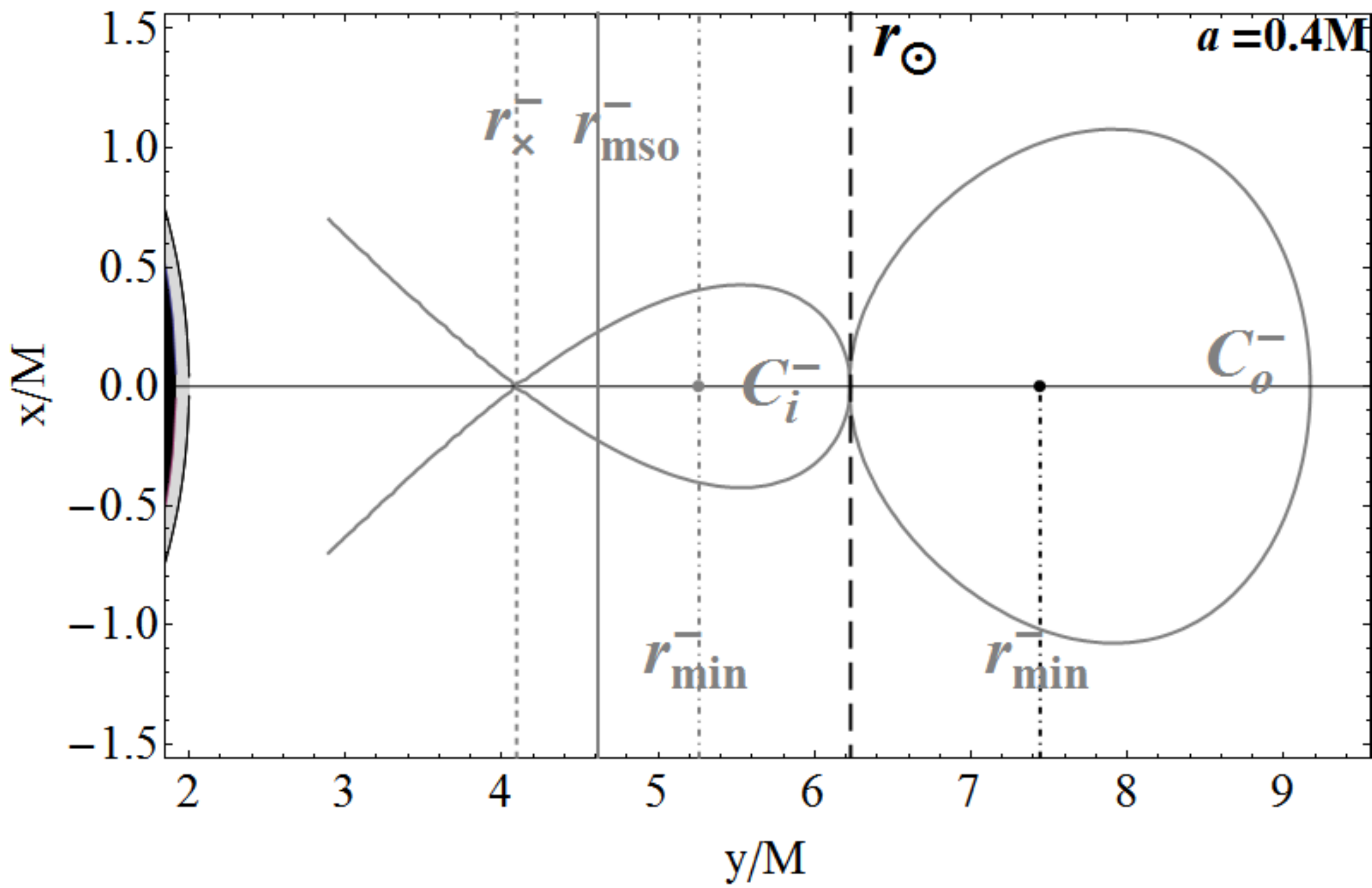}
\\
\includegraphics[width=.42\textwidth]{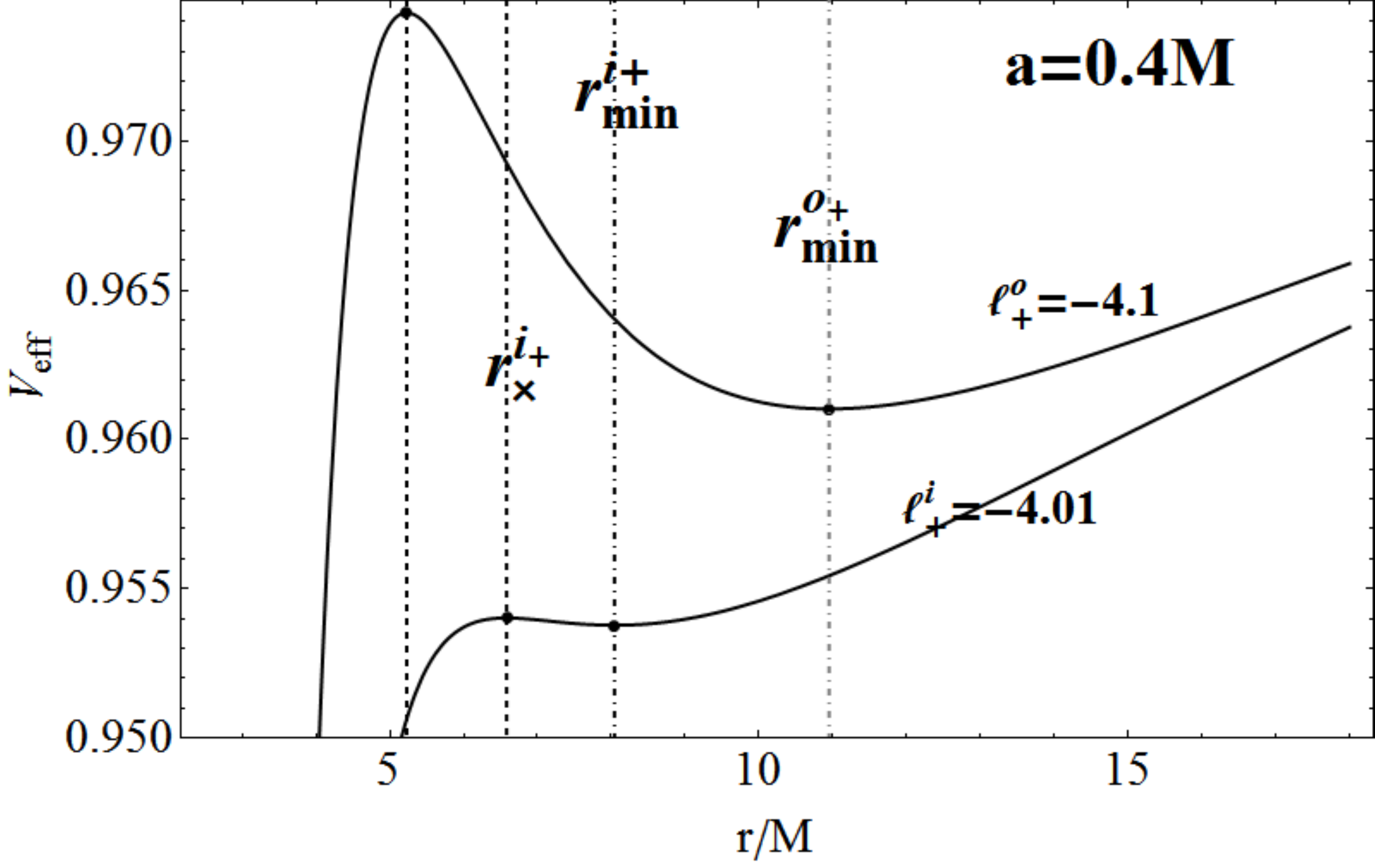}
\includegraphics[width=.42\textwidth]{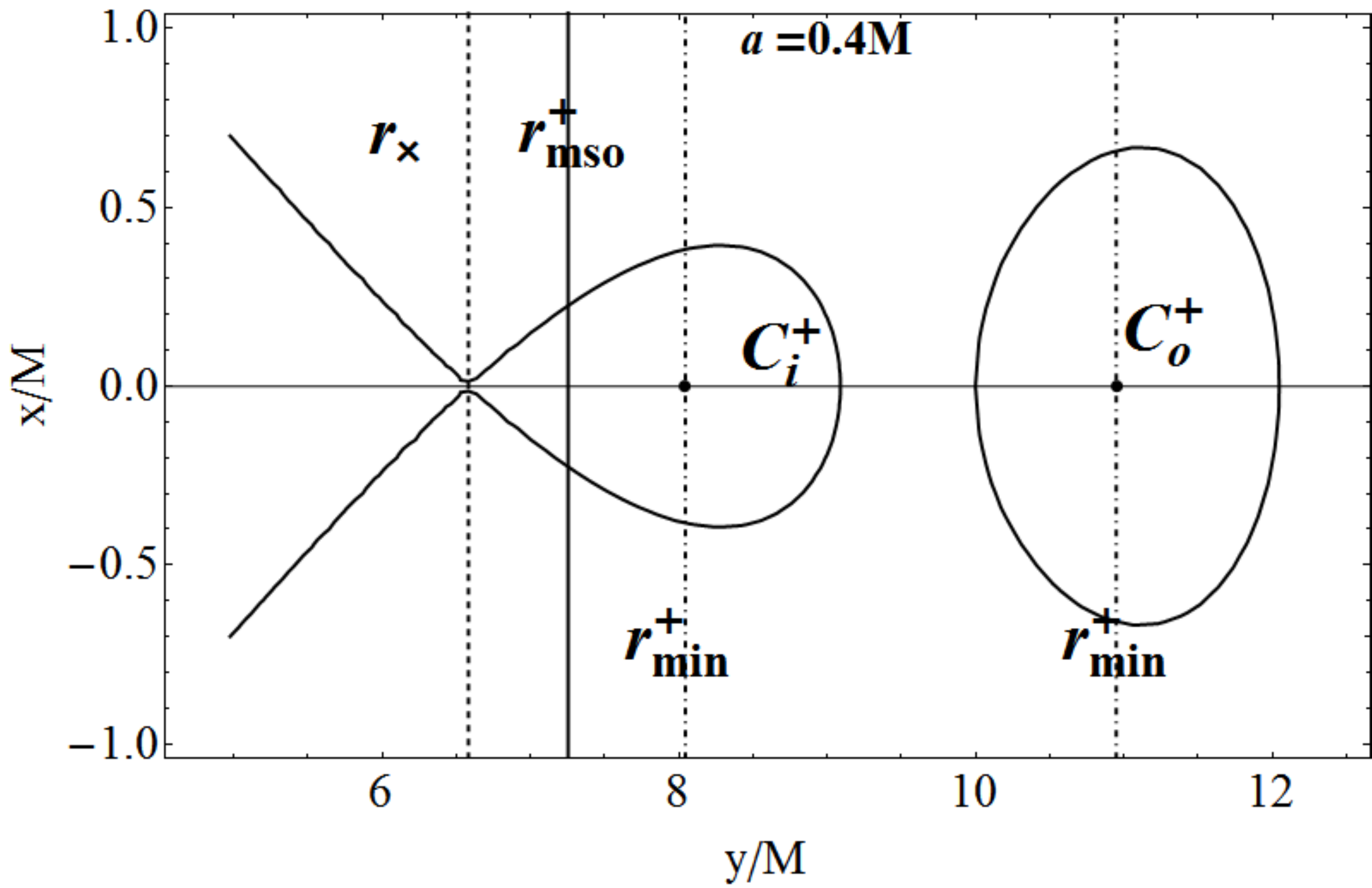}
\end{tabular}
\caption{\footnotesize{$\ell$corotating couples of accretion disk orbiting a central Kerr black hole   attractor with spin $a=0.4 M$. Effective potentials (\emph{left-panels}), and  cross sections on the equatorial plane of the outer  Roche lobes  for a couple of  counterrotating disks  $(\cc_{i}^+,\cc_{o}^+)$ (\emph{right-below}),  and  corotating disks $(\cc_{i}^-,\cc_{o}^-)$ (\emph{right-upper}), correspond to \textsl{{{scheme I}}} and {\textsl{II}} of Fig.\il(\ref{Table:Torc}) respectively.  $(x, y)$ are Cartesian coordinates and  $r_{\mso}^{\pm}$ are the marginally stable circular orbits for counterrotating and corotating matter respectively Accretion for an $\ell$corotating couple (from the $r_{\times}$ point) may emerge only from the inner disk. Collision (at contact point $r_{\odot}$) between the outer Roche lobes of the disks, here shown for the couple $(\cc_{i}^-,\cc_{o}^-)$ (\emph{upper}), is  possible  for any $\ell$corotating couples--see also Fig.\il(\ref{Fig:doub-grap-ll-cor}).}}\label{Fig:Goatms}
%\end{tabular}
\end{figure}
We will always intend relations between the magnitude  of the specific angular momentum, if not otherwise specified. However, we should consider that  for the corotating fluids there could occur  the penetration  of the geodesic corotating structure   into the  ergoregion, which does not occur  for the  counterrotating rings,\footnote{ The instability point
$r_{\jj}^-\in \Sigma_{\epsilon}^+ $ for attractors  $a\in]a_1, a_b[$, where $\Sigma_{\epsilon}^+=]r_+,2M]$  is the ergoregion on the equatorial plane of the Kerr geometry, and  $r_{\jj}^-\in ! \Sigma_{\epsilon}^+$ for the faster attractors with $a\in[a_b, M]$; at $a\in]a_2,a_b[$ there is $r_{\times}\in\Sigma_{\epsilon}^+$ and at $a\in]a_b,M]$ there is $r_{\times}\in!\Sigma_{\epsilon}^+$, where
${a}_1/M\equiv1/\sqrt{2}\approx0.707107;\quad{a}_b/M\equiv 2 (\sqrt{2}-1)\approx0.828427 ;\quad
a_1:\;r_{\gamma}^-=r_{\epsilon}^+;\quad a_b^-:\;r_{\mbo}^-=r_{\epsilon}^+
$
 and $ a_2:\; r_{\mso}^-(a_2)=r_{\epsilon}^+$ where $a_2/M\equiv {2 \sqrt{2}}/{3}\approx0.942809$--see
 \cite{pugtot,ergon} and Fig.\il(\ref{Figs:Ptherepdiverg}). } and by the different behaviour   $\partial_{a/M}r_{\mathrm{N}}^{\pm}\gtrless0$ and $\partial_{a/M}\bar{\mathfrak{r}}_{\mathrm{N}}^{\pm}\gtrless0$ --see Figs\il(\ref{Figs:Ptherepdiverg}) and Figs\il(\ref{Fig:OWayveShowno}).

\begin{figure}[h!]
\centering
\begin{tabular}{cc}
\includegraphics[width=.42\textwidth]{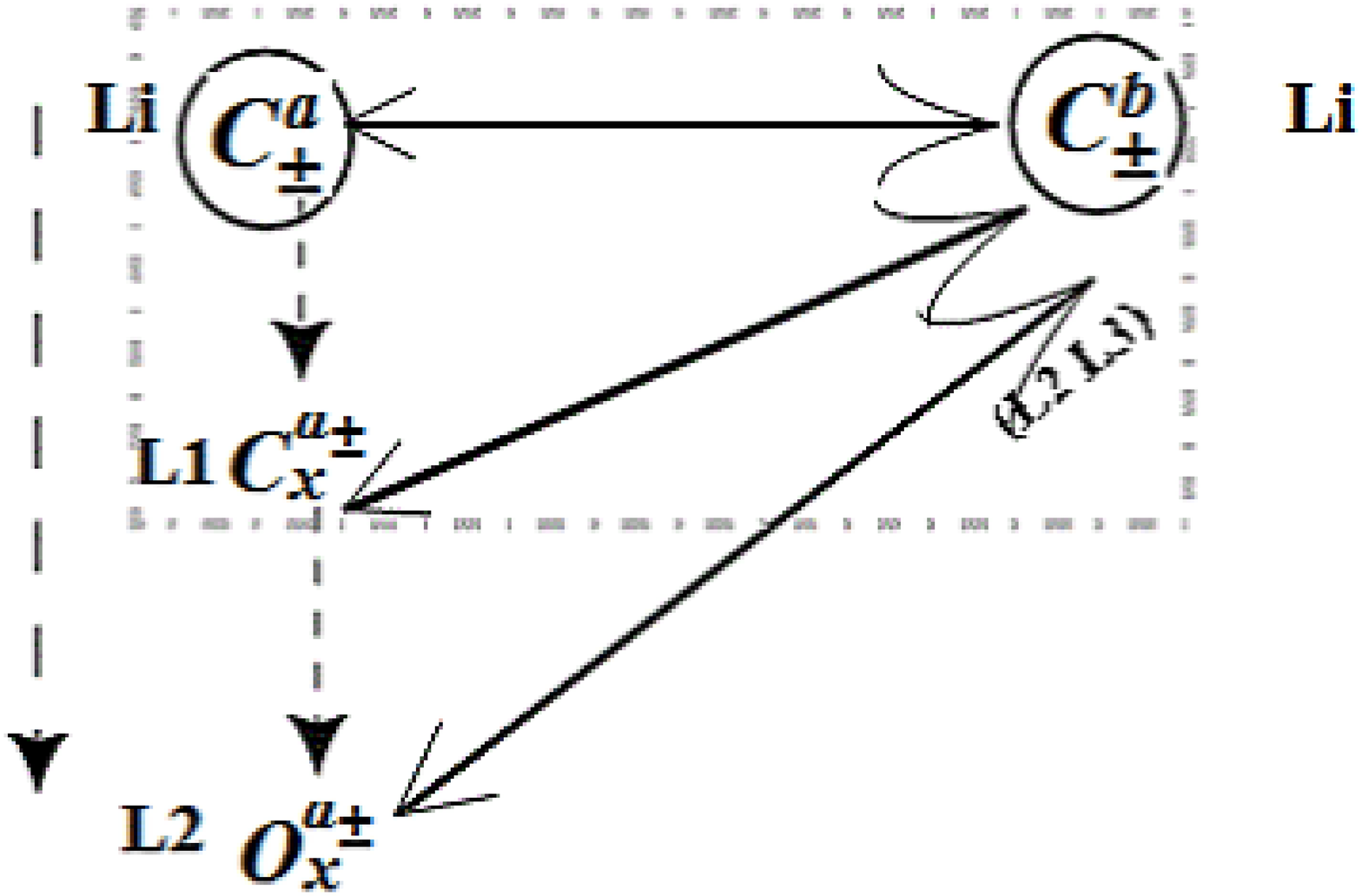}
\end{tabular}
\caption{\footnotesize{$\ell$corotating couples corresponding to \textsl{{scheme I}} and {\textsl{II}}  of Fig.\il(\ref{Table:Torc}) see also Figs\il(\ref{Fig:Goatms}). Evolutive graph of a double accretion disk system made by a couple of $\ell$corotating toroidal disks (monochromatic graph). This case  also describes an  $\ell$counterrotating  couple of  toroidal disks  (bichromatic graph) in a \emph{static} spacetime $(a=0)$. The initial state is assumed to be the couple of configuration in equilibrium $\pp_{\pm}-\pp_{\pm}$. Description of  graph blocks are in Fig.\il(\ref{Table:Graphs-models}). State lines for  $\ell$corotating  couple are listed in  Fig.\il(\ref{Table:statesdcc})--see Sec.\il(\ref{Sec:lc-or}). }}\label{Fig:doub-grap-ll-cor}
%\end{tabular}
\end{figure}

In terms of  the graph models, introduced in Fig.\il(\ref{Table:Graphs-models}) and Sec.\il(\ref{Sec:Taa-DISK}), the $\ell$corotating couples are represented by {monochromatic} graph of Fig.\il(\ref{Fig:doub-grap-ll-cor}).

We set up our analysis by considering  an  initial state of equilibrium, formed by a couple of tori   in equilibrium. The possible  (initial or final) states   for this case   are listed in  Fig.\il(\ref{Table:statesdcc}). This state represents  also the graph center in Fig.\il(\ref{Fig:doub-grap-ll-cor}), which have  therefore only a subsequent section or loops.
Except  the case of  evolutive loops,  in general we will  deal  with a system which is  evolving  from an initial state of equilibrium towards  unstable configurations.
In other words, for convenience  of discussion  we adopt  here the arbitrary   assumption of existence of a  phase in the formation of the double tori  system   in which both  tori are in an equilibrium state, and   the system can  eventually   evolve towards an  instability  phase   $\mathbf{C_{\times}}^{2}$, $\mathbf{C_{\odot}}^{2}$  or also  $\mathbf{C_{\odot}^{\times}{}^{2}}$. It is   easy to see  that  choice of a different initial state and therefore a different graph center does not   change qualitatively the graph which will be just centered in  a different state. % We

We  start our analysis of the couple by  focusing on the state lines representing  all possible couples of disks orbiting a Kerr attractor with dimensionless spin $a/M\in[0,1]$, according to the constraints imposed by the geodesic structure and the  condition of non-penetration of matter.
Then  we discuss the system   evolution connecting, whenever possible, the different state lines with the evolutive lines. Graphs construction    in the $\ell$corotating case is detailed in Sec.\il(\ref{Sec:graph-app}). We discuss  the possible evolutive  loop for  the $\ell$corotating  couples  and  the  $\ell$counterrotating couples in the static spacetimes  in Sec.\il(\ref{App:deep-loop}).
State lines, represented here in  Fig.\il(\ref{Table:statesdcc}),  were introduced in  \cite{open}. We  here report  the principal results adapted to the case of  ringed disks of the order $n=2$--see also \cite{letter}.

\medskip

\textbf{Couple evolution from equilibrium to instability}

\medskip

We start our considerations by
assuming that the initial state for a tori pair provides two equilibrium topologies, $\cc_i<\cc_o$,  then we shall consider  possible evolution towards an  instability  or a  sub-configuration $\pp_{\times}$ of the pair, and  possible collision, where the emergence of configurations $\mathbf{C}_{\oplus}^2$ will be deepened  in Sec.\il(\ref{Sec:non-rigid}).

Considering  the no loop evolution, a couple of tori in a Schwarzschild spacetime and  the $\ell$corotating systems   around a Kerr geometry are  completely described by the graph in Fig.\il(\ref{Fig:doub-grap-ll-cor}).
In any monochromatic  graph (or bichromatic graph in a static spacetime), all the state lines  are \emph{oriented}  in the same direction--Fig.\il(\ref{Fig:doub-grap-ll-cor}).
Because of assumptions of  the unique geodesic structure of the background geometry, there is no evolutive phase  in which the outer disk of the couple  is accreting, but \emph{only} the inner configuration of the doubled  system can accrete onto the source. Only the inner disk of the couple could  evolve towards the unstable topology, and the subsequent section  is formed  by the evolutions of the  inner vertex only. The evolution of the final state, due to the  inner vertex, can affect   all the  state lines and their evolutions.
Two tori, which are  both corotating or counterrotating with respect to the central attractor, can admit only the inner configuration accreting and if, for some processes, the outer accretion disk would approach its unstable phase, then the double ($\ell$corotating) system would be   destroyed for collision or merging before the outer ring would effectively reach its unstable topology. This  reduces   the possibility of existence of the double tori system and its stability--as shown in \cite{letter}  the  possible states with an instability are   $\cc^{\pm}_{\times}<\cc^{\pm}$ for  $a\in[0,M]$ (and $\cc^{\pm}_{\times}<\cc^{\mp}$ for  $a=0$).
This is property of any couple of $\ell$corotating tori  regardless  the  dimensionless spin of the attractor and  largely also of a single  accretion disk model, as long as it is assumed that the accretion occurs  at the (stressing) inner margin of the disk which is located at  $r_{in}\in]r_{\mbo},r_{\mso}]$.
Therefore the outer ring have to be considered  to be \emph{quiescent}--i.e. in  equilibrium . It can however grow, increasing the  $K$ parameter, or it can change the specific angular momentum. Thus also changes in the  ring morphology  may cause  an instability  of the entire ringed disk for ring collision. Moreover, even with a quiescent outer ring, an  accretion phase occurring in the  inner ring could induce a  ring collision, during the earliest stage of the  {accretion}, the inner  ring reaches, according to its specific angular momentum, its maximal elongation on the equatorial plane, i.e.  $\lambda_i=\lambda_{\times}^i=\max{\lambda}$ where $K=K_{\max}$, and the inner disk outer margin  moves outwardly.
On the other hand,  the outer ring   could collide with  the inner one, and eventually merging with this  leading to an accretion or  inducing an evolutive loop.
Therefore, we need to discuss these two different, competitive phenomena for \emph{accretion} of the inner ring and \emph{collision} among the rings and the subsequent three fates this may induce.
It is therefore interesting to discuss the emergence of loops for monochromatic graphic and the possibility of merging  of tori--
see  Sec.\il(\ref{App:deep-loop}).
\subsection{The $\ell$counterrotating couples}\label{Sec:lcounterrsec}
\begin{figure}[h!]
\centering
\begin{tabular}{cc}\hline\hline
\includegraphics[width=.37\textwidth]{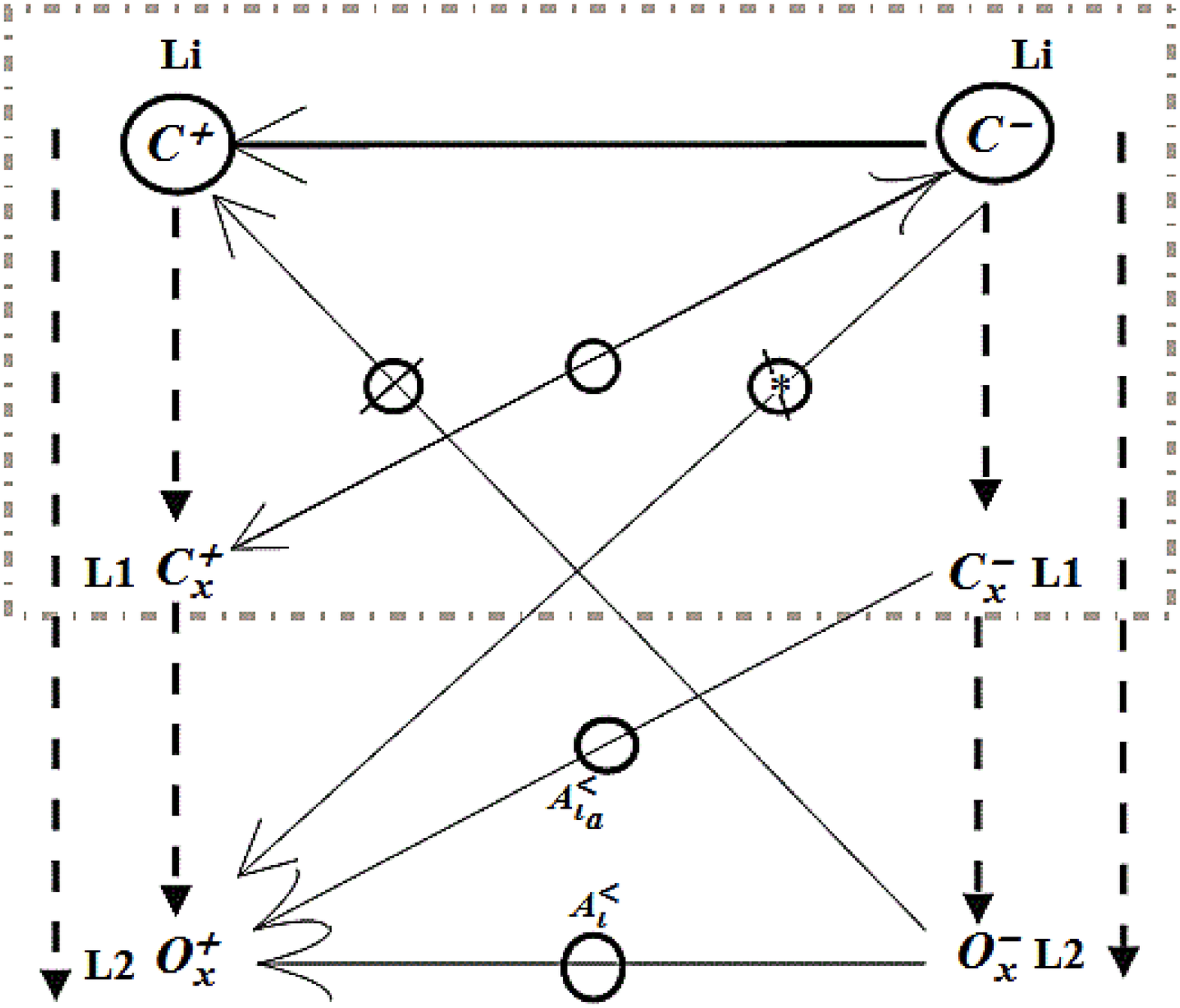}&
\includegraphics[width=.37\textwidth]{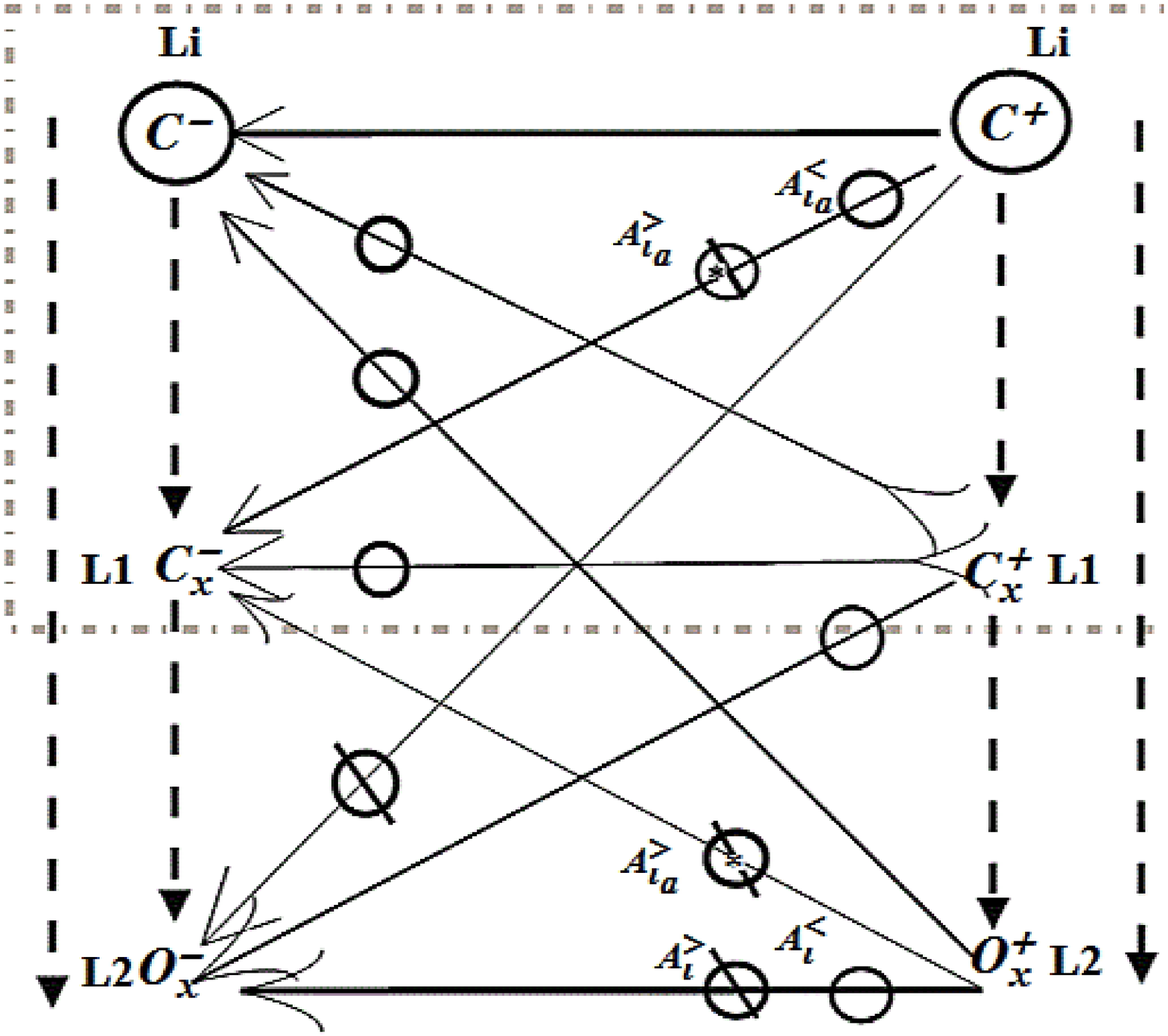}\\\\
\hline\hline
\end{tabular}
\caption{\footnotesize{Graph of a double $\ell$counterrotating tori  in a Kerr black hole spacetime $a\in]0,M[$ (bichromatic graph). Left:  graph  centered in the initial state $\cc^+<\cc^-$--see also {\textsl{scheme III}} of Fig. (\ref{Table:Torc}) and Fig.\il(\ref{Figs:ApproxPlo}). Right:  graph  centered in the initial state $\cc^-<\cc^+$--see also {\textsl{scheme IV}} of Fig.\il(\ref{Table:Torc}) and Figs\il(\ref{Figs:Kind-End}). The initial state is assumed to be the couple of configuration in equilibrium.  Description of  graph blocks is in Fig.\il(\ref{Table:Graphs-models}). State lines for  $\ell$counterrotating couples are in Fig.\il(\ref{Table:statescc}).}}\label{Fig:CC-CONT}
%\end{tabular}
\end{figure}

We consider the $\ell$counterrotating couples   orbiting  a Kerr attractor with dimensionless spin  $a/M\in]0,1]$  corresponding to   bichromatic graphs--see {\textsl{schemes III}} and {\textsl{IV}} of  Fig.\il(\ref{Table:Torc}).
In comparison to    $\ell$corotating couples (monochromatic graph), or   $\ell$counterrotating  torii orbiting a Schwarzschild black hole, this case is    complex, being diversified   for classes of attractors, and  for the   disk spin orientation with respect to the central attractor--\cite{open}.
 It is therefore necessary to consider separately the case $\cc^+<\cc^-$ (inner counterrotating  torus and outer corotating torus),  discussed in Sec.\il(\ref{Sec:coun-co}) from the $\cc^-<\cc^+$ case (inner corotating ring and outer counterrotating torus), investigated  in Sec.\il(\ref{Sec:tex-dual}).
The  graph formalism   can significantly simplify the analysis of the evolution of this particular double system. Using the results of  \cite{open} we build the Fig.\il(\ref{Table:statescc})  collecting the main states for the graph in Fig.\il(\ref{Fig:CC-CONT}).
The most relevant effect distinguishing these {pairs}   from  the  $\ell$corotating couples  is  that  the (counterrotating) outer torus of the couple  in general may undergo a P-W  instability phase with the emergence of an instability point eventually giving rise  to a feeding  of material towards its companion inner torus.

The double sequentiality  according to the  configuration  and  criticality indices, respectively, of some lines of  Table \il(\ref{Table:statescc}) and  the states of Fig.\il(\ref{Fig:CC-CONT}) are not specified, depending
   on the   vertex  decoration in terms of   the angular momentum.
In fact, as demonstrated in   \cite{ringed,open},
  the sequentiality of centers of the   $\ell$counterrotating  couples in equilibrium   does not necessarily constrain  the critical sequentiality    ($\ell\non{\in}\mathbf{L3}$):  there are special cases where, at fixed   $\ell_i\ell_o<0$,  with $\pp_i<\pp_o$, there can be $\pp_i\prec \pp_o$, which corresponds to $\bar{\mathfrak{C}}_0$    in Eq.\il(\ref{C0})  (if there is  $\cc^-<\cc^+$),
or otherwise it corresponds to
$\bar{\mathfrak{C_{1_a}}}$
 in   Eq.\il(\ref{C1a})
 within the conditions Eq.\il(\ref{Eq:so1})
 or (\ref{Eq:so2}), or, conversely, there can be   $\pp_i\succ \pp_o$
 i.e. a couple of the  $\bar{\mathfrak{C_{1_b}}}$ class,  in Eq.\il(\ref{C1b}), which includes also the
 $\ell$corotating couples.

Then the outer  vertex  of  the $\bar{\mathfrak{C_{1_a}}}$ and  $\bar{\mathfrak{C_{1_b}}}$  couples \emph{must} be in  equilibrium or destroyed: this means that  before the outer torus reaches its unstable phase  the ringed disk will be destroyed for collision,   prohibiting any subsequent evolutive lines.

Such a situation may be prevented if a change  of criticality  order occurs, which means a \emph{transition} from a $\bar{\mathfrak{C_{1_a}}}$ or  $\bar{\mathfrak{C_{1_b}}}$ class to $\bar{\mathfrak{C_{0}}}$ class.
However, as the  inversion of the  configuration sequentiality is not permitted, such a transition  could happen \emph{only} for the couples
$\pp^-<\pp^+$,  detailed in Sec.\il(\ref{Sec:tex-dual}).
\begin{figure}[h!]
\centering
\begin{tabular}{c}
\includegraphics[width=.51\textwidth]{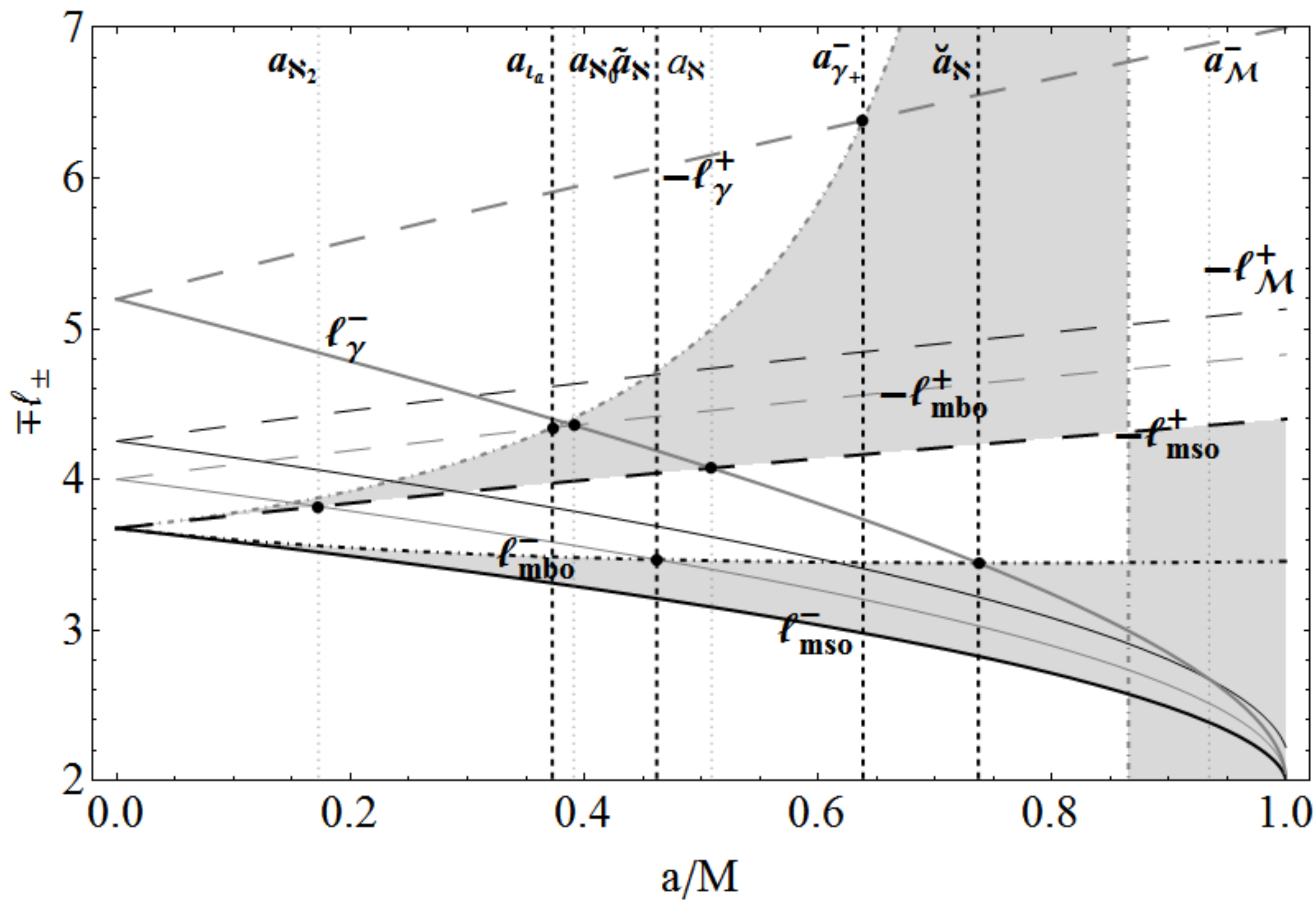}
\end{tabular}
\caption{\footnotesize{ Fluid specific angular momentum $\ell^{\pm}_i=\ell^{\pm}(r^{\pm}_i)$   where $r^{\pm}_i\in \{\mathbf{R}_{\mathrm{N}}^{\pm},r_{\mathcal{M}}^{\pm}\}$,  $r_{\mathcal{M}}^{\pm}$ is the maximum point    of
derivative $\partial_r(\mp \ell^{\pm})$ for $a/M$ respectively and  $\mathbf{R}_{\mathrm{N}}\equiv \{r_{\gamma}^{\pm}, r_{\mbo}^{\pm},r_{\mso}^{\pm}\}$. Some notable spacetime spin-mass ratios are also plotted,  a list can be found in Table\il(\ref{Table:nature-Att}). Dotted-dashed curves are respectively $-\ell^+(r_{mso}^-)\geq \ell^-(r_{mso}^+)$.
}}\label{Fig:SAPJoke}
%\end{tabular}
\end{figure}
In fact, in the isolated  disks--attractor systems, the couple evolution is strongly  determined by the decoration of  the initial state.
However, conditions for occurrence of this class transition   are very complex, depending on the relation between  characteristic values of the specific angular momentum $\{\mp\ell_{\mso}^{\pm}, \mp\ell^{\pm}(r_{\mso}^{\mp})\}$  which determine the boundaries of the ranges in Eq.\il(\ref{Eq:(15)})--see also Fig.\il(\ref{Figs:Ptherepdiverg}).

More generally, for $\ell\non{\in}\mathbf{L3}$ we can discuss the state sequentiality according to the   arguments presented in \cite{ringed,open}. We distinguish two cases according to the magnitude of the specific angular momentum.
\begin{enumerate}
\item
\bea\label{Eq:fac-tus}
\mbox{For} \quad|\ell_-/\ell_+|>1\quad\mbox{ \emph{there must be }} \quad \bar{\mathfrak{C}}_{1b}\quad\mbox{of Eq.\il(\ref{C1b})} \quad\mbox{or}\quad \pp^+<\cc^-\quad \mbox{and}\quad \pp^+\succ \cc^-.
\eea
This case, analyzed in
Sec.\il(\ref{Sec:coun-co}) and represented by the \emph{scheme III} of Fig.\il(\ref{Table:Torc}), is  described by the graph of Fig.\il(\ref{Fig:CC-CONT})-left.
As confirmed by the graph Fig.\il(\ref{Fig:CC-CONT})-left, only the  inner counterrotating ring can accrete onto the source.
On the other hand, the condition $\pp^+<\cc^-$ does not necessarily imply  the relation of Eq.\il(\ref{Eq:fac-tus}) for the angular momentum.
Moreover, the condition ${\ell_->-\ell_{\mso}^+}$   implies strong constraints on the initial state  of the outer corotating torus of the couple. In fact, due to the constraint $|\ell_-/\ell_+|>1$,
if the attractor belongs to the class  $a<a_{\aleph_2}$ (where $a=a_{\aleph_2}:\;-\ell_{\mso}^+=\ell_{\mbo}^-$),   the outer corotating  torus can belong to
  one of the ranges  $\mathbf{Li}$.  For $a\in]a_{\aleph_2},a_{\aleph}[$,  the outer corotating torus  specific angular momentum has to be in $\mathbf{L2}$ or $\mathbf{L3}$, and for $a >a_{\mathcal{M}}$, the torus is centered at { $r>r_{\mathcal{M}}^-$}.  For the faster  attractors  with $a>a_{\aleph}$ (at $a=a_{\aleph}:\; -\ell_{\mso}^+=\ell_{\gamma}^- $) the corotating torus  has to be located far from the attractor,  having  the specific angular momentum in $\mathbf{L3}$; the corresponding effective potential has thus no  maximum points--Fig.\il(\ref{Figs:Ptherepdiverg}).
 \item
\bea&&
\mbox{For}\quad {|\ell_-/\ell_+|<1}\quad\mbox{there is }\quad
 \pp^-\lessgtr \pp^+,\quad\mbox{and}\quad r_{\min}^-<\bar{\mathfrak{r}}\quad\mbox{ where  } \bar{\mathfrak{r}}>r_{\mso}^-:\;\ell_-=-\ell_+ >\ell_{\mso}^+
 \\\label{Eq:p-mpscott}&&
\mbox{if}\quad  \ell_-\in]\ell_-(r_{\min}^+),\ell_-(\bar{\mathfrak{r}})[\quad\mbox{then there is }\quad \pp^+<\pp^-,
\\\label{Eq:p-mpscott1}&&
\mbox{if}\quad \ell_-\in]\ell_{\mso}^-, \ell_-(r_{\min}^+)[ \quad\mbox{then there is }\quad \pp^-<\pp^+.
 \eea
 The sequentiality according to the   criticality has been fixed  in  the first column  of Table\il(\ref{Table:REDUCtIN}),
combining   the additional restrictions provided by the angular momentum and   the constraints from  the  complementary geodesic structure of spacetime $\bar{\mathbf{R}}_{\mathrm{N}}$, represented in Fig.\il(\ref{Figs:Ptherepdiverg}).

The case $\pp^+<\pp^-$ of (\ref{Eq:p-mpscott}) is described in Fig.\il(\ref{Fig:CC-CONT})-right and represented in \textsl{scheme IV} of Fig.\il(\ref{Table:Torc}), see also Fig.\il(\ref{Figs:ApproxPlo}).
We note that in general, a small range of  angular momentum  in  the case   $\pp^+<\pp^-$ with   $|\ell_-/\ell_+|<1$,  is associated to a  limited orbital  region which decreases as the torus distance    from the attractor  increases, or its dimensionless spin decreases, i.e., in the $R\equiv r/a\gg r_{\mso}/a$\footnote{The emergence of the Newtonian limit is  discussed in \cite{ringed,open}. Here we could consider either   $R\equiv r/a\geq \bar{\mathfrak{r}}_{\gamma}/a$ or $R\equiv r/a\geq r_{\mathcal{M}}/a$.} limit.  In fact this behavior could be interpreted   as a consequence of   the rotational effects of the attractor which  disappear in the Newtonian limit.
The  existence of such  a $\cc^->\cc^+$  couple   is very constrained  since the extension  of the orbital difference, $r_{\min}^--r_{\min}^+$, in the case $\cc^->\cc^+$ is very limited and
 depends on the $r_{\mathcal{M}}^{\pm}$; such toroidal configurations are
 more prepared  to collide with subsequent possible  merging  of the tori.

The case $\pp^-<\pp^+$ of (\ref{Eq:p-mpscott1})  is shown in
 Fig.\il(\ref{Fig:CC-CONT})-left and \textsl{ scheme III} of Fig.\il(\ref{Table:Torc})--see also Figs\il(\ref{Figs:Kind-End}).
 The possible specific angular momentum of the  $\cc^-$ configuration with  $\ell_-<-\ell_+$ (for  $\cc^+<\cc^-$ or $\cc^-<\cc^+$) depends on  the unstable topology  of the   $\pp^+$ torus.
  The instability of the $\pp^+<\pp^-$ couple  must take place on $\pp^+$.  When  $\ell_+\in \mathbf{L1}$, then there is the maximum possible separation between the centers of the  couple.
  Therefore, it is necessary to  consider the angular momentum  $\ell_-(r_{\min}^+)$, which
  is the  lower limit  of the range  in Eq.\il(\ref{Eq:p-mpscott}) and  the specific angular momentum $\ell_{-}(\bar{\mathfrak{r}})=-\ell_+(r_{\min}^+)$ which is the upper limit of this range.
  We can establish the   upper-bound  by considering  the topology of the $\pp^+$  configuration and the constraints  on the ranges of the angular momentum. Whereas,  the  lower bound
 satisfies the relation $
  \ell_-(r_{\min}^+)<\ell_-<\ell_{-}(\bar{\mathfrak{r}})=-\ell_+(r_{\min}^+)$--see Table\il(\ref{Table:REDUCtIN}) and   \cite{open}.
 In fact, it is necessary to know the radius  $r_{\min}^+$, for fixed range of  $\ell_+$, and  then   establish the range of angular momentum  $\ell_-$ of  the corotating torus  centered in  $r_{\min}^+$--see also \cite{ringed}.
Then, we can combine the restrictions provided by the angular momentum range and those   derived from the condition on the relation of the two angular momenta  with results of Table\il(\ref{Table:REDUCtIN}).
Therefore, having a { $\cc^+_3$}  torus then the $\pp^-$ torus   can be in any topology, analogously, but with some restrictions, for
  $\cc^+_2$, mainly for $\cc_3^-$  and  $\cc_2^-$  for  slow  attractors ($a<a_{\aleph_0}=0.390781M\;\ell_{\mso}^+=-\ell_{\mbo}^-$).
For a  $\cc^+_1 $ torus  there is only $\cc^-\neq \cc_3^-$ for
  $a<a_{\aleph_0}$.
  If $\cc^-=\cc_1^-$   or  $\cc_2^-$,  then $\cc^+$  can be in any angular momentum range , but subjected to  several  restrictions if orbiting around slower attractors. However, these results have to be combined with those presented in Table\il(\ref{Table:REDUCtIN}).
 If $\cc^-=\cc^-_2 $, we have only the  constraints provided by the complementary geodesic structure given in Table\il(\ref{Table:REDUCtIN}),
while if  $\cc^-=\cc^-_3$ then, for  $a<a_{\aleph_0}$, we have $\cc^+=\cc_2^+$ or $\cc_3^+$.
%%%%%%%%%%%%%%%%%%%%%%%%%%%%%%%%%%%%%%%%%%%%%%%%%%%%%%%%%%%%%%%
%%
%
\end{enumerate}
Sec.\il(\ref{Sec:coun-co}) and Sec.\il(\ref{Sec:tex-dual}), are dedicated to    the  $\ell$ounterrotating couples, focusing on the sequentiality. As conclusion on this analysis we summarize  the situation in the following points:
\begin{enumerate}
\item
\bea\label{Eq:conf-decohe}
&&\mbox{For}\quad \pp^-<\pp^+\quad\mbox{there is}\quad|\ell_-/\ell_+|<1\quad\mbox{and }\quad
\ell_-\in]\ell_{\mso}^-, \ell_-(r_{\min}^+)[,
\eea
see Sec.\il(\ref{Sec:tex-dual}) and Eq.\il(\ref{Eq:p-mpscott1}).
This case is detailed in Sec.\il(\ref{Sec:app-submmp})  where additional restrictions  are discussed.
The relation between the  instability  points (for $\ell\non{\in}\mathbf{L3}$), which is  the critical sequentiality,   is fully addressed  in Sec.\il(\ref{App:comm-con-crit}) and Table\il(\ref{Table:REDUCtIN}).

In fact, the angular momenta of the tori in  $\pp^-<\pp^+$, are fixed   in second-column of Table\il(\ref{Table:REDUCtIN}), while the location of the eventual instability points  has been established  in Fig.\il(\ref{Table:statescc}) and  first-column of Table\il(\ref{Table:REDUCtIN}).
Since the distance between the radii $r_{\mso}^+>r_{\mso}^-$ increases  with increasing spin of the attractor,   for some ranges of angular momentum the critical  points of the  outer counterrotating configuration are  $r_{\mso}^-<r_{\max}^+<r_{\mso}^+$, and the couple  of (\ref{Eq:conf-decohe}) are $\pp^-<\pp^+$ with $\pp^-\prec\pp^+$ as shown in Table\il(\ref{Table:REDUCtIN}). Therefore, this couple is    a  $\mathfrak{\bar{C}_{1_a}}$   one within the conditions of Eq.\il(\ref{Eq:so2}) for $r_{\mso}^-<r_{\max}^+<r_{\min}^-<r_{\min}^+$,
or a $\mathfrak{\bar{C}_{0}}$  of Eq.\il(\ref{C0}) if  $r_{\mso}^-<r_{\min}^-<r_{\max}^+<r_{\mso}<r_{\min}^+$.
Therefore, the situation depends  on the angular momentum of the outer configuration and the class of the attractor.
 On the other hand, if   $r_{\max}^+<r_{\mso}^-$, then assuming  $\tilde{\ell}_-\equiv\ell_-(r_{\max}^+)\in]\ell_{\mso}^-,\ell_-(r_{\min}^+)[$ this angular momentum separates the configurations with
$ \ell_-\in]\ell_{\mso}^-,\ell_-(r_{\max}^+)[$ where $r_{\max}^->r_{\max}^+$ implying  $\pp^-<\pp^+$ with $\pp^-\succ\pp^+$ which is a $\bar{\mathfrak{C}}_{1b}$ of Eq.\il(\ref{C1b}),
 from those with
$ \ell_-\in]\ell_-(r_{\max}^+),\ell_-(r_{\min}^+)[$  where  $r_{\max}^-<r_{\max}^+$
 implying  $\pp^-<\pp^+$ with $\pp^-\prec\pp^+$ which is a $\bar{\mathfrak{C}}_{1a}$  one of Eq.\il(\ref{C1a}).
\item
Conversely for the couple of tori
\bea\label{Eq:fro-plus-intere}
&&\pp^->\pp^+\quad\mbox{ there is  either}\quad
\\\nonumber
&&
\bar{\mathfrak{C}}_{1a}:\;  \cc^->\pp^+\quad\mbox{and}\quad   \cc^-\succ \pp^+\quad\mbox{within  conditions of Eq.\il(\ref{Eq:so2}) or} \\\nonumber
&&
\bar{\mathfrak{C}}_{1b}:\;  \cc^->\pp^+\quad\mbox{and}\quad  \cc^-\prec \pp^+.
\\\nonumber
&&\quad\mbox{There can be}\quad
|\ell_-/\ell_{+}|>1\quad\mbox{if}\quad r_{\min}^->\bar{\mathfrak{r}}_-, \quad\mbox{then we obtain  }\quad\bar{\mathfrak{C}}_{1b}\quad\mbox{according to Eq.\il(\ref{Eq:fac-tus}), }
\\\nonumber
&&\quad\mbox{or there is }\quad|\ell_-/\ell_{+}|<1 \quad\mbox{if}\quad r_{\min}^-\in]r_{\min}^+,\bar{\mathfrak{r}}_-[,
\quad\mbox{where}\quad
\bar{\mathfrak{r}}_-:\;\ell_-(\bar{\mathfrak{r}}_-)=-\ell_+.
\eea
The outer corotating torus  of this couple cannot be unstable, as shown
in Sec.\il(\ref{Sec:coun-co}).
\end{enumerate}
Finally we conclude this section mentioning  the couples with
\bea
\ell_{i/o}=-1\quad \mbox{which corresponds to the case}\quad  \bar{\mathfrak{C}}_{1_b}\quad \mbox{considered in Eq.\il(\ref{C1b})},
\eea
discussed also  in
 \cite{ringed} as limiting cases   for  the perturbation analysis  and as limiting situation of $|\ell_{i/o}|\lessgtr1$.

%%%%%%%%%%%%%%%%%%%%%%%%%%%%%%%%%%%%%%%%%%%%%%%%%%%%%%%%%%%%%%%%%%%%%%%%%%%%%%%%%%%%%%%%%%%%%%%%%%

The evolution of these systems is fully described  in the graphs of Fig.\il(\ref{Fig:CC-CONT}).
Comparing the graphs  of Fig.\il(\ref{Fig:doub-grap-ll-cor}) and Fig.\il(\ref{Fig:CC-CONT}),
it is clear that in  the $\ell$counterrotating case   both vertices  of a state may evolve. As a consequence of this  a change of the central state  of the graph  (which is also  the initial state, the graph having only a subsequent section) generally heavily deform the entire  graph, being  strongly dependent on the initial data (the decorations of the state vertices).  The evolution of a state  line   is highly  constrained by  the  initial decoration, as can be seen by comparing Fig.\il(\ref{Table:statesdcc}) for the $\ell$corotating states and Fig.\il(\ref{Table:statescc}) and Table\il(\ref{Table:REDUCtIN}) for the  $\ell$counterrotating states.
Consequently  we have only a limited number of possible   states and evolutive lines for a    $\ell$counterrotating  system:  fixing the range of angular momentum for the separated initial couple (implying  constraints on the  $K$-parameters--see Sec.\il(\ref{Sec:non-rigid}) and also \cite{ringed}),  we obtain    rather stringent constraints from which it might be  possible to predict in  large extension  the existence and stability of the (isolated) couple  of rotating tori around a spinning  central black hole.

For completeness, we also consider the configurations $\oo_{\times}$ whose existence implies a relaxation of the condition of non-penetration  of matter--we refer to Sec.\il(\ref{Sec:graph-app}) for further discussion.
The following Sec.\il(\ref{Sec:coun-co}) is focused on the $\pp^+<\pp^-$  double system  introduced
in Eq.\il(\ref{Eq:fro-plus-intere}), while in Sec.\il(\ref{Sec:tex-dual}) we investigate  $\pp^-<\pp^+$  couples introduced in Eq.\il(\ref{Eq:conf-decohe}).

\begin{table}[h!]
\caption{$\ell$counterrotating  couples: decoration of bichromatic vertices with angular momentum classes $\mathbf{Li}$ of the Kerr geometries with dimensionless spin $a\in]0,M]$. State lines are in Fig.\il(\ref{Table:statescc}), graphs are in Fig.\il(\ref{Fig:CC-CONT}). Comments can be found in Sec.\il(\ref{App:notes-tables})--see also Fig.\il(\ref{Fig:OWayveShowno}). Definitions of spins are in Table\il(\ref{Table:nature-Att}).}\label{Table:REDUCtIN}
\centering
\begin{tabular}{|l|llllll|}
\hline
\textbf{ Criticality:} & \textbf{Couples}:&  &$\pp^+<\pp^-$ &\textbf{Couples:}& &$\pp^-<\pp^+$
 \\\hline
$\mbox{\footnotesize{$\blacktriangleright$}}\;\; \cc_{\times}^-\prec \oo_{\times}^+:\;a>a_{\gamma_{+}}^-$&$\mbox{\footnotesize{$\blacktriangleright$}}$  $\pp^+_3<\pp^-$ & $\mapsto$ &$\pp^+_3<\pp_3^-$&$\mbox{\footnotesize{$\blacktriangleright$}}$  $\pp_3^-<\pp^+$&$\mapsto$& $a>a_{{}_u}$:$\;$%\rtb{ $!\non{\exists}!$ }
$\pp_3^-<\pp^i_+$
\\
&&&&&&$a<a_{{}_u}:\; \pp_3^-<\pp_3^+$ $\pp_3^-<\pp_2^+$
 \\\\
$\mbox{\footnotesize{$\blacktriangleright$}}\; \oo_{\times}^-\prec \oo_{\times}^+:\;a>a_{\iota}$
&$\mbox{\footnotesize{$\blacktriangleright$}}$  $\pp^+<\pp_3^-$ & $\mapsto$ &$\pp^+_i<\pp_3^-$&$\mbox{\footnotesize{$\blacktriangleright$}}$  $\pp^-<\pp_3^+$&$\mapsto$&$\pp_i^-<\pp_3^+$
\\
$\mbox{\footnotesize{$\blacktriangleright$}}\; \oo_{\times}^-\prec \cc_{\times}^+$&$\mbox{\footnotesize{$\blacktriangleright$}}$  $\pp_2^+<\pp^-$ & $\mapsto$ & $a>a_{{}_u}:\; \pp^+_2<\pp_3^-$&$\mbox{\footnotesize{$\blacktriangleright$}}$  $\pp_2^-<\pp^+$&$\mapsto$&$\pp_2^-<\pp_i^+$
\\
&&&$a<a_{{}_u}:\; \pp^+_2<\pp_3^-$, $\pp^+_2<\pp_2^-$&&&
\\\\
$\mbox{\footnotesize{$\blacktriangleright$}}\;\cc_{\times}^-\prec \cc_{\times}^+$&$\mbox{\footnotesize{$\blacktriangleright$}}$ $\pp^+<\pp_2^-$ &$ \mapsto$ &$ a>\breve{a}_{\aleph}:\; \non{\exists}$& $\mbox{\footnotesize{$\blacktriangleright$}}$ $\pp^-<\pp_2^+$&$\mapsto$&$\pp_i^-<\pp_2^+$
\\
&&&$a\in]a_{{}_u},\breve{a}_{\aleph}[:\; \pp^+_1<\pp_2^-$&&&
\\
&&&$a<a_{{}_u}:\;\pp^+_1<\pp_2^-$ $\pp^+_2<\pp_2^-$&&&
\\\\
&$\mbox{\footnotesize{$\blacktriangleright$}}$  $\pp_1^+<\pp^-$ & $\mapsto$ & $a>\breve{a}_{\aleph}:\; \pp_1^+<\pp_3^-$&$\mbox{\footnotesize{$\blacktriangleright$}}$  $\pp_1^-<\pp^+$&$\mapsto$ &$\pp_1^-<\pp_i^+$
\\
&&&$a\in]\tilde{a}_{\aleph},\breve{a}_{\aleph}[:\; \pp^+_1<\pp_2^-$ $\pp^+_1<\pp_3^-$&&
\\
&&&$a<\tilde{a}_{\aleph}:\;$ $\pp^+_1<\pp_i^-$&&&
\\\\
&$\mbox{\footnotesize{ $\blacktriangleright
$}}$ $\pp^+<\pp_1^-$ & $\mapsto$ & $a>\tilde{a}_{\aleph}:\;\non{\exists}$&$\mbox{\footnotesize{$\blacktriangleright$}}$ $\pp^-<\pp_1^+$&$\mapsto$&$a>a_{{}_u}:\; \pp_i^-<\pp_1^+$
\\
&&&$a<\tilde{a}_{\aleph}:\;$ $\pp^+_1<\pp_1^-$&&&$a<a_{{}_u}:\; \pp_1^-<\pp_1^+\;  \pp_2^-<\pp_1^+$
\\
\hline\hline
\end{tabular}
\end{table}
\subsubsection{The $\ell$counterrotating configurations I: $\cc^+<\cc^-$}\label{Sec:coun-co}
%f
We start by exploring the  bichromatic graph centered on the initial $\cc^+<\cc^-$ state in equilibrium, sketched  in  \textsl{scheme III} of Fig.\il(\ref{Table:Torc}), examples of Boyer surfaces are in Fig.\il(\ref{Figs:ApproxPlo}).
The  second column of Fig.\il(\ref{Table:statescc}) shows  the set of the possible states  of these configurations, and  details on the sequentiality are provided in Table\il(\ref{Table:REDUCtIN}).
We discussed the configuration sequentiality following  Eq.\il(\ref{Eq:conf-decohe}).
The graph  in  Fig.\il(\ref{Fig:CC-CONT})-left  describes all the possible evolutive phases of the centered  $\pp^+<\pp^-$ system.
 We can therefore give some conclusions, comparing   with the graph of Fig.\il(\ref{Fig:doub-grap-ll-cor}) for the $\ell$corotating  couples, describing   also  a bichromatic graph  in a static ($a=0$) spacetime.
   Similarly to the $\ell$corotating  case, the state lines and their evolution are essentially  independent from  the class of attractors.

Considering  the cases where the equilibrium state  may evolve towards the
 $\cc_{\times}$ topologies associated to  accretion, we conclude  that, if the   inner torus is accreting  then, similarly to the $\ell$corotating torus and to the bichromatic graph in the static geometry, the system can evolve \emph{only} into a state where the {outer} torus is  in its equilibrium topology (the  vertex  $\cc_{\times}^+$ is connected to only one state line). {Moreover,  as  collision between the two equilibrium tori  is  in general  possible,  any instability of the outer torus is inevitably preceded by the destruction of the macro-configuration}.
In fact an inversion in the critical sequentiality is not possible for this couple.
The case  of a bichromatic  graph with the central state  $\cc^+<\cc^-$  is indeed similar to the  bichromatic  graph  representing a $\ell$corotating couple (or the case of static spacetime): mono or bichromatic graphs  in static spacetime and the  bichromatic  ones where  $\cc^+<\cc^-$  for  a  Kerr geometry are indistinguishable on many aspects on states properties and evolution.  In the investigation of  the collision for the  bichromatic graph at $a\neq0$   we should  consider the  opposite relative rotation of the  tori.
Finally, we note that since the inner torus is counterrotating  with respect to the attractor, this system will be confined in the orbital range $r>r_{\mbo}^+$, because for some topologies, as clear from \cite{open}, the inner margin of the torus may be in $]r_{\mbo},r_{\mso}]$, while the tori must be centered at $r>r_{\mso}^+$.

If $\ell\in \mathbf{L1}$ or  $\mathbf{L2}$  all of these configurations are described by  Eq.\il(\ref{Eq:fro-plus-intere}), and therefore they cannot constitute a  $\mathfrak{C}_0$ system. It is therefore evident from  Eq.\il(\ref{Eq:fro-plus-intere}), also
for the peculiar sequentiality of the couples, that $\mathfrak{C}_0$ configurations    show strong similarities with the couple described by the monochromatic graphs.
Besides, from  Table\il(\ref{Table:REDUCtIN})  and considering also Eq.\il(\ref{Eq:fro-plus-intere}),
we find that  the  $\mathfrak{\bar{C}}_{1_b}$- $\cc^+<\cc^-$ couples are
\bea
\mathfrak{\bar{C}}_{1_b}:\quad\pp_1^+<\pp_1^-\quad \forall a,\quad\mbox{and}\quad \pp_2^+<\pp_2^- \quad\mbox{for}\quad a\in]a_{\iota},a_{{}_u}[.
\eea
However, a vertex could also be a $\cc_3$ configuration and it may  be associated to the  first phases of the torus formation. being  far enough from the attractor ($r_{\min}>\bar{\mathfrak{r}}_{\gamma}$) and with large specific angular momentum ($\ell>\ell_{\gamma}$). The torus would, during its evolution, decrease magnitude of specific angular momentum. In  this last  case a decrease of specific angular momentum magnitude, from a $\cc_3$ configurations,   could  be preceded by an  $\oo_{\times}$ topology, for the specific angular momentum transition would be $\mathbf{L3}$ to $\mathbf{L1}$ through $\mathbf{L2}$.
We see that this configuration should be  the most difficult to observe because its  formation  is strongly constrained by the  attractor spin.
More generally, from the  second column Table\il(\ref{Table:REDUCtIN}), we can draw the following  evolutionary schemes while more discussion regarding loops for these couples are in Sec.\il(\ref{App:deep-loop}):
\begin{enumerate}
\item
\textbf{Accretion: $\cc_{\times}^+<\cc^-$ and final states of evolution}

\medskip

The macro-configuration with state $\cc_{\times}^+<\cc^-$  \emph{must} be a  $\bar{\mathfrak{C}}_{1b}$ one,
unless the outer corotating disk is in $\cc_3^-$, which is only possible for the attractors with {$a<\tilde{a}_{\aleph}$}
(this can be seen by considering the first and second column of  Table\il(\ref{Table:REDUCtIN})
 and results of Eq.\il(\ref{Eq:fro-plus-intere})). Therefore, the couples of $\bar{\mathfrak{C}}_{1a}$ \emph{cannot} lead to accretion, and any instability of one torus of the couple will destroy the couple.
Then, in the fields of the  faster attractors, the specific  angular momentum  of the outer disk cannot decrease to $ \mathbf{L1}$ without   destruction of the macro-configuration.
Consequently,  we arrive  to the remarkable conclusion that  for the slow attractor   with $a<\tilde{a}_{\aleph}$   there {must be}  $\pp_1^+<\pp_1^-$--see Fig.\il(\ref{Figs:ApproxPlo}).

This means that   such a double system  is possible \emph{exclusively}  in the geometry of  the slow rotating attractors, with tori centered in $]r_{\mso}^+,\bar{\mathfrak{r}}_{\mbo}^+[$ and $]r_{\mso}^-,\bar{\mathfrak{r}}_{\mbo}^-[$ respectively.
The second notable result is that such a couple is the \emph{only} possible with $\pp^+<\pp_1^-$  and must exist  in the fields of these slow attractors.
Assuming that the inner torus has been formed before or simultaneously with the  formation of the outer torus,
 the \emph{final} states\footnote{This is an arbitrary assumption. We  assume that the torus evolution takes place following a possible decrease, due to  some dissipative processes, of its specific  angular momentum  magnitude  towards the  $\mathbf{L1}$ range where accretion is possible, although fluctuations with increase of the angular momentum are also possible. Then it is reasonable in this framework   to assume that the ringed  disk final state is the  one in which both configurations of the couple are in $\pp_1$. On the other hand, as we have seen, these states can be reached only in few cases and under particular conditions (according to the  sequentiality of the  configurations and magnitude of  dimensionless spin of the attractor). This means that in many cases  before this happens, the macro-configuration would be destroyed for example because  collision.} $\pp_1^+<\pp_1^-$  of their evolution  can be reached only around attractors with
 $a<\tilde{a}_{\aleph}$, when the outer torus  reaches the angular momentum $\mathbf{L1}$.
Thus, the outer corotating torus may be in its last stage of evolution only if the inner counterrotating one is $\pp_1^-$, otherwise the ringed accretion disk would be destroyed due to  merging of the two tori. Any instability of the outer torus  would   in any case lead to the destruction of the macro-configuration, which therefore seems to be unlikely to exists in the ``old'' systems where the tori have reached their last evolutive stages, but they should be a feature of relatively ``young''  systems. This can be seen as a strong indication that the $\pp^+<\pp^-$ couples may not be frequent as double tori  systems  with the exclusion of the  recent population of Kerr black hole attractors.
\item
\textbf{Accretion: $\cc_{\times}^+<\cc^-$ and initial states of evolution towards accretion}

\medskip

If the attractor is slow  enough, i.e. $a<\tilde{a}_{\aleph}$, then the outer torus $\cc^-_o$ can be anywhere according  to the  range of specific angular momentum,  being  part of a system where the inner counterrotating torus is $\pp_1^+$.  This means that   formation of such a double system   is most likely in those geometries.
On the other hand,  if the tori orbit a fast attractor with  $a>\breve{a}_{\aleph}$, then the
 couple can form  only during  the earliest stages when the corotating torus has large angular momentum, i.e., $\cc^-_+=\pp_3$ for $a>\breve{a}_{\aleph}$, or $\cc^-_o=\pp_3$ or $\cc^-_o=\pp_2$ for $]\tilde{a}_{\aleph},\breve{a}_{\aleph}[$--see details in Table\il(\ref{Table:REDUCtIN}).
\item

\textbf{Formation of the couple and the  early stages of evolution}

\medskip

During the  evolution from an equilibrium torus $\cc$ to the unstable (accretion) topology $\cc_{\times}^1$,   magnitude of the torus specific   angular momentum generally decreases preserving the state sequentiality.
Then we can provide constraints on the formation of these couples identifying  conditions  for appearance of these couples form in some geometries  at some stages in the evolution of the inner counterrotating  torus towards the accretion.
To carry out these arguments,  we assume   three hypothetical stages of the torus evolution: an early stage formed as a $\cc_3^+$, an intermediate  $\cc^+_2$ one, and  the final $\cc_1^+$ stage eventually leading to $\cc_{\times}^+$. On the other hand, a torus may  be  formed in any of these stages.
We prove  that these couples may  be formed only in certain stages of the inner torus evolution  for some Kerr attractors. This analysis in turn sets    significant limits on  the observational  investigation of these  systems, providing  constraints on  the tori--attractor system, and it  is  able   to  impose     some constraints on the central attractor of an observed  couple.

 From  Table\il(\ref{Table:REDUCtIN}) we see that  configurations formed very  far  from the attractor
and with large angular momentum in magnitude are strongly constrained.
If the  inner torus is formed as a $\cc_i^+=\cc_3^+$ one, then  at this stage the  outer torus  must be a $\cc_3^-$ one necessarily, having    a large angular momentum magnitude; any other solution would inevitably lead to the  collision of  the two tori. This means that possible  formation of a second corotating torus in the early stages of formation of the  counterrotating one is severely limited.
Conversely,  it is clear that a torus with  large angular momentum may be formed under any circumstances not undermining the  evolution of the first vertex of the state and therefore its  evolutive line.

A more complicated situation  occurs,   if the inner torus  is in its intermediate stage with {$\ell=\ell_2$}.
Then the outer  torus  must in all cases undergo  stringent conditions on its specific angular momentum and  the situation depends also on the attractor spin: if $a>a_{\iota} $, then only an outer $\cc^-_3$ torus may be
 formed,  reducing thus the possibility  of the formation of  the double torus  around  the fastest attractors. In the geometry of the  slower attractors where $a<a_{\iota}$,  the  outer torus may be in $\pp_2$.

When  the outer corotating   torus is $\cc_2^-$,
the   double tori systems cannot  orbit
the faster attractors with
$a>\breve{a}_{\aleph}$,
while
for dimensionless spin $a\in]a_{{}_u}, \breve{a}_{\aleph}[$, the inner torus
must be in
$\pp^+_1$. For the slower attractors with $a<a_{{}_u}$   the inner counterrotating torus must be  a
$\pp^+_1$ or a $\pp_2^+$ one.
\end{enumerate}
\subsubsection{The $\ell$counterrotating couple II: $\cc^-<\cc^+$}\label{Sec:tex-dual}
This section is focused on the $\ell$counterrotating configurations with $\pp^-<\pp^+$,   sketched in
\textsl{scheme IV} of Fig.\il(\ref{Table:Torc}).

Bichromatic graph, centered in the initial equilibrium state $\cc^-<\cc^+$, is in Fig.\il(\ref{Fig:CC-CONT})-right.
Possible states are listed in Fig.\il(\ref{Table:statescc}),
details on the sequentiality can be found in Table\il(\ref{Table:REDUCtIN}).

\begin{figure}[h!]
\centering
\begin{tabular}{cc}
\includegraphics[width=.48\textwidth]{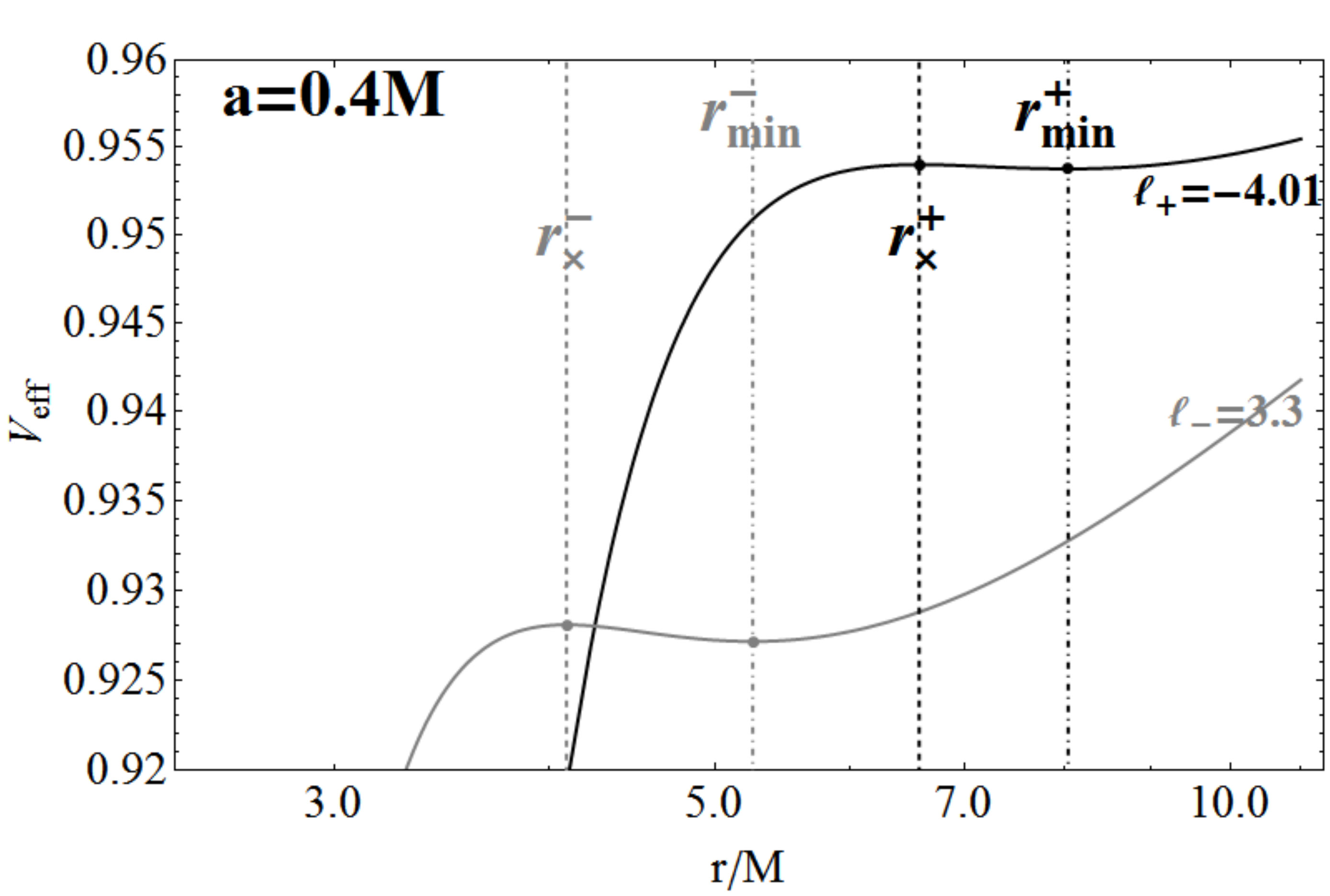}
\includegraphics[width=.48\textwidth]{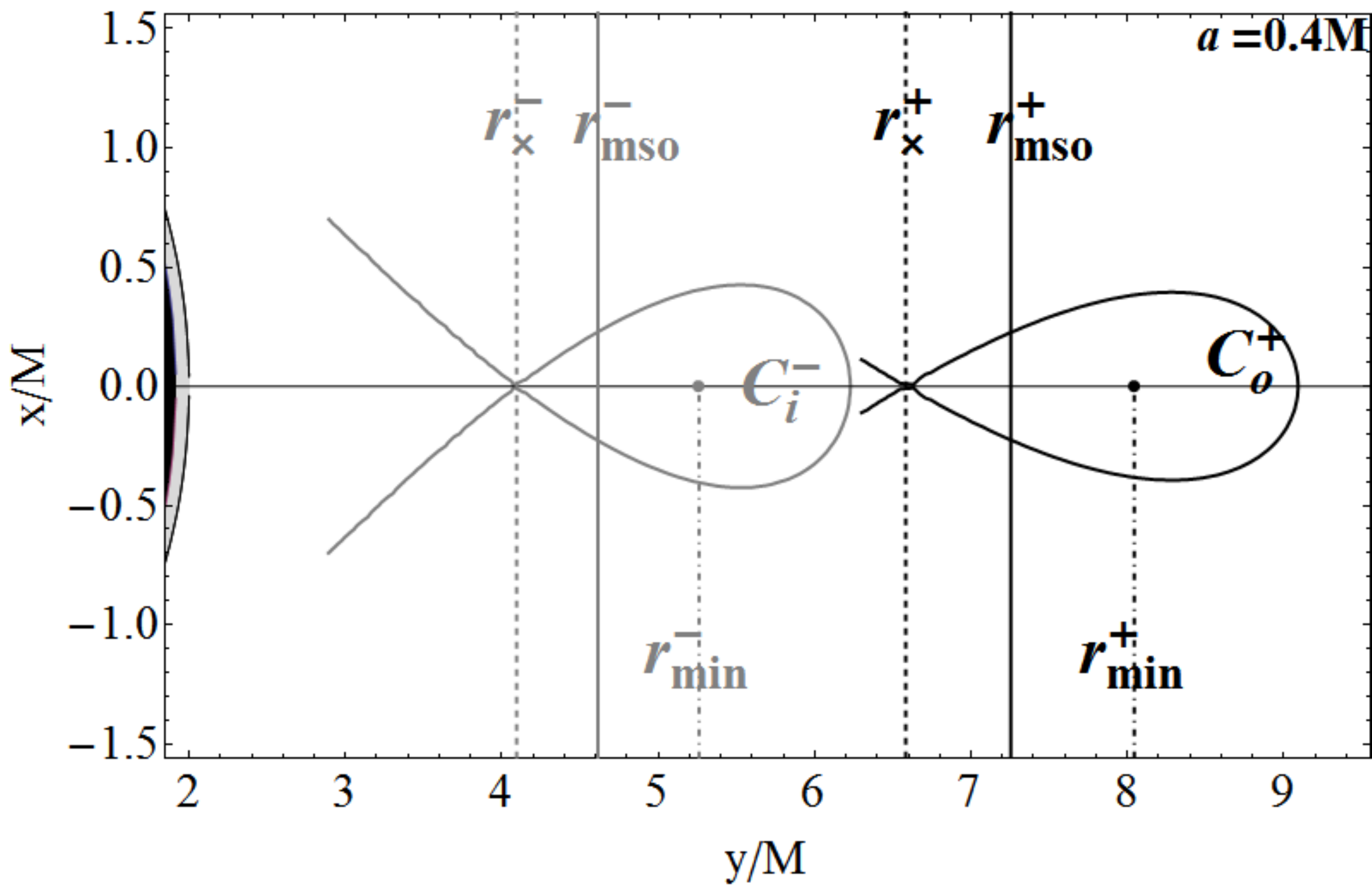}
\end{tabular}
\caption{$\ell$counterrotating  couple  $(\cc_{i}^-,\cc_{o}^+)$ made by an inner corotating accreting  torus and outer counterrotating torus  in accretion orbiting a central Kerr black hole   attractor with spin $a=0.4 M$. Effective potentials (\emph{left-panel}), and  cross sections on the equatorial plane of the outer  Roche lobes  (\emph{right-panel}) corresponding   to \textsl{scheme IV} of Fig.\il(\ref{Table:Torc}).  $(x, y)$ are Cartesian coordinates and  $r_{\mso}^{\pm}$ are the marginally stable circular orbits for counterrotating and corotating matter respectively, $r_{\min}^{\pm}$ are the center of the outer Roche lobe (point of minimum of the fluid effective potentials). Accretion for this couple (from the $r_{\times}$ point) may emerge  from the inner  or the outer torus or even for both the toroidal structure--Sec.\il(\ref{Sec:tex-dual}) and Figs\il(\ref{Fig:CC-CONT})-\emph{left}}\label{Figs:Kind-End}
%%\end{tabular}
\end{figure}

This case   significantly  differers from the  $\pp^+<\pp^-$ one, as illustrated by  graph of  Fig.\il(\ref{Fig:CC-CONT})-right and discussed in Sec.\il(\ref{Sec:coun-co}).
The major difference  for a $\pp^-<\pp^+$ system is  due to the distinctive double geodesic structure  of the Kerr spacetime in the case of corotating inner torus, where in fact the   critical sequentiality is not uniquely  determined by the  configuration sequentiality.
By comparing the two graphs of Fig.\il(\ref{Fig:CC-CONT}), we can note that for $\pp^-<\pp^+$  state,  there are
more state lines   connected by the evolutive lines  for  the inner vertex then  the outer vertex--see also Figs\il(\ref{Figs:Kind-End}).
This  means  that the evolution towards instability  may occur for the $\bar{\mathfrak{C}}_0$ system also from the second counterrotating vertex   or even  from both  the vertices:  this variety  of solutions makes this case  less restrictive than the    $\pp^+<\pp^-$ one, allowing   different  evolutive paths  and favoring several possibilities for the couple  tori formation.

On the other hand,  by observing the third column of Table\il(\ref{Table:REDUCtIN}), providing   \emph{necessary} conditions of  the fixed configuration sequentiality, we note that the states $\pp^-<\pp^+$ are  distinguished  for different Kerr attractors only  in the case of a  $\cc^-_3<\pp_1^+$ couple, which may be formed  \emph{only}  in the geometries of the fast Kerr attractors with   $a>a_{{}_u}$. As we shall discuss at the end of this section,  this has an important consequence on  the formation of  the couple  and the  eventual evolution  towards the accretion, implying  that in  these geometries, also in the early stages of formation of the corotating inner torus, an outer counterrotating  torus  can be formed  evolving  finally  into a  $\pp_1^+$  topology  leading eventually to  accretion--Fig.\il(\ref{Figs:Kind-End}).
Furthermore, as mentioned at the beginning of Sec.\il(\ref{Sec:lcounterrsec}), the
couples $\pp^-<\pp^+$ may give rise to a \emph{class transition} from a $\bar{\mathfrak{C_{1_a}}}$ or  $\bar{\mathfrak{C_{1_b}}}$ class (where instability of the outer torus is forbidden) to a $\bar{\mathfrak{C_{0}}}$ class, with a consequent change in the critical sequentiality--see Eqs\il(\ref{C0},\ref{C1a},\ref{C1b}).
Such a  transition implies the  final state   fulfills   the condition in   Eq.\il(\ref{Eq:(15)}) for  the \emph{specific angular momenta} of the two tori.

On the other hand, we should  consider the  arrangement of the angular  momenta as given in Fig.\il(\ref{Fig:SAPJoke}) and the decoration of the initial state neglecting the size of the torus (the $K$ parameter). More specifically, the class of specific angular momentum  for this couple of configurations depends on the class of attractors  and the  constraints of   Eq.\il(\ref{Eq:(15)})  for a   $\bar{\mathfrak{C}}_{0}$.
Then we need to consider the Kerr attractors  where both the  conditions are met on the   torus specific angular momentum and the definitive  constraints on the radii   $\mathfrak{C}_{1_a}$ or  $\mathfrak{C}_{1_b}$ and the  final state $\mathfrak{C}_{0}$ given by Eq.\il(\ref{C1a}) or Eq.\il(\ref{C1b}) respectively, and the last one of  Eq.\il(\ref{C0})--see also Figs\il(\ref{Figs:Ptherepdiverg},\ref{Fig:OWayveShowno},\ref{Fig:SAPJoke}).

Assuming the transition $\mathfrak{C}_{1_a}\dashrightarrow \mathfrak{C}_{0}$, in accord with  Eq.\il(\ref{C1a}) and Eq.\il(\ref{Eq:so1}), the state in $\mathfrak{C}_{1_a}$ must necessarily be  equilibrium or a  $\cc^-<\cc^+$ couple, which means that if the inner torus is accreting onto the attractor,  it cannot lead  to a class transition.

Considering  apart the possibility of $\cc_3$ states, we focus on the toroidal configurations covered by the classification  in Eqs\il(\ref{C0},\ref{C1a},\ref{C1b}), and we list  here the states $\pp^-<\pp^+$   in these  different classes.
From Table\il(\ref{Table:REDUCtIN}), and considering also Eqs\il(\ref{C0},\ref{C1a}), we obtain
\bea\label{Eq:nat-bh}
 &&
\bar{\mathfrak{C}}_{1b}:\quad \pp_1^-< \cc_2^+\quad\mbox{for}\quad
 a<a_{\gamma_+}^-\quad\mbox{or}\quad\pp_2^-<\cc_2^+ \quad\mbox{for}\quad a<a_{\iota},
 \\
&&\label{Eq:wa-o-kno}
\bar{\mathfrak{C_0}}:\quad\pp_1^-<\pp_1^+ ,\quad  \pp_2^-< \pp_1^+\quad\quad\mbox{for}\quad a>a_{\iota},\quad
 \pp_1^-< \pp_2^+\quad \mbox{for}\quad a>a_{\gamma_+}^-,\;\mbox{according to}\; \mbox{Eq.}\il(\ref{Eq:(15)}),
\\
 &&\label{Eq:wa-o-knso}
 \bar{\mathfrak{C}}_{1a}:\quad \pp_1^-<\cc_1^+,\quad  \pp_2^-< \cc_1^+,\;
 \pp_2^-<\cc_2^+ \quad\mbox{for}\quad a>a_{\iota},\quad
 \pp_1^-< \cc_2^+\quad \mbox{for}\quad a>a_{\gamma_+}^-.
\eea
where  we used property of  $\bar{\mathfrak{C_{1_b}}}$ for
Eq.\il(\ref{Eq:nat-bh}), property Eq.\il(\ref{C0}) for Eq.\il(\ref{Eq:wa-o-kno}),  and  property
 Eq.\il(\ref{C1a}) for Eq.\il(\ref{Eq:wa-o-knso}) and finally in Eq.\il(\ref{C1b})   the  first and third column of Table\il(\ref{Table:REDUCtIN}) has been taken into account.

In the following
 we will concentrate  primarily on the  $\cc$ and  $\cc_{\times}$ topologies referring  to   Eq.\il(\ref{Eq:conf-decohe}) for the relation between specific angular momentum and having in mind results of Table\il(\ref{Table:REDUCtIN}). Further  discussions regarding loops  in these $\ell$counterrotating  couples are in Sec.\il(\ref{App:deep-loop}).
\begin{enumerate}
\item
\textbf{Accretion  and final states of evolution}

The following
accretion states are possible:
\bea
\cc^-<\; \prec \cc_{\times}^+\quad (\bar{\mathfrak{C_0}}), \quad
\cc_{\times}^- <\; \prec \cc_{\times}^+ \quad (\bar{\mathfrak{C_0}}); \quad \cc_{\times}^-<\cc^+ \quad \mbox{with an antecedent state}  \quad \cc^-<\cc^+,
\eea
--see Fig.\il(\ref{Fig:CC-CONT}).
Geometrical correlation,  and then collision, is generally possible.  The
critical sequentiality  of  the couple remains  undetermined if  the   outer vertex   is in equilibrium--see Table\il(\ref{Table:REDUCtIN}). If the outer vertex is unstable in fact, then it must be a $\bar{\mathfrak{C}}_{0}$ of Eq.\il(\ref{Eq:wa-o-kno}) (for $\ell_-\in \mathbf{L1}$ or $\mathbf{L2}$), if the outer torus is in equilibrium, then it may  be a $\bar{\mathfrak{C}}_{1_b}$, $\bar{\mathfrak{C}}_{1_a}$ or also $\bar{\mathfrak{C}}_{0}$, according to Eq.\il(\ref{Eq:wa-o-kno}) and Eq.\il(\ref{Eq:wa-o-knso}).
Considering the third column of Table\il(\ref{Table:REDUCtIN}),
if the inner torus  is in a final stage of evolution, eventually accreting onto the black hole, then the outer torus could  acquire  any angular momentum.\footnote{
We note that  the inner corotating   torus, orbiting Kerr attractors with  $a> a_2=(2\sqrt{2}/3)M\approx 0.942809M$
($r_{\mso}^-(a_2)=r_{\epsilon}^+$), can  be centered inside  the ergoregion or also partially or  totally contained in this, being therefore    not correlated with  the  counterrotating tori \cite{pugtot,ergon}.}

Differently,  if the outer counterrotating  torus is $\pp_1^+$  and    it is   in its  last evolutive phase, according to the evolutive framework assumed here, then the inner corotating ring  could be in any evolutive stage (as long as  the  constraint of no  penetration of matter is  fulfilled)   if orbiting  the fast attractors with $a>a_{{}_u}$. The formation of a $\pp_1^+$  outer torus is in principle possible at any stage of evolution of the inner torus (i.e. for any $\ell_i$).
On the other hand, for the slow attractors with   $a<a_{{}_u}$,  the corotating ring must be in an intermediate or in its last evolutive phase. As mentioned earlier,  the existence of a couple $\cc_3^-<\pp_1^+$ is possible only for Kerr attractors with   $a>a_{{}_u}$-- Table\il(\ref{Table:REDUCtIN}).

Finally, the accretion from the outer configuration may be possible only in the class $\bar{\mathfrak{C}}_{0}$ of Eq.\il(\ref{C0}) and,  in accordance with the constraints of  Eq.\il(\ref{Eq:(15)}), could be also consequence of transition from an equilibrium state in $\bar{\mathfrak{C}}_{1_a}$  or  $\bar{\mathfrak{C}}_{1_b}$.

We focus on the emergence of an unstable phase for the outer vertex corresponding to the
last configuration to be formed\footnote{ Eventually in a very simplified scenario one can assume the inner torus with elongation range $\Lambda$, may even be formed   after or simultaneously with formation of an outer torus  from some local material.}.
Remarkably, the outer configuration can be in accretion
for each attractor, but for  slow attractors it  is limited only to the final stages of evolutions $\pp_2$ and $\pp_1$, for the
 corotating inner ring which cannot be  $\cc^-_3$.
 For slow attractors,  $a<a_{{}_u}$, the outer  torus cannot accrete on an inner  corotating torus in the early stages of development  $\cc_3^-$   this is prohibited due to  Table\il(\ref{Table:REDUCtIN}).
We achieve the remarkable result that for  an accreting  torus  corotating with the Kerr attractor
 there is \emph{no} inner, corotating or counterrotating  torus being  between the accreting torus and the  attractor.
On the other hand, there can be only an  inner \emph{corotating} torus
if the outer accreting  ring is  counterrotating with respect to the attractor (an outer torus in accretion is forbidden also in the couples formed by  the $\ell$corotating surfaces with counterrotating  tori).

 Finally we note that the class of the angular momentum of the inner torus can be inferred from results of Table\il(\ref{Table:REDUCtIN}).
\item
\textbf{Accretion:  intermediate phases $\pp_2$}

We now focus  on the intermediate $\pp_2$ evolutive phases.  The considerations outlined  in  Eq.\il(\ref{Eq:nat-bh}) hold.
We note that this phase  is the one requiring in general  fewer constraints   on the vertex decoration. In fact,   for both $\pp_2^{\pm}$ cases  each of the two vertices may be, independently on the spin  attractor, in any evolutive stage considered the other in $\pp_2$. For all these reasons we can say that the formation and stability of such a couple with a $\pp_2$ configuration is  the  less constrained,  while the formation and stability of  a $\pp^-<\pp^+$ couple  would be hampered  in the earliest or  latest evolutive stages.

\item

\textbf{Couple formation and  the  early stages of evolution}

\medskip

We now consider the case of toroidal configurations which are  far  away of the attractor ($r>r_{\gamma}^{\pm}$)  in a first phase of their formation,  with large magnitude of  specific  angular momentum ($\mp\ell_{\pm}>\mp\ell_{\gamma}^{\pm}$). First, from Table\il(\ref{Table:REDUCtIN}) we see that  the  inner vertex can be in any topology, if the outer  configuration  is  $\cc_3^+$, which is associated with the earliest stages of formation.

If the inner ring is    $\cc_3^-$,
then  for large spin,  $a>a_{{}_u}$, a double system may be formed with  the outer torus   having  different angular momentum allowed (see also the  case of $\cc^+<\cc^-$ in second column of  Table\il(\ref{Table:REDUCtIN})),
 whereas  for slow attractors, $a<a_{{}_u}$,  only  configurations $\pp^+_3$  and $\pp^+_2$ could be  considered for  the formation of  a double system and therefore as initial states towards the accretion. This may be important when  formation of the double system  occurs almost simultaneously, or if the double tori can be formed in later  phases of the life of  the inner torus-attractor system. In this evolutive scheme   we could say that these tori can also be formed almost simultaneously in  any Kerr geometries,  but in the spacetimes of the  slow attractors,   the outer torus must have  sufficiently large specific angular momentum. Thus  these couples are probably  formed around the faster attractors.

 \item
\textbf{Evolution paths towards the  accretion }
The evolution of the  $\cc$  topology towards  the accretion phase $\cc_{\times}$  might generally  happen along several different paths according  to the initial specific angular momentum of the equilibrium configuration. In fact, it could possibly give rise to an composite evolutive line, involving more then two state lines and determined by the composition of two intermediate states in which, for example, from an initial $\cc_3$ configuration    the torus  reaches, due to  loss of the  specific angular momentum magnitude, the  $\cc_{\times}^1$ topology of the  accretion. Then, referring to Fig.\il(\ref{Fig:CC-CONT})-right   and avoiding   to discuss the possible loops,
we concentrate on  the part of graph formed by the only vertices
$(\cc_{\pm}, \cc_{\times}^{\pm})$.
Loops are discussed in Sec.\il(\ref{App:deep-loop}).
We suppose that  the accreting \emph{inner} torus, reaching its maximum
elongation on the equatorial plane  $\lambda=\lambda_{\times}$,    does not collide with the outer vertex (the related conditions are addressed in Sec.\il(\ref{Sec:non-rigid})).
Then we obtain  two possible processes:
\bea\label{Eq:a-bbh}
\mathbf{(a)}:\;\cc^-<\cc^+\, \dashrightarrow \, \cc_{\times}^-< \cc^+\quad \mbox{or}\quad \mathbf{(b)}:\; \cc^-<\cc^+\, \dashrightarrow\, \cc_{\times}^-< \cc^+\, \dashrightarrow \, \cc_{\times}^-< \cc_{\times}^+,
\eea
demonstrated in  Fig.\il(\ref{Fig:CC-CONT}). The  process  $\mathbf{(a)}$ of Eq.\il(\ref{Eq:a-bbh})  may not involve  an evolution of the outer configuration remaining, in accordance with the constraints discussed in Table\il(\ref{Table:REDUCtIN}), in the equilibrium topology.
On the other hand, for a state in $\bar{\mathfrak{C}}_{0}$,   the outer torus can reach   the stage of accretion prior to or together with the inner torus,  according to the evolutive lines of the graph in  Fig.\il(\ref{Fig:CC-CONT}). In this case, by considering also  Eq.\il(\ref{Eq:nat-bh}), we obtain the following two evolutive paths:
\bea\label{Eq:non-sile-voic}
\mathbf{(c)}:\;\cc^-<\cc^+\, \dashrightarrow \, \cc^-< \cc_{\times}^+\quad \mbox{or}\quad \mathbf{(d)}:\; \cc^-<\cc^+\, \dashrightarrow\, \cc^-< \cc_{\times}^+\, \dashrightarrow \, \cc_{\times}^-< \cc_{\times}^+.
\eea
The $\mathbf{(b)}$ path of Eq.\il(\ref{Eq:a-bbh})  and $\mathbf{(d)}$ path  of Eq.\il(\ref{Eq:non-sile-voic}) represent   an extension  of the  paths $\mathbf{(a)}$ and $\mathbf{(c)}$ respectively. Assuming that \emph{after} (or simultaneously)  with the emergence of the  instability of one  vertex, an instability also in the other vertex of state  may occur, which may be independent. This is  contrary to  situations  as of the   $\pp^+<\pp^-$ couples, where such an extension is not possible because it must be preceded by  merging with destruction of the couple\footnote{This can result in an evolution towards both the accretion and  the  $\oo_{\times}$ configuration, only if the outer torus is
 in $\oo_{\times}^{+}$ or  in equilibrium, which means that large specific angular momentum is required. We note that if the outer torus  cannot be in equilibrium,  then in some cases there is no correlation. Instead, the outer torus can grow up to  $\oo_{\times}^+$ only in sufficiently slow spacetimes, $\mathbf{A}_{\iota_a}^<$,  where a  correlation is  possible, and the torus  can be  a  $\oo_{\times}^-$ configuration  for  slow attractors of the class  $\mathbf{A}_{\iota}^<$.}.
 %\\
 %
\end{enumerate}
\subsection{Collisions,  emergence of the $\mathbf{C}_{\odot}^{2}$ macro-configuration and merging}\label{Sec:non-rigid}
Collision  among the tori may  take place as consequence   of the following  mechanisms:   In the couple $\pp^-<\cc_1^+$, it may occur only as impact of the inner Roche lobe of the outer accreting torus on the inner torus. Conversely, collision may involve  only an evolution  of the outer Roche lobe of the inner torus,  even for the two equilibrium tori, with the formation of a  $\mathbf{C}_{\odot}^{2}$  or $\mathbf{\cc^x}_{\odot}{}^2$ macro-configuration where   $y_1^i=y_3^o$. The effective potential for such a ringed disk is
\bea
\label{Eq:on-t-Grav}
&&\left.V_{eff}^{\mathbf{\cc_{\odot}^2}}\right|_{K_i}=
V_{eff}^{i}(\ell_i)\Theta({y}_{\odot}-y)\bigcup
V_{eff}^{o}(\ell_o)\Theta(y-{y}_{\odot}),\quad\mbox{where}\quad y_{\odot}\equiv y_{3}^o=y_1^i,\quad \bar{\lambda}_{\mathbf{C}_{\odot}}=0,\quad {\lambda}_{\mathbf{C}_{\odot}}=\lambda_i+\lambda_o,
\eea
see Fig.\il(\ref{Fig:Goatms}).
Such a  system may arise  as a consequence of the accretion of the  inner ring which,   reaching  the  maximum elongation $\lambda_{\times}$ on the equatorial plane at the emergence of   the instability,  impacts on the outer equilibrium torus. It is necessary therefore   to  change   the inner torus parameters  only.  On the other hand, it is clear that the  condition  $y_1^i=y_3^o$ could follow also from   a change in the outer torus morphology only,  not involving an instability in any ring of the couple.
More generally,   for the occurrence of such
collisions  between  two tori where the  outer  tori is quiescent,
 conditions for  correlation must be matched \cite{open}. From  Fig.\il(\ref{Table:statescc})  and  Fig.\il(\ref{Table:statesdcc}) we can infer    necessary conditions  for the  states of $\ell$counterrotating  and $\ell$corotating couples to  be separated (not correlated) preventing the emergency  of collision.
 However, for all the other states,   we  look for  the relations between the specific angular momenta $(\ell_i, \ell_o)$ of the two tori  such that it is possible to find a couple   $(K_i, K_o)$ for which a geometrical correlation can occur    implying   condition $y_1^i=y_3^o$ in terms of relations between the couples of parameters $(K_i, K_o)$ and  $(\ell_i, \ell_o)$.
 First,  the  \emph{necessary} conditions on the \emph{outer} torus for collisions in the macro-configuration to occur read:
\bea\label{Eq:w-ful}
&&\mbox{for}\quad \pp_o\neq\pp_1^o\quad \mbox{there has to be a well defined effective potential}\;\;
V_{eff}(\ell_o)<1\quad\mbox{in}\quad [y^i_1, r_{\min}^o[,
\\\nonumber
&&\mbox{which particularly means:}
 \\
 &&\label{Eq:w-ful1}
\qquad \mbox{for }\quad \pp_o=\pp_2^o \quad\mbox{there has to be }\quad y^i_1\in]\bar{\mathfrak{r}},r_{\min}^o[\quad\mbox{where}\quad \bar{\mathfrak{r}}\in[r_{\mbo}^o,r_{\min}^o[:\; V_{eff}(\ell_o,\bar{\mathfrak{r}})=1
 \\\label{Eq:w-ful2}
&& \mbox{while  for}\quad  \pp_o=\pp_o^1 \quad\mbox{this condition has to be supplied with }\quad y_1^i\in]r_{\max}^o,r_{\min}^o].
\eea
In general these conditions hold for   $r_{\min}^i<r_{\max}^o$ or $r_{\max}^i<r_{\max}^o<r_{\min}^i$
or also for $r_{\max}^o<r_{\max}^i<r_{\min}^i $.
However,  Eqs\il(\ref{Eq:w-ful}--\ref{Eq:w-ful2}) imply\footnote{A relation as $r_{\bullet} \in \pp$ stands for the inclusion of a radius $r_{\bullet}$ in the configuration $\pp$ (location of $\pp$ with respect to $r_{\bullet}$) according to some conditions; viceversa
  $\non{\in}$ non inclusion.}   \textsl{\textbf{1.}} $r_{\mbo}^o\in \cc_i$ or \textsl{\textbf{2.}} $r_{\mbo}^o< y_3^i$. The first condition holds only for the $\ell$counterrotating   couples as described  in  \cite{open}, the second condition instead,   includes  also  the $\ell$corotating couples,  and considers also the case  $r_{\mbo}^o\non{\in}\cc_i$  which, for example, is always verified for  $\cc^+<\cc^-$  where $r_{\mbo}^o<r_{\mbo}^i$.

The \emph{necessary} conditions of  the \emph{inner} torus   for collision to occur  read
\bea\label{Eq:theo-sit}
&&\mbox{for}\quad  \pp_1^i:\;V_{eff}(\ell_i,y_3^o)\leq K_{\max}^i,
\\\label{Eq:theo-si-1t}
&&\mbox{for}\quad  \pp_2^i\quad  \exists\quad\mbox{always}\quad  K_i: \; \mbox{(\ref{Eq:theo-sit}) is satisfied and, particularly the following condition also always holds,}
\\\nonumber
&&\mbox{for}\quad   \cc_3^i:\;  \exists \; \bar{\mathfrak{r}}<r_{\min}^i: V_{eff}(\bar{\mathfrak{r}},\ell_i)= V_{eff}(y_3^o,\ell_i)\quad\mbox{and a well defined effective potential  } \;V_{eff}(\ell_i)<1 \;\mbox{in}\;[\bar{\mathfrak{r}},r_{\min}^i].
\\
\eea
Note that there  is $r_{\mbo}^{\pm}\non{\in}\cc_{\pm}$   \cite{open}.
Condition (\ref{Eq:theo-si-1t}) implies that for a $\cc^i_2$ torus it is always possible to find a proper $K_i$ parameter such that
  $y_1^i=y_3^o$--see also Table\il(\ref{Table:REDUCtIN}).
Since  for a $\cc_2^o$ there is $\sup{K_o}=1$, the condition ensuring that   collision  does \emph{not} occur reads $y_1^i<r^o_{\sup}$,  where $r_{\sup}^o:\, V_{eff}^o(r_{\sup}^o)=1$, or   that the potential is not well defined. This  last condition holds also for a  $\cc^o_3$ ring.
  However, for the characterization of  collision in a  $\mathbf{C}^{2}_{\odot}$ macro-configuration we should consider simultaneously the conditions on  the outer edge of the inner ring   and  on  the morphology of  the outer torus. For a  $ \pp_1^i$ ring, because collision does \emph{not} occur, it has to be  $K_o:\, K_{\max}^i<V_{eff}^i(y_3^o)<1$.
  Besides, a parameter   $K_i<1:\; y_1^i=y_3^o$ can always exist for a $\cc_2^i$  disk, as there is $K_{\max}^i\geq1$ with   the effective potential  well defined for $r>r_{\min}$ and asymptotically    $V_{eff}=1$.
 These  necessary but not sufficient conditions  for the collision imply  a precise relation on the ring sequentiality,  according to the constraints provided by  Table\il(\ref{Table:REDUCtIN}). Further restrictions can be found by comparing the inclusion relations of the  notable radii addressed  in \cite{open}.
As  the  $\ell$corotating couples form always a $\bar{\mathfrak{C}}_{1_b}$  ringed disk, these configurations are more likely  leading to collision, particulary for the couples made by  two  $\cc_1^{\pm}$  tori, where collision is always possible.
Other cases as the  $\cc_1^{\pm}<\cc_2^{\pm}$  couples imply  satisfaction of the property $ V_{eff} (y_{3}^1,\ell_2)<1$, which is
favored in the case  where   $\ell_2/\ell_1\approx1$ and  $K_1/K_2\gg1$, or  as  $|\ell_2|= |\ell_{\mbo}|+\epsilon_+$ and  $|\ell_1|= |\ell_{\mbo}|-\epsilon_-$  where $\epsilon_{\pm}\gtrapprox0$.
Analogous relations hold for  $\cc_2^{\pm}<\cc_2^{\pm}$, $\cc_2^{\pm}<\cc_3^{\pm}$ and $\cc_3^{\pm}<\cc_3^{\pm}$.
Furthermore,  from the analysis of critical and configuration sequentiality,
Table\il(\ref{Table:REDUCtIN}) shows  some  necessary but not sufficient conditions for collision emergence  constraining  also the $\ell$corotating configurations  which, according  to  the only constraints of  Eq.\il(\ref{Eq:on-t-Grav}),   in principle   may lead to collision.
However,  by considering  the effective potential  in Eq.\il(\ref{Eq:def-partialeK}), we can obtain an immediate relation  for   colliding  configurations in $\mathbf{C}_{\odot}^{2}$  in the  $\ell$corotating  case:
 \bea\label{Eq:emp-evid-not}
 K_i\in]K_{\min}^i,{V_{eff}(\ell_i, r_{\min}^{o})}]\quad\mbox{and}\quad
K_o\in]K_{\min}^o, V_{eff}(\ell_o, y_1^{i})] \quad\mbox{for }\quad  \ell_i\ell_o>0,
 \eea
\cite{ringed}.  Then  Eq.\il(\ref{Eq:emp-evid-not}) is a  necessary condition for  a ringed  disk  of order two, represented in the
  \textsl{schemes I } and \textsl{II} of Fig.\il(\ref{Table:Torc}), to  evolve into a   $\mathbf{C}_{\odot}^{2}$ configuration.

On the other hand, the situation for a  $\ell$counterrotating  couple is particularly complex, depending on  state correlation and the possible sequentiality as sketched in Table\il(\ref{Table:REDUCtIN}).
Moreover, for these couples we cannot easily write down a condition analogue to Eq.\il(\ref{Eq:emp-evid-not}). This is   due to the fact that, as seen in  Sec.\il(\ref{Sec:lcounterrsec}), for the $\ell$counterrotating couples the order relation between magnitude of specific angular momenta and the location of the disks and the effective potential at minimum points are not straightforwardly traced.
In the following we shall focus mainly on the  $\ell$corotating couples.

\medskip

\textbf{Collision after growing of the \emph{outer} ring}

\medskip

We  focus first on the  $\ell$corotating couples.
 From Fig.\il(\ref{Table:statescc}) we know that  $\cc^i_{\pm}<\cc^o_{\pm}$ implies  $\cc^i_{\pm}\succ \cc^o_{\pm}$, then we obtain that for
 $\ell_i<\ell_o$, when  $\ell_i $ and  $\ell_o$ are  in  $\mathbf{L1}$ or $\mathbf{L2}$, it is  possible to find a proper
$ K_o$ for  the emergence of a $\mathbf{C}_{\odot}^{2}$ configuration.
Therefore, in an   initial  separated couple with $\{\ell_i,\ell_o\}$ in $\mathbf{L1}$ or $\mathbf{L2}$,   the outer ring can always grow to a proper $K_o$ to impact on the inner ring and, according with the state selection of Fig.\il(\ref{Fig:doub-grap-ll-cor}), the two tori shall collide  before the \emph{outer} $\ell$corotating ring is accreting.
In fact, there is
$r_{\max}^o<r_{\max}^i<r_{\mso}<r_{\min}^i<r_{\min}^o$, and then  $\sup{y_3^o}=r_{\max}^o<r_{\min}^i<y_{1}^i$.
This is immediate to infer if both the configurations are  $\pp_1$.
If, on the other hand, we have $\cc_o=\cc_2^o$, then there is
\bea\label{Eq:fis-1}
&&
r_{\gamma}<r_{\max}^o<r_{\mbo}\leq\sup{y_3^o}=\bar{\mathfrak{r}}_{1}^o<r_{\max}^i\leq y_{3}^i<r_{\min}^i,\quad \mbox{where}\quad  \bar{\mathfrak{r}}_{1}^o:\; V_{eff}(\ell_o,\bar{\mathfrak{r}}_{1}^o)=1, \quad \bar{\mathfrak{r}}_{\mbo}<r_{\min}^o<\bar{\mathfrak{r}}_{\gamma},
\\
&&
\mbox{if}\quad \cc_i=\cc_2^i\quad\mbox{ then there is also a}\quad    K_i:\; \bar{\mathfrak{r}}_{\mbo}<r_{\min}^i<y_{1}^i\leq y_3^o<r_{\min}^o<\bar{\mathfrak{r}}_{\gamma},
\\
&&
\mbox{if}\quad   \cc_i=\cc_1^i \quad\mbox{then there is also }\quad r_{\gamma}<r_{\max}^o<r_{\mbo}<r_{\max}^i<{r}_{\mso}<r_{\min}^i<\bar{\mathfrak{r}}_{\mbo}<r_{\min}^o<\bar{\mathfrak{r}}_{\gamma}.
\eea
In Eq.\il(\ref{Eq:fis-1}) we considered the fact that
$\ell_{\mbo}$ is the inferior  (in magnitude) of the  $\mathbf{L2}$ range, and there is $V_{eff}(\ell_{\mbo},r_{\mbo})=1$, but $1=V_{eff}(\ell_{\mbo},r_{\mbo})<V_{eff}(\ell_{o},r_{\mbo})$ for any
$\ell_o\in]\ell_{\mbo},\ell_{\gamma}[$. Therefore, there exists  $\bar{\mathfrak{r}}_{1}^o<r_{\min}^o:\; V_{eff}(\ell_o,\bar{\mathfrak{r}}_{1}^o)=1<V_{eff}(\ell_{o},r_{\mbo})$ that is, of course, possible if and only if
$\bar{\mathfrak{r}}_{1}^o\in]r_{\mbo},r_{\min}^o[$ (for there is $\partial_r V_{eff}(r)<0$ at $r<r_{\min}$).
Moreover, we used the property that $r_{\mbo}^{\pm}\non{\in}\cc_j^{\pm}$ for any $\ell_j$.
%On the other hand we have that:
%%
%\bea
%\mbox{if}\quad \cc_o=\cc_3^o\quad  <\bar{\mathfrak{r}}_{\gamma}\leq r_{\min}^o
%\eea
%%

\medskip

\textbf{Collision after growing or  accretion of the \emph{inner} ring}

\medskip

In the argument above, we considered  only the role of the outer configuration of a couple in the emergence of collision,
 it is however clear that in finding  out the condition for  $y_1^i=y_3^o$, we should consider   the couple of parameters
$(K_i, K_o)$.
Particularly,  we  need to investigate  the elongation  of the inner torus  in the equatorial plane,  up to the extreme limit  of the configuration $\oo_{\times}$ or  $\cc_{\times}^1$, eventually  colliding with the outer ring.
First we report here some  immediate considerations  holding  also for the  $\ell$counterrotating couples. We consider, to fix the ideas,  an  $\ell$counterrotating  couple  $(\cc_{i}^-,\cc_{o}^+)$ made by an inner corotating accreting  torus and an outer counterrotating accreting torus  as in
Fig.\il(\ref{Figs:Kind-End}), with maximum elongation $\lambda_{\times}$ in the equatorial plane, and the radii   $\bar{\mathfrak{r}}^{+_i}_{\max}< \bar{\mathfrak{r}}^{+_o}_{\max}$, solutions of $V_{eff}(\ell_+,r_{\max}^+)\equiv K_{\max}^+= \bar{K}_{\max}^+$  for $\ell_+\in \mathbf{L1}_+$, where
 $\bar{\mathfrak{r}}^{+_i}_{\max}$ is the accretion point and $\bar{\mathfrak{r}}^{+_o}_{\max}$ is the  outer edge of the counterrotating torus. The existence of the double system $\cc_{\times}^-<\cc_{\times}^+$ is ensured by the condition
 $\bar{\mathfrak{r}}^{+_o}_{\max} \leq\bar{\mathfrak{r}}_{\max}^-$
 where the equality  holds as condition for the  collision as shown, for example, in Fig.\il(\ref{Fig:Goatms}).
Then, considering
 $\bar{\mathfrak{r}}_{\max}^-:V_{eff}(\ell_-,r_{\max}^-)\equiv K_{\max}^-= \bar{K}_{\max}^-$,
we need to choice
 $(\bar{K}_{\max},\bar{K}_{\max}]^-)$  on the curves
$ V_{eff}(\ell_{\pm},r)$, which is always possible to find  as we can have $-\ell_+<-\bar{\ell}_+: \bar{\mathfrak{r}}^{+_o}_{\max} = r_{\mso}^-$
 and then $-\ell_+\in]-\ell_{\mso}^+,-\bar{\ell}_+[$ \cite{open}.

\begin{figure}[h!]
\centering
\begin{tabular}{lr}
\includegraphics[width=.4\columnwidth]{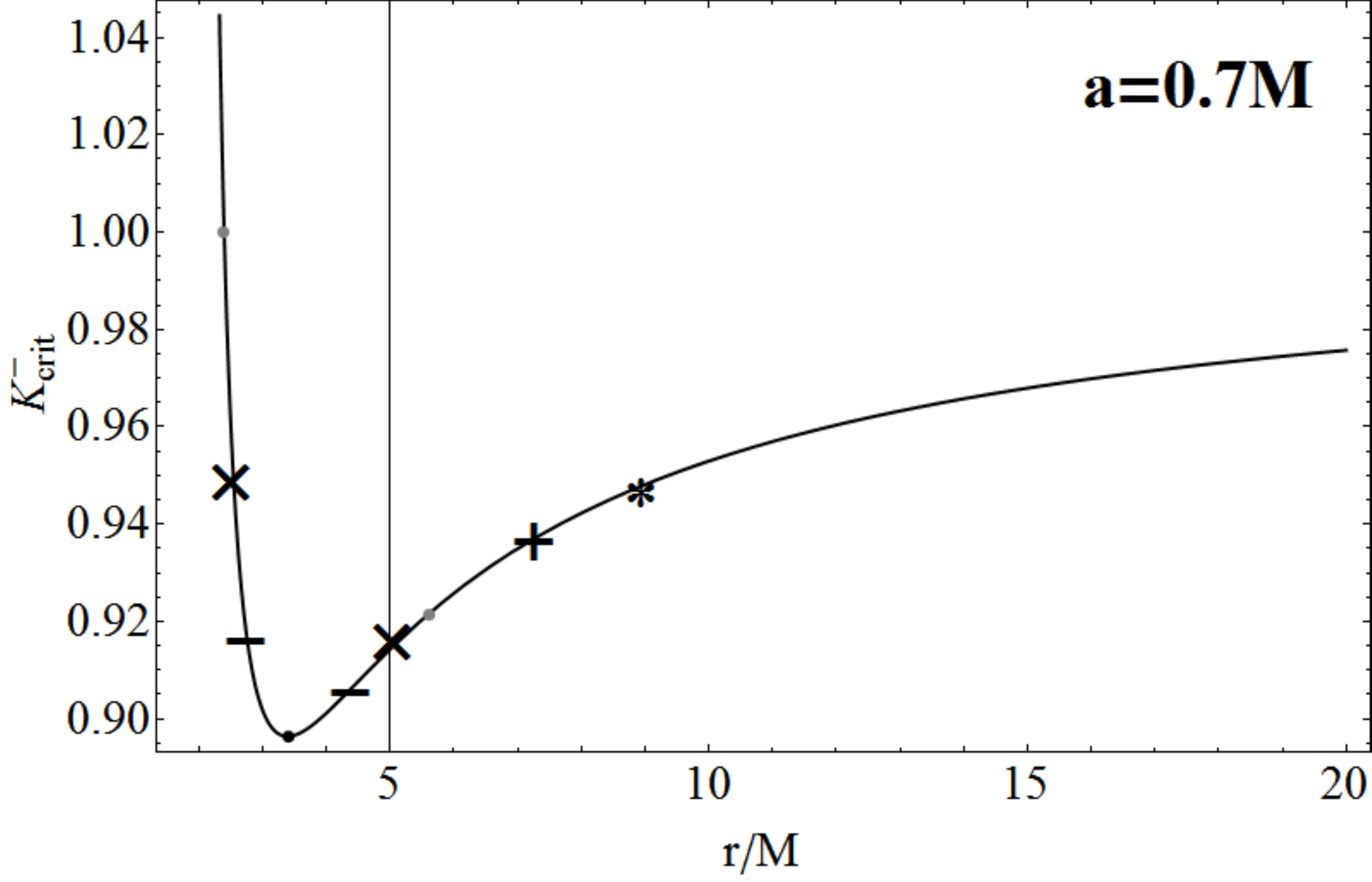}
\includegraphics[width=.4\columnwidth]{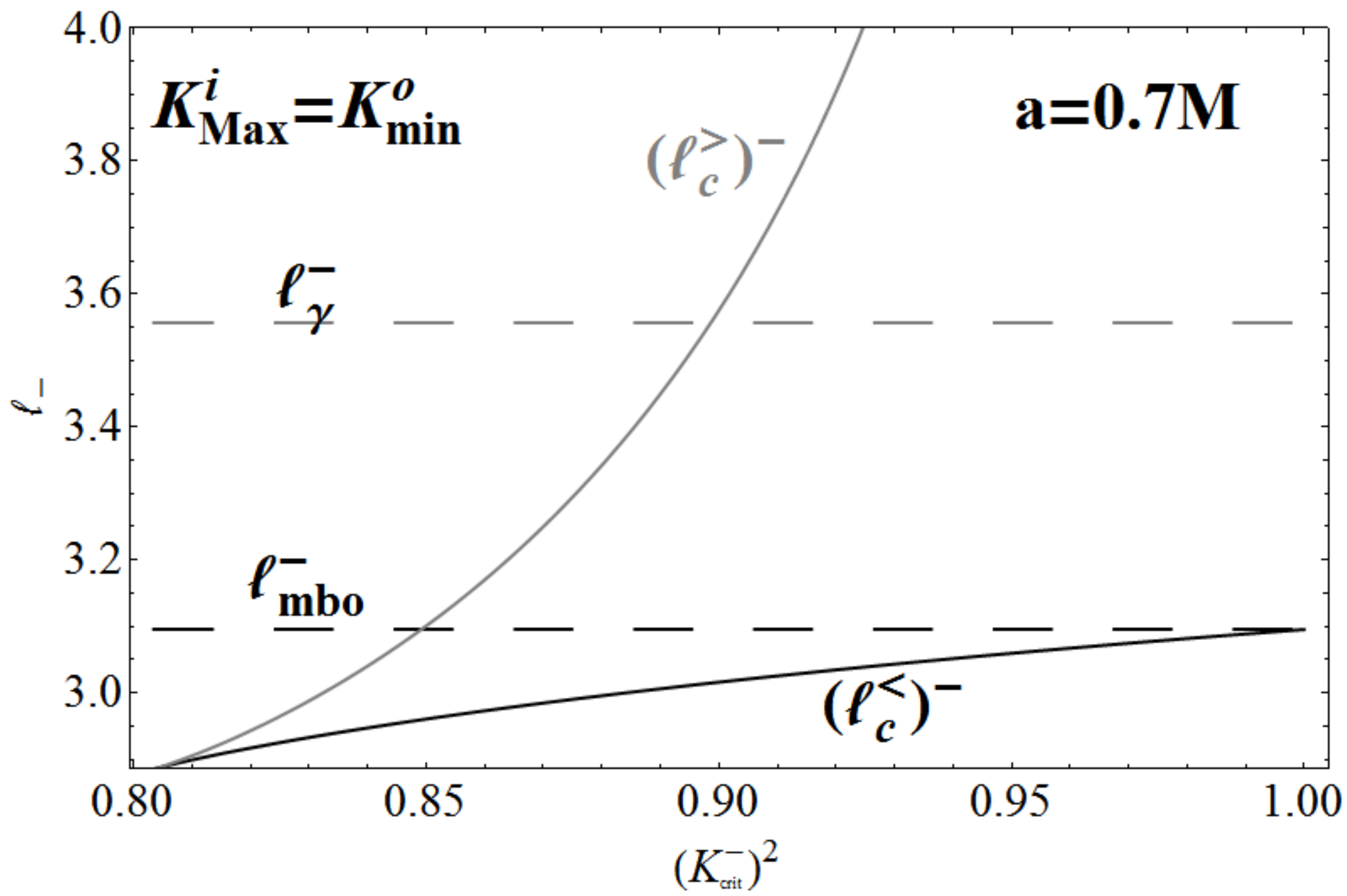}
\\
\includegraphics[width=.4\columnwidth]{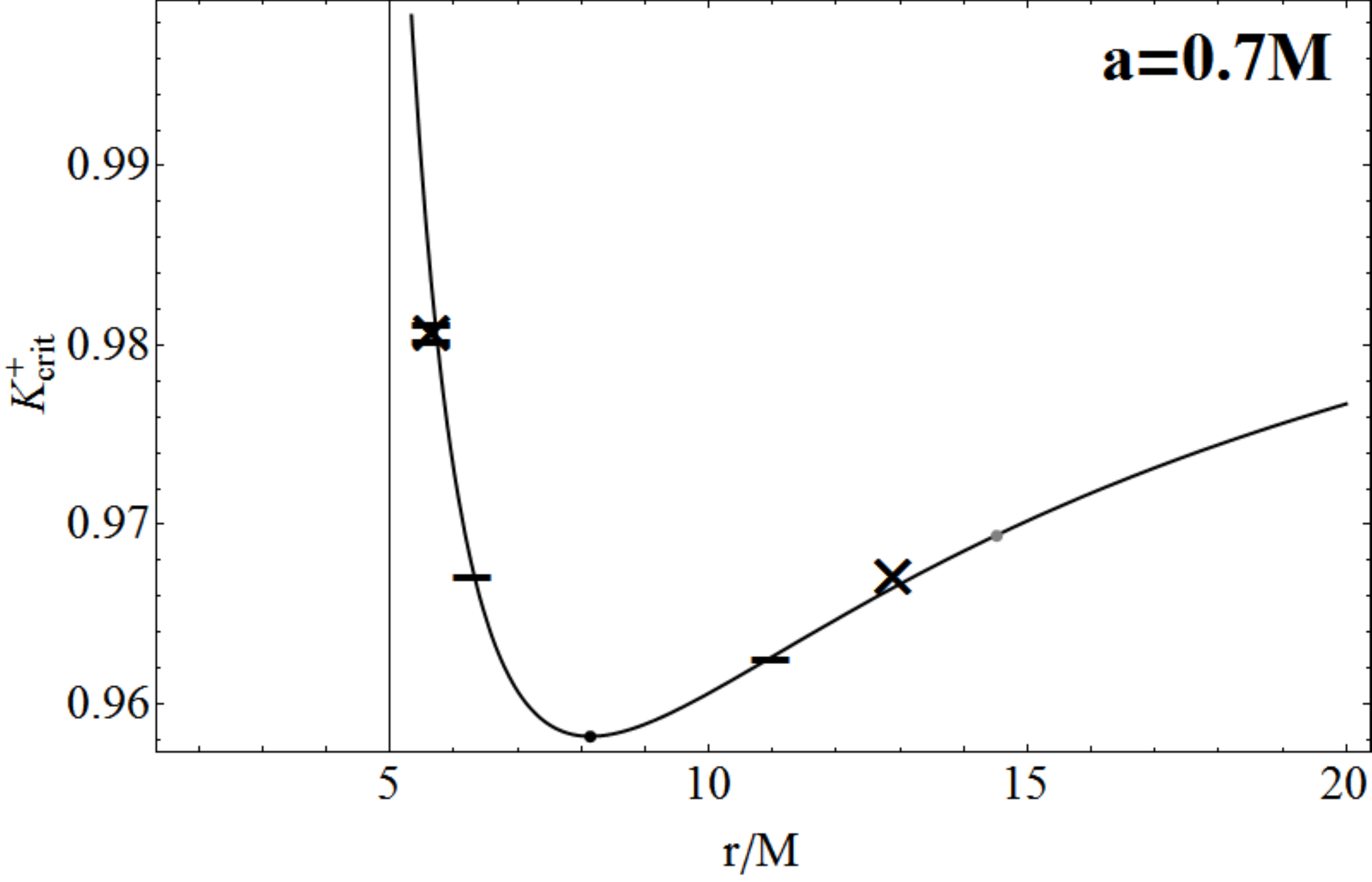}
\includegraphics[width=.4\columnwidth]{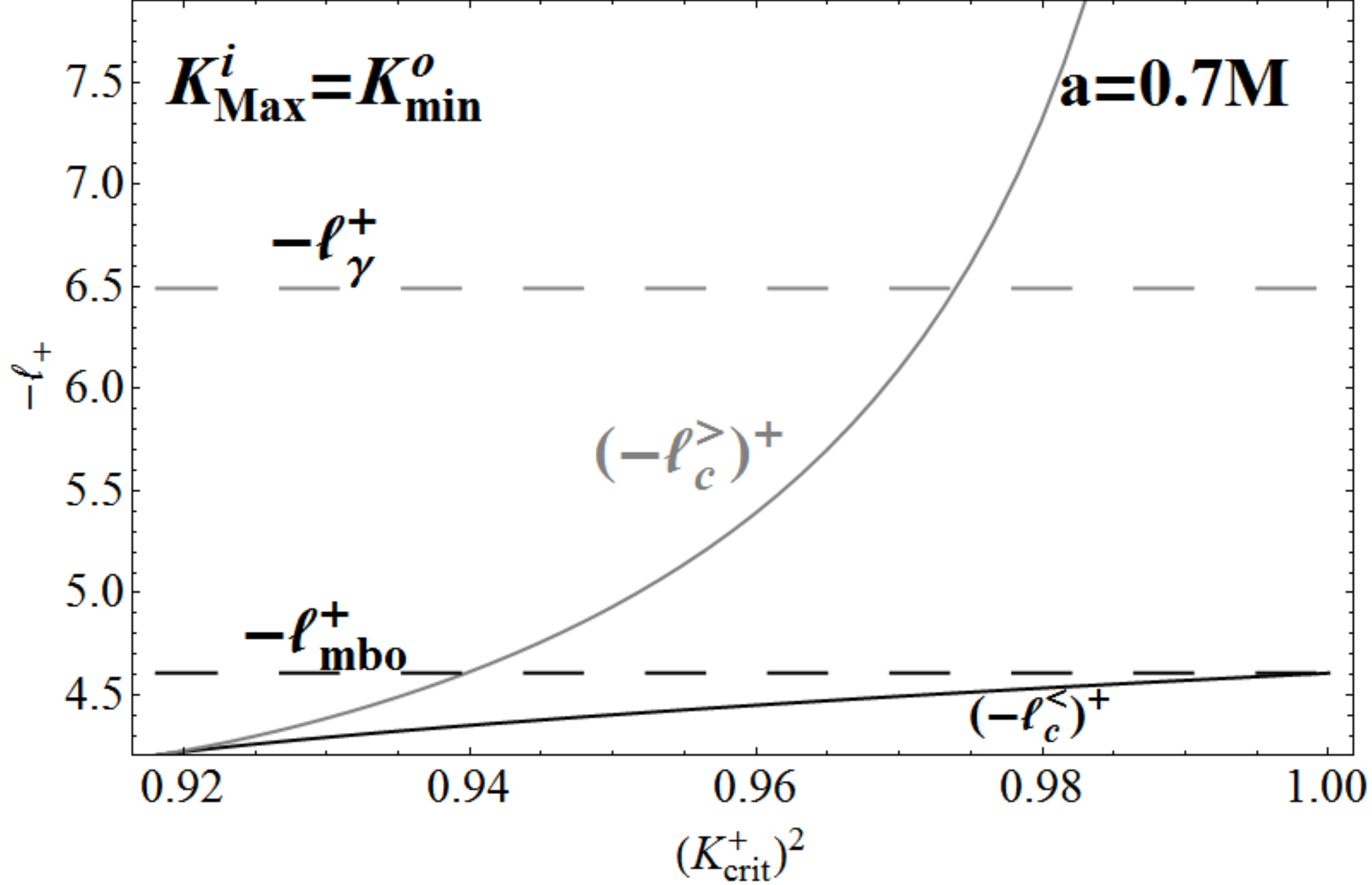}
\end{tabular}
\caption{\footnotesize{$\ell$corotating couples $\cc^{-}_i<\cc^-_o$ (upper panels) and $\cc^+_i<\cc^+_o$ (lower panels). Case $K_{\max}^i=K_{\min}^o$. The specific angular momenta $\ell_{c}^{\lessgtr}:\; V_{eff}(\ell_i,r_{\max}^i)\equiv K_{\max}^i=K_{\min}^o\equiv V_{eff}(\ell_o,r_{\min}^i)$ versus $K_{crit}^2\in\{K_{\max}^2,K_{\min}^2\}$, where $\ell_{c}^{<}=\ell_i$ and $\ell_{c}^{>}=\ell_o$, in the region $\mathbf{K0}$ and  $\ell>\pm\ell_{\mso}^{\mp}$--see Sec.\il(\ref{Sec:non-rigid}). The specific angular momentum $\ell_{\mbo}^{\pm}\equiv\ell_{\pm}(r_{\mbo}^{\pm})$, $\ell_{\gamma}^{\pm}\equiv\ell_{\pm}(r_{\gamma}^{\pm})$ for the marginally bounded orbit, $r_{\mbo}$, and the marginally circular orbit or photon orbit, $r_{\gamma}$, are also plotted. }}\label{Fig:doub-grap-ll-cor-for.al}
%\end{tabular}
\end{figure}

Focusing on the couples  $\cc_{\times}^1<\cc_o$, and using Eq.\il(\ref{Eq:emp-evid-not}), we find
 \bea\label{Eq:emp-evid-not-Lore}
&& K_i\in]K_{\min}^i,{V_{eff}(\ell_i, r_{\min}^{o})}]\quad\mbox{and}\quad
K_o=V_{eff}(\ell_o, y_1^{i})>K_{\min}^o \quad\mbox{for }\quad  \ell_i\ell_o>0,
\\\label{Eq:ang-axes}
&&\mbox{or}\quad K_{\min}^i<K_i=V_{eff}(\ell_i, y_1^{i})\leq {V_{eff}(\ell_i, r_{\min}^{o})}<{V_{eff}(\ell_o, r_{\min}^{o})}=K_{\min}^o<V_{eff}(\ell_o, y_1^{i})=K_o<K_{\max}^o,
 \eea
and assuming an  unstable  inner configuration,  Eq.\il(\ref{Eq:emp-evid-not-Lore}) implies
\be\label{Eq:pos}
K_{\min}^i<K_i=K_{\max}^i\leq {V_{eff}(\ell_i, r_{\min}^{o})}<K_{\min}^o<V_{eff}(\ell_o, y_1^{i})=K_o<K_{\max}^o,
\ee
confirming  that $\pp_{\times}^i\neq \oo_{\times}^2$ \cite{ringed}. However, the necessary condition, Eq.\il(\ref{Eq:pos}),  and particularly
the relation $K_{\max}^i<K_{\min}^o$ is satisfied only in  some special cases. The following two cases may occur
\bea\label{Eq:y-coo}
&& \mbox{\textsl{\textbf{I}}}:\;\; K_{\min}^i<K_{\max}^i<K_{\min}^o<K_{\max}^o\quad\mbox{or}\quad\Delta K_{crit}^i<\Delta K_{crit}^o \quad\mbox{and}\quad \Delta K_{crit}^i\cap\Delta K_{crit}^o=0,
\\\label{Eq:n-coo}
&&\textsl{\textbf{II}}:\;\;  K_{\min}^i<K_{\min}^o<K_{\max}^i<K_{\max}^o\quad \mbox{or}\quad \Delta K_{crit}^i\cap\Delta K_{crit}^o=K_{\min}^o.
\eea
The first follows Eq.\il(\ref{Eq:emp-evid-not-Lore}) and therefore satisfies the necessary condition for collision, the second instead forbids any collision after  instability of the inner ring (which is preceded by merging). The two cases are regulated by the ratio of the specific angular momenta of the tori of  the couple.
Then there has to be a specific  angular momentum  ratio  $\non{\ell}_{c}\equiv\ell_o/\ell_i:K_{\min}^o=K_{\max}^i$, which is the discriminant case between Eq.\il(\ref{Eq:y-coo}) and Eq.\il(\ref{Eq:n-coo})--see \cite{ringed}.
The discriminant case has to satisfy the relation $V_{eff}(\ell_i,r_{\max}^i)=V_{eff}(\ell_o,r_{\min}^o)=V_{eff}(\ell_i,y_1^i)$.
 However in \cite{ringed} it was shown that in the case  $K_{\max}^i=K_{\min}^o$
 the \emph{outer} ring $\cc_o$ cannot be unstable,
%as it follows from discussion in %Sec.\il(\ref{Sec:minimacoinc},%\ref{Sec:procedure}).
 and the inner  ring $\cc_i$ cannot be in accretion. This is because if  $\lambda^i=\lambda^i_{\times}$, then  there is  $y_1^i=y_{\min}^o$. In conclusion, the ringed disk  cannot be unstable according to a P-W instability, neither it can form a $ \mathbf{C}_{\odot}^{2} $ with an inner accreting  torus.  Then the two rings collide before the inner disk will reach the accretion phase.
 At fixed  $\ell_o$, the specific angular momentum in magnitude, for this case there is  $\ell_i=\non{\ell}_i\equiv\ell_o/\non{\ell}_c\in \mathbf{L1}$
 or  $\non{\ell}_c \ell_{\mso}<\ell_o<\non{\ell}_c \ell_{\mbo}$. As $\non{\ell}_c>1$ by definition,
 this  case  will hold for  a part of $\cc_2^o$ and at lest for a part of $\cc_1^o$ and possibly a $\cc_3^o$  tori, according for the condition
$\ell_o>\ell_i (\ell_{\gamma}/\ell_{\mbo})$.
This last condition distinguishes the $\ell$corotating couples of corotating  or counterrotating rings where
$\ell_{\gamma}^+/\ell_{\mbo}^+>\ell_{\gamma}^-/\ell_{\mbo}^-$ and $\partial_a(\ell_{\gamma}^{\pm}/\ell_{\mbo}^{\pm})\gtrless0$.
Then the necessary condition   for colliding $\mathbf{C}_{\odot}^{\times}$ for $\cc_{\times}^1$ may be
 rephrased by saying that  $\ell_i<\non{\ell}_i$ in magnitude.
Consistently,  condition in Eq.\il(\ref{Eq:n-coo})    prohibits also  the inner torus  to be in accretion.
Focusing on the discriminant case,
 with reference to Fig.\il(\ref{Fig:doub-grap-ll-cor-for.al}), we consider the angular momenta associated to this case.
 We find  the solutions of the problem  $K_{crit}^{\pm}=$constant  for the  $\ell$corotating  couples of counterrotating or corotating  tori respectively.
 The  solution provides   the angular momenta related  to the constant  surfaces of the curve $K_{crit}$, as functions of the constant value  $c\geq K_{\mso}^{\pm}$   in the range  $K_{crit}\in \mathbf{K0}$, say $\pm\ell_{c}^{\mathbf{>}}{}^{\mp}\geq\pm\ell_{c}^>{}^{\mp}$.
The two panels are to be read as follows:
the (horizontal)  lines $K_{crit}=$constant  on the first panel provide
$ K_{crit}=K_{\max}^i=K_{\min}^o$ and the associated two radii $r_{\max}^i<r_{\min}^o$, for the two $\ell$corotating configurations with unknown $\pm\ell_{\mp}^i<\pm\ell^o_{\mp}$. Symbols on the curve indicate the couple of $K_{crit}$ associated to an equal $\ell$ and therefore to one ring. On the other hand we could, as done in \cite{ringed,open}, use the curve $\ell(r)$ to find out $(\ell_i, \ell_o)$, through $r_{crit}$ obtained by this first panel, however here we  can get this information alternately by using the second panel of Fig.\il(\ref{Fig:doub-grap-ll-cor-for.al}).
Then, by taking this   $K_{crit}$  on the second panel (vertical line), we select  the two angular momenta $\ell_i=\ell_{c}^<$ and $ \ell_o=\ell_c^>$ respectively  associated to the two  $\ell$corotating rings   with
$K_{\max}^i=K_{\min}^o$.
Thus, on this second panel the (horizontal dashed) lines  $\ell_i=\ell_{c}^<$=constant   and     $ \ell_o=\ell_c^>=$constant   set, crossing  the  curve $\ell_{c}^>$  and $\ell_{c}^<$ respectively,  the
couples $K_{\min}^i$ and  $K_{\max}^o$ (details with configurations with special  $K$  can be found in \cite{ringed}).

Investigating  the topology of the couple, we note that   $\ell_{c}^<$ is well defined also for $K\in\mathbf{K1}$, therefore  $\ell\in\mathbf{L2}$,  and it is  associated  with the instability points (in the region $r<r_{\mso}$).
 The curve associated with the minimum points,  $\ell_{c}^>$ (region $r>r_{\mso}$), instead correctly extends only in   $\mathbf{K0}$ up to  $\mathbf{L3}$,  confirming that  the discriminant case occurs only in $\mathbf{K0}$.
Fixed a $K_{crit}=\bar{K}_{crit}$, then a torus $\mathbf{\cc}_{\odot}^{\times}$ can be formed if  $\ell_o=\ell_{c}^>>\bar{\ell}_{c}^>$ or $\ell_{i}=\ell_{c}^<<\bar{\ell}_{c}^<$, where the values $\bar{\ell}_{c}^{\gtrless}$ are associated to the line  $K_{crit}=\bar{K}_{crit}=$constant. More precisely, we obtain, at fixed  $\ell_o=\bar{\ell}_{c}^>$ (and a proper choice of $K_o$), collision after accretion of the inner ring  \emph{only} if its specific angular momentum is small enough in magnitude i.e.  $\ell_i=\ell_{c}^<< \bar{\ell}_{c}^{<}$, while for larger values of the magnitude of the specific angular momentum, $\ell_i=\ell_{c}^<\geq \bar{\ell}_{c}^{<}$, for any $K_o\in ]\bar{K}_{crit}, K_{\max}^o[\subset \mathbf{K0}$ (where  $ K_{\max}^o:\, \bar{\ell}_{c}^>={\ell}_{c}^<$ if $\bar{\ell}_{c}^>\in \mathbf{L1}$ or  we can take $ K_{\max}^o\equiv1$  if $\bar{\ell}_{c}^>\in \mathbf{Li}$ with $\mathbf{Li}\in\{\mathbf{L2}, \mathbf{L3}\}$ as we are interested  only to the closed $\cc_o$ equilibrium configurations), the possible phase of accretion of the inner torus must be preceded by the collision and possibly merging  of the two rings. Collision by accretion,  for $K_o$ small enough, can take place only for $\ell_i<\bar{\ell}_c^{<}{}^-$. We can therefore also provide an upper boundary for the angular momentum of the inner surface as $\bar{\ell}_{c}^>\in \mathbf{L1}$. In fact  as clear from  Fig.\il(\ref{Figs:Ptherepdiverg}), when the angular momentum of the inner torus reaches this limit, then the two tori overlap completely $(r_{\min}^i=r_{\min}^o)$, if $\bar{\ell}_{c}^>\in \mathbf{L1}$.
 Whereas, if $\bar{\ell}_{c}^>\in \mathbf{Li}>\mathbf{L1}$, then $\ell_i$  is    bounded  by $\ell_{\mbo}$ from above.
 This trend is   qualitatively independent from the spin of the attractor and the  direction of rotation of the tori with respect to the attractor.
Finally, condition (\ref{Eq:pos}) may also hold for a  $\cc^o_3$ ring, because the maximum  of the outer configuration is  not actually  involved.  Therefore, the problem will in turn be
how small should $\ell_3$  to be for the elongation of the inner ring in accretion matching  the outer ring.
Further  discussion regarding  the possible  loops in the double systems are in Sec.\il(\ref{App:deep-loop}).
\section{Observational evidence of doubled tori disks and their evolution}\label{Sec:a-ew-one}
{The systems  investigated here and   in  \cite{ringed,open} offer a  methodological challenge of describing a set of virtually separated  sub-systems as an entire configuration.
The double  tori of the ringed accretion disk  may have different topologies and  geometries, being  characterized by different rotation laws, giving rise to   four different spin-spin alignments with respect to the spin of the central attractor, as sketched in  Fig.\il(\ref{Table:Torc}).
 The evolution of the entire macro-configuration then would result from the evolution of    each sub-configuration reaching an interacting phase when  the two configurations reach  contact eventually.
Tori in a double system   may  collide and merge,  or, eventually, turn  to generate  some feeding--drying processes:
the accreting matter from the outer torus of the couple can  impact on the  inner torus, or    the outer torus may be inactive  with  an active inner torus  accreting  onto the \textbf{BH}, or both tori  may be active. We demonstrate that      some configurations  will   collide  for some initial conditions  and attractors dimensionless spin. This process  likely    ends in  the formation of a single orbiting  toroidal accretion disk. Our studies may provide information also  on the \textbf{SMBH} accretion disk  formation  due to \textbf{BH}   interaction  with the environment  in different stages of its life.
   The   phenomenology associated with these systems   may  therefore be  very wide and we believe that this study  could  open up a new field of investigation in astrophysics,  leading, as proposed also in   \cite{S11etal,KS10,Schee:2008fc},  to reconstruct  the interpretive framework of  some phenomena in AGNs environments, thought so far  in terms of a single  accretion  disk,   in terms of  the   multiple accretion disks.
   The inner edge of the outer ring  and the role played by the outer edge of the inner disk should be crucial.
   }

   {
   As discussed in  \cite{open} a strong overflow of matter from a torus  of the double configuration, could be also  related to a jet formation.  Emission may be released in  high energy collisions.
The enormous energy emitted by the accretion disks in quasars or AGN, in the form of electromagnetic radiation and jets, is generally  attributed to the strong gravity of the central black hole when the gravitational binding energy of accreting matter is transformed into radiation.
}

{
Finally, we stress that the study of the equilibrium tori could  be  the  starting point for a future analysis of the oscillation modes in the structure of the relativistic ringed disks  which can be related to various astrophysical phenomena. The radially oscillating tori of the ringed disk could be related to the high-frequency quasi periodic oscillations observed in non-thermal X-ray emission from compact objects (QPOs), a still obscure feature of the  X-ray astronomy related to the inner parts of the disk.
 }
\section{Summary and Conclusion}\label{Sec:conclusion}
We investigated evolutionary schemes of  ringed  accretion disks constituted by two   toroidal axi-symmetric tori  orbiting  on the equatorial plane of  a  central super-massive Kerr black hole. We discussed the emergence of  the instability phases for  each ring of the macro-configuration in the  full general relativistic treatment by considering the effects of the geometry of the Kerr spacetimes on the systems.
 As results of this analysis   we   identified particular classes  of  central  Kerr attractors   in dependence of   their dimensionless spins and the constraints imposed on the evolutionary  schemes  of the double  toroidal system. The schemes outline  the topological transition of the tori  from an   equilibrium topology    to the instability   depending   on the   rotation of the tori relative to each other and  to  the  central Kerr  black hole.
  States  representing the  pair of tori for the four macro-configurations listed in   Fig.\il(\ref{Fig:Goatms}) are summarized in Table\il(\ref{Table:REDUCtIN}).
We used these blocks  to construct   the  evolutionary  schemes  in Fig.\il(\ref{Fig:doub-grap-ll-cor}) for a couple of tori in  the  Schwarzschild spacetime  and   the  $\ell$corotating pairs in  a Kerr spacetime, while   Figs\il(\ref{Fig:CC-CONT}) show  the case of  $\ell$counterrotating pairs orbiting a Kerr attractor. These couples may  be formed only in certain stages of the inner torus evolution  for some Kerr attractors. Our analysis in turn sets    significant limits on  the observational  evidence  of these  systems, providing  constraints on  the tori--attractor systems   imposing    limits on the central attractor spins, and relating the attractor to the couple of tori and their evolution from formation to final stage towards accretion or collision.
The presented analysis of evolutionary schemes  is related to the case of ``frozen''  Kerr geometry.
We here do not follow evolution of ($M$, $a$) parameters of Kerr spacetime.
In some cases inclusions of evolution of $a/M$ parameter could introduce some instabilities  when some critical  values of  $a/M$ will be crossed   to accretion of corotating  or counterrotating  matter.
 The mutation of the geometry determines in general  a change of the dynamical properties of the tori,  eventually resulting, as argued in \cite{ringed}, also in an iterative process, which could give rise even to the  runaway instability \citep{Abra83,Abramowicz:1997sg,Font:2002bi,Rez-Zan-Fon:2003:ASTRA:,Lot2013,Hamersky:2013cza,ergon}.
 This is a theme for future work.
Finally, we considered the situation in which the pair of tori may collide either remaining quiescent, or after   the emergence of instability from  one of the sub-configuration, discussing the  mechanisms that, during the tori evolution,  could lead to collision. The possible  scenarios  may, eventually end in the   merging of the tori with destruction of the  macro-configuration. {Note that in the case of $\ell$corotating tori,   the outer torus   collides,    with the inner torus, eventually the tori merge \emph{before}   it can reach  the accretion phase,  with the consequent destruction of the system   (therefore it could not be give rise to a configuration  $\mathbf{C}_{\odot}^x{}^2$ with accretion point  $r_{\times}=y_1^i=y_3^o$).
Similarly, according to  the  double geodesic structure of the Kerr spacetime there is no  outer corotating torus  in accretion in a couple with  an inner counterrotating  torus, as discussed in Sec.\il(\ref{Sec:coun-co}) for the couples $\pp^+<\pp^-$.}

{ Our interest in this investigation was   justified by a series of studies and  observational evidences   supporting  the existence  of super-massive \textbf{BHs}  characterized by  multi-accretion episodes  during their life-time.
 These facts justify questioning the relevance of the ringed accretion disks theoretically and changeling  them phenomenologically.
 The presence of such structures modifies substantially  we believe,  so far assumed   scenario of a single disk, taken as the basis of the High Energy  Astrophysics, connected with the accretion.
  New observational effects may then be associated with these  complex structures,  showing their existence unequivocally.
 We have different  pieces of evidence suggesting what such a situation  might be the case.
From a theoretical perspective, there are a number of
possible physical mechanisms for a ringed disk formation.
 Our work is related  to their  dynamics  around \textbf{SMBHs}, where the curvature effects become relevant, and the general relativistic treatment adopted here is the most appropriate.
There are indeed suggestions   for these objects are  hosted in the geometries of \textbf{SMBHs}, therefore we  assume these systems as most probable  environments.
 It is  generally accepted  then that the nuclei of most galaxies
contain \textbf{SMBHs}, this picture  is  supported by several studies  and  agrees with
several observational facts.
 In these environments, \textbf{SMBHs} life may report  traces  of  its  host galaxy dynamics.
Repeated galaxy mergers may  constitute  one  mechanism
for a diversified feeding   of a \textbf{SMBH}.
Probably, the  more immediate  situation to think  where a ring of matter may be formed is in
binary \textbf{BH} systems, and  this applies in many
 astrophysical contexts : X-ray binaries or \textbf{SMBHs} binary systems which
are characterized  by diversified accretion episodes,
feeding  \textbf{BHs} with matter and  angular momentum.
Concerning then  possibility of counterrotating   disks, which we  fully address in the present paper, we refer to well known studies  providing  strong evidences,  and a
 further fascinating hypothesis  of    misaligned disks.
On this last possibility,  there is a quite large literature;  we refer for example to  \cite{Aly:2015vqa,Dogan:2015ida,Lodato:2006kv}.
Another mechanism, perfectly fitting particularly with our model of $\ell$corotating tori, foresees  a splitting of  one  accretion  disk,  resulting in  the formation of a tori couple.
In other words a ringed accretion disk may result from  a
fragmentation of a prior  single  accretion disk, due to a local self-gravitational instability.
All these facts lead us to support the suggestion that the existence of such structured objects are likely to
be considerably  significant in  \textbf{AGNs}. Therefore,
from the observational point of view,
    we expect
our results have implications in a number of
different observational features of \textbf{AGNs} and we have marked some before.
On the other hand, it is  opening  possibilities of  different observational evidences for  new intriguing phenomena induced by the   tori interactions or  oscillations.
The presence of an inner tori may also enter as a new unexpected ingredient in the accretion-jet puzzle.
From a methodological view-point, the study of such systems clearly  open an incredible amount of possibilities to be investigated.
 The  analysis  introduced here shows the huge number   of cases  that occurred even within a  simple three parameters  model (the specific angular momentum  $\ell$ the $K$ parameter and the attractor spin-mass ratio). Thus, we have now two possibilities  to approach the analysis:  by solving numerically in a diversified scenario the equations for a very specific case fixing  the disks and attractors parameters. In this way we may also include more ingredients to each  disk model, but we lose a general overview of the situation, needing  moreover to set some initial configurations. The investigation developed in this work  is fixing  these two issues: we substantially reduce the parameter  space of our model, providing range of variations of the variables and parameters which may fit also to  some extent for other  disk models, providing also  attractor classes on the bases of the tori features. In the end, indicating the attractor which we should chase to find evidences for. We were able  to  provide also indication of the initial  disk couple  evolution.
Any numerical analysis of more complex situations,  sharing  the same symmetry of one at last disk, will be compelled with the results presented here.}

Particularly, this study, Sec.\il(\ref{Sec:non-rigid}), provides  strong constraints for the model parameters of the evaluation of the center-of-mass energy in  collision between rings  which was first evaluated,  within the  test particle approximation, in \cite{letter}. It was proved   that  energy  efficiency of the collisions increases  with increasing  dimensionless black hole spin, giving  very high values  for near-extreme black holes--such  systems  can  be  significant for the high energy astrophysics related especially to accretion onto super-massive black holes, and the extremely energetic phenomena in quasars and AGN.

{{In conclusion we believe our results may   be  of significance for the high energy astrophysical phenomena,  such as  the
shape of X-ray emission spectra,  the X-ray obscuration and absorption
by one of the ring, and the extremely energetic  radiative phenomena in quasars and AGN that could be observable by the planed X-ray observatory ATHENA \footnote{http://the-athena-x-ray-observatory.eu/}},
}

\acknowledgments
D. P. acknowledges support from the Junior GACR grant of the Czech Science Foundation No:16-03564Y.
Z. S.  acknowledges the Albert Einstein Center for Gravitation and Astrophysics supported by the Czech Science Foundation grant No. 14-37086G.
{{The authors have benefited during the preparation of this work of discussion with a number of colleagues, particularly we  thank Prof. J. Miller,
Prof. M. A. Abramowicz and  Prof. V. Karas.  {We	 would	 like	 to	 thank also 	 the	 	 anonymous	 reviewer	for the  useful  suggestions	 and
constructive comments, which helped us
to improve the manuscript.}}}
\appendix
\section{Graphs}\label{Sec:graph-app}
\begin{figure}[h!]
\begin{center}
\begin{tabular}{c}
\includegraphics[width=0.71\columnwidth]{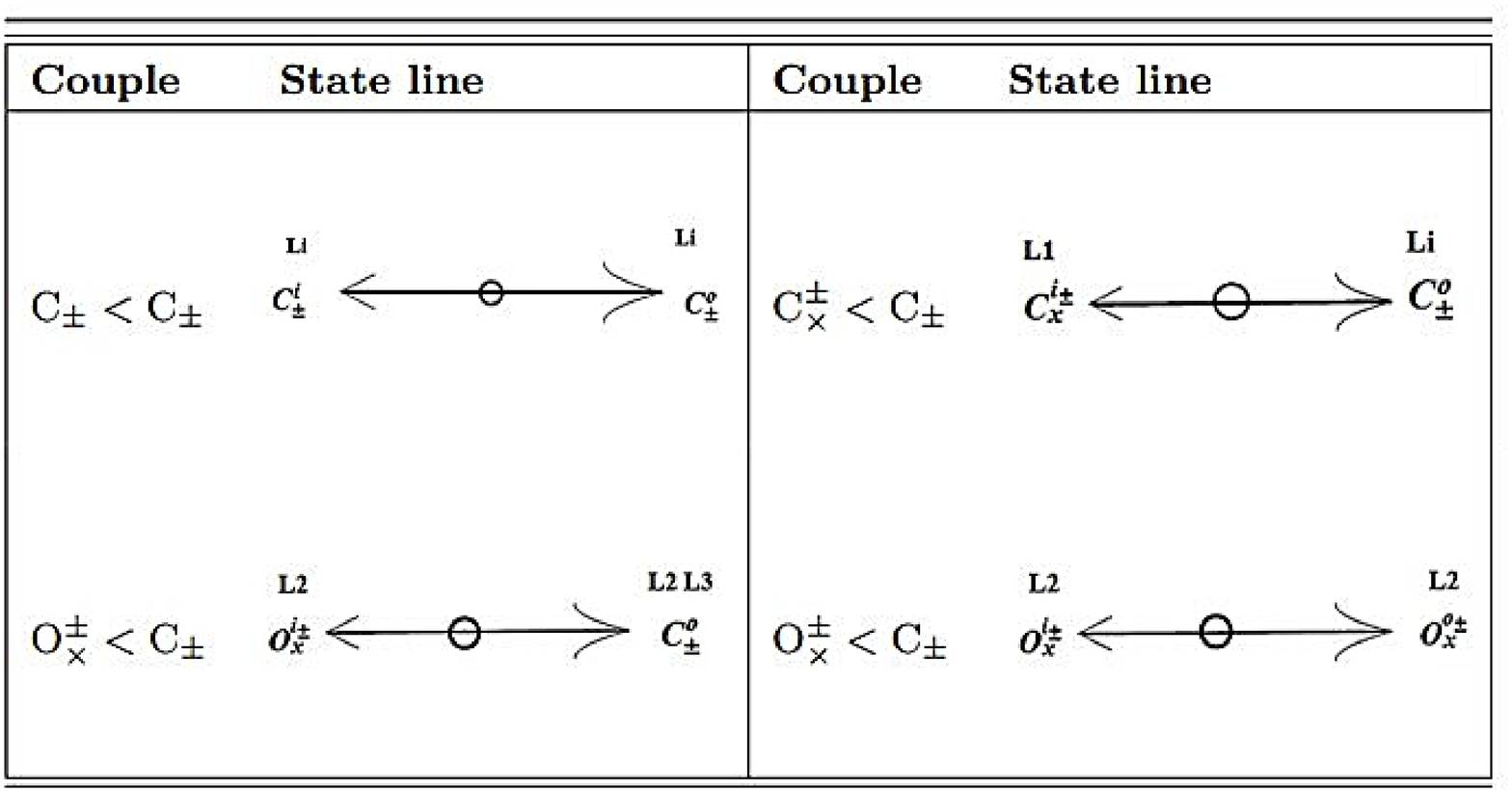}
\end{tabular}
\caption{$\ell$corotating couples: state lines of a monochromatic graph (or a bichromatic graphs for a doubled system in a static spacetime) of    Fig.\il(\ref{Fig:doub-grap-ll-cor}), corresponding  the \textsl{scheme I} and the  \textsl{scheme II} of Fig.\il(\ref{Table:Torc}). Main blocks are introduced in Fig.\il(\ref{Table:Graphs-models}). }\label{Table:statesdcc}
\end{center}
\end{figure}

\begin{figure}[h!]
\begin{center}
\begin{tabular}{c}
\includegraphics[width=0.8\columnwidth]{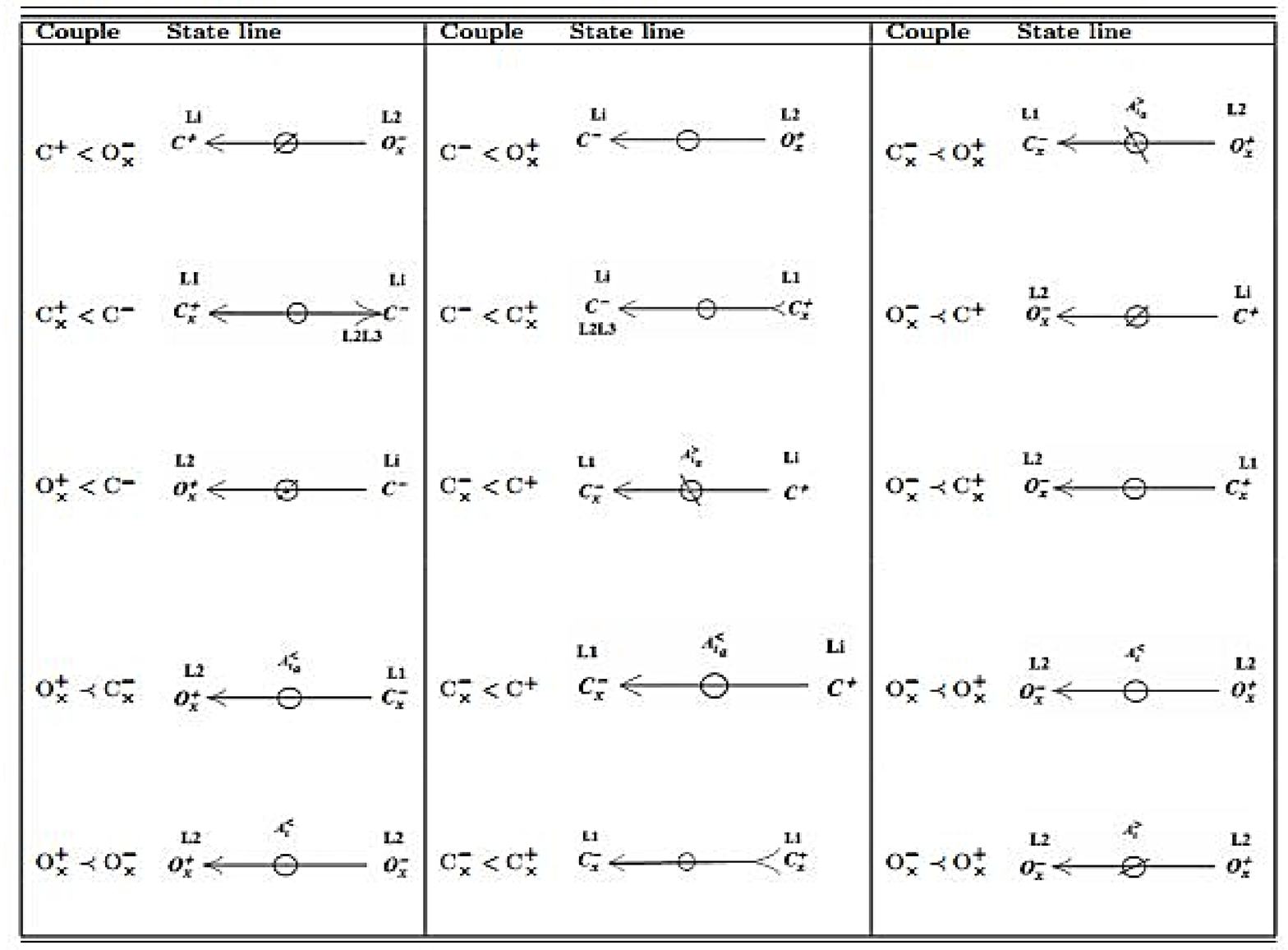}
\end{tabular}
\caption{$\ell$counterrotating couple: state lines of   bichromatic graph of Figs\il(\ref{Fig:CC-CONT}) for a couple of $\ell$counterrotaing disks   in a Kerr black hole $a\neq0$ spacetime, corresponding to the \textsl{scheme III} and  the  \textsl{scheme IV} in Fig.\il(\ref{Table:Torc}). Main blocks are introduced in Fig.\il(\ref{Table:Graphs-models}).}\label{Table:statescc}
\end{center}
\end{figure}

\begin{figure}[h!]
\begin{center}
\begin{tabular}{c}
\includegraphics[width=0.71\columnwidth]{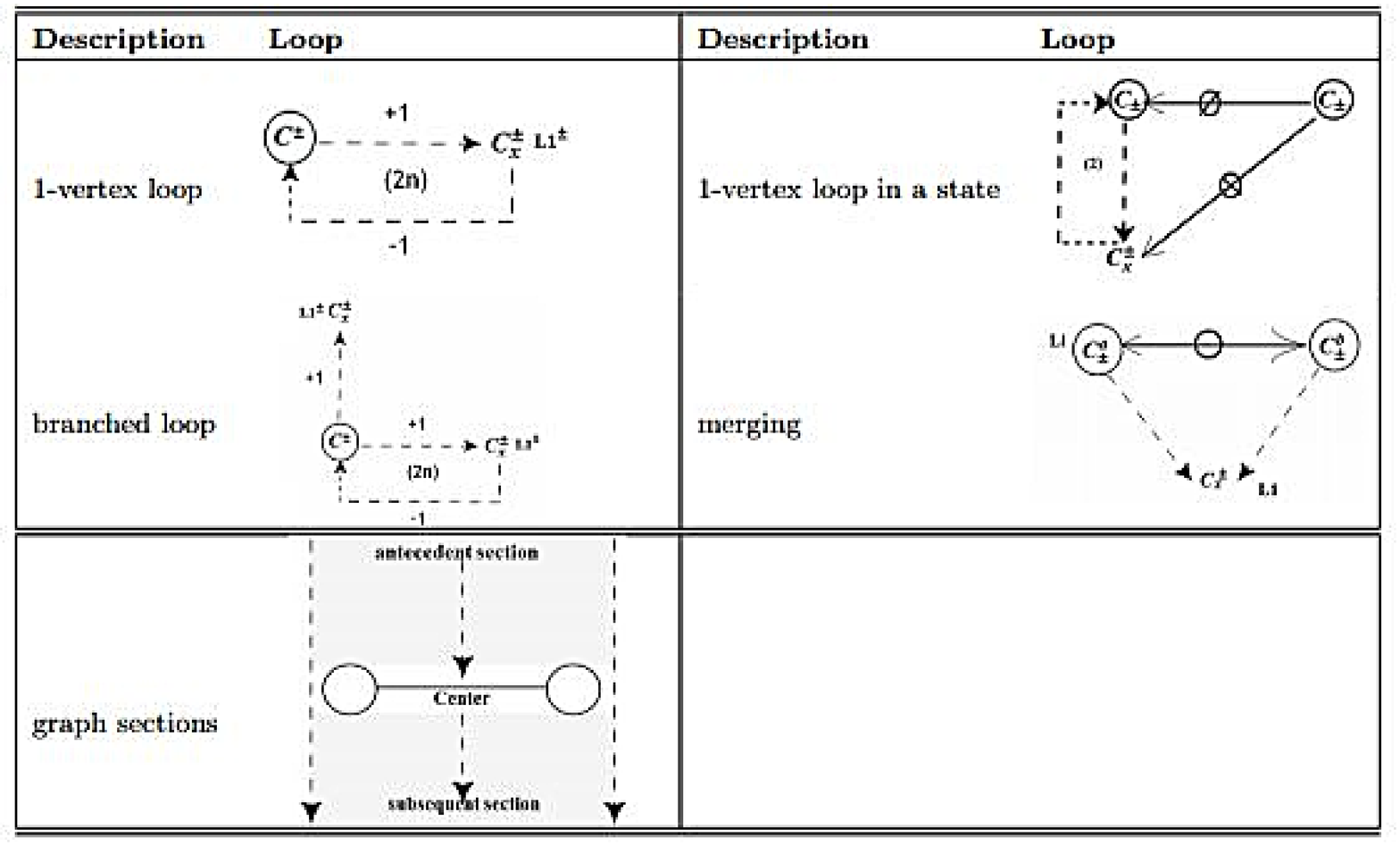}
\end{tabular}
\caption{
\emph{Upper}: a  loop closed on an equilibrium vertex.
The cycle is completed  $n$  times with $2n$ evolutive lines (\emph{left})- $\pm 1$ is accordingly attached to oriented evolutive lines. \emph{Right panel} shows an example of such a loop with $n = 2$ attached to a state line.
\emph{Middle}: a cycle with an open evolutive line on an   topology  of accretion represented as a  loop  with  a ``branch''(\emph{left panel}). \emph{Right panel} shows the  merging of two configurations through a closed loop of evolutive lines  in which a state line is ``contracted'' on a vertex.
\emph{Bottom}:  graph  sections. }\label{Table:loops}
\end{center}
\end{figure}

In this section we clarify some aspects of the graphs formalisms used in this work.
 State lines of the  $\ell$corotating  couples, used in  the monochromatic  graph of Fig.\il(\ref{Fig:doub-grap-ll-cor}), are  listed in Fig.\il(\ref{Table:statescc}), while the state line for the $\ell$counterrotating couples,  used in the bichromatic graphs of    Figs\il(\ref{Fig:CC-CONT}), are listed in  Fig.\il(\ref{Table:statesdcc}).
  Samples of   loop  graphs discussed in this works are in Table\il(\ref{Table:loops}).
%
%\begin{table}[h!]
%\caption{$\ell$corotating couples: state lines of a monochromatic graph (or a bichromatic graphs for a doubled system in a static spacetime) of    Fig.\il(\ref{Fig:doub-grap-ll-cor}), corresponding  the \textsl{scheme I} and the  \textsl{scheme II} of Fig.\il(\ref{Table:Torc}). Main blocks are introduced in Fig.\il(\ref{Table:Graphs-models}). }\label{Table:statesdcc}
%\centering
%%\resizebox{.471\textwidth}{!}{%
%\begin{tabular}{|ll|ll|}
%\toprule\hline\hline
%\textbf{Couple} &$\;$ \textbf{State line} & \textbf{Couple} & $\;$\textbf{State line}\\
%\hline
%$\cc_{\pm}<\cc_{\pm}$ & \raisebox{-7ex}{\includegraphics[width=.2\textwidth]{de1}}&$\cc_{\times}^{\pm}<\cc_{\pm}$ & \raisebox{-7ex}{\includegraphics[width=.2\textwidth]{de2}}\\
% $\oo_{\times}^{\pm}<\cc_{\pm}$ &\raisebox{-7ex}{\includegraphics[width=.2\textwidth]{de3}}&$\oo_{\times}^{\pm}<\cc_{\pm}$ & \raisebox{-7ex}{\includegraphics[width=.2\textwidth]{de4}}\\
%\hline\hline
%\end{tabular}%}
%\end{table}
%
%
The evolutive lines generally split  the graph into two parts centered  around the center: the \emph{antecedent section},
from which the heads of the arrows converging at the center start, and the \emph{subsequent section} which is the one onto which  the evolutive lines,
starting from or  crossing the graph center converge with head in the antecedent section--Table\il(\ref{Table:loops}). A graph may also have only one section.
As pointed out in Sec.\il(\ref{Sec:basic-intro}), the  evolutive lines may  connect state line with different    critical sequentiality  but not different configuration sequentiality which is preserved during the evolution.
 In this discussion, vertices which are connected or crossed  by an evolutive  line, pertain at  equal chromaticity. Thus,   if the graph center   is  {monochromatic}  (bichromatic) then the entire graph is monochrome (bichromatic).
An evolutive loop  is defined as the union of evolutive lines and the vertices they cross, closing  on an initial  topology as in Table\il(\ref{Table:loops}), The  triple vertix transition $\cc\dashrightarrow \cc_{\times}\dashrightarrow \cc$ is a loop example.
 %%%
%
%
%
%\begin{table}[h!]
%\centering
%\caption{
%\emph{Upper}: a  loop closed on an equilibrium vertex.
%The cycle is completed  $n$  times with $2n$ evolutive lines (\emph{left})- $\pm 1$ is accordingly attached to oriented evolutive lines. \emph{Right panel} shows an example of such a loop with $n = 2$ attached to a state line.
%\emph{Middle}: a cycle with an open evolutive line on an   topology  of accretion represented as a  loop  with  a ``branch''(\emph{left panel}). \emph{Right panel} shows the  merging of two configurations through a closed loop of evolutive lines  in which a state line is ``contracted'' on a vertex.
%\emph{Bottom}:  graph  sections. }\label{Table:loops}
%\begin{tabular}{|ll|ll|}
%\hline\hline
%\textbf{Description} &$\;$ \textbf{Loop} & \textbf{Description} & $\;$\textbf{Loop}\\
%\hline
%1-vertex loop & \raisebox{-7.7ex}{\includegraphics[width=.2\textwidth]{Loop-cyclen}}&1-vertex loop in a state  & \raisebox{-7.7ex}{\includegraphics[width=.2\textwidth]{loop-indepen}}\\
% branched loop  & \raisebox{-7.7ex}{\includegraphics[width=.2\textwidth]{loop-oriet}}&%oriented loop II &\raisebox{-7.7ex}{\includegraphics[width=.3\textwidth]{looporiet2}}\\
%merging &\raisebox{-7.7ex}{\includegraphics[width=.2\textwidth]{merging}} \\\hline\hline
%graph sections &\raisebox{-7.7ex}{\includegraphics[width=.25\textwidth]{graph-sec}}& &\\
%\hline\hline
%\end{tabular}
%\end{table}
%\begin{figure}[h!]
%%%%%%%%%%%%%%%%%%%%%%%%%%%%%%%%%%%%%%%%%%%%%%%%%%%%%%%%%%%%%%%%%%%%%%%\input{D-loop-Apart}
\subsection{Consideration on the loop emergence}\label{App:deep-loop}
% \btb{*****}
 A loop is a closed evolutive line on the graph vertex  representing  a topological transition of a configuration which would finally  restore  the initial topology. An example  is the  drying-feeding process  introduced in \cite{ringed,open}. In this section we briefly discuss possible evolutive scenarios based for the couple of tori on the analysis of the equilibrium and instability states, leaving a more careful and detailed study of the mechanism for such topological transition for future analysis.
A  loop in general   could  take place as a consequence of the  evolution of one vertex  independently by the evolution of the other vertex of the state, or it can be due to a  collision among the state vertices.
Considering particularly  the $\ell$corotating couples   we see
that   separate  configurations may  be favored for the counterrotating tori  having  larger distance between the orbital regions and therefore orbiting in the spacetimes of the  faster attractors,   while  the  separate  states of  corotating  couples  are favored in the geometries of   slow attractors--Figs\il(\ref{Figs:Ptherepdiverg}).
     In the case of the   counterrotating tori,  for     larger dimensionless  spin  of the attractor, the    closer and smaller corotating tori   should be more frequent since   the angular momenta range  decreases  with the  \textbf{BH} spin,   reducing therefore the orbital   ranges  and the distance from the central attractor --\cite{ringed,open}.
The  disk extension (elongation on the equatorial plane and the   density) depends primarily on the  $ K$ parameter. On the other hand, the closer to $1$ is the ratio $\ell_{i/o}\gtrapprox1$,  and the smaller are the separated  tori,   the greater is the possibility of a loop following a   geometric correlation and the merging.

To fix the ideas we consider here  a simple set up within a   simplified  schemes based on the following two  processes.

\textbf{\textsl{1.} Loops after independent evolution of the   inner ring towards accretion}
{Accretion} onto the black hole is    associated with loss of matter  and angular momentum. In our model  both $K$  and the magnitude of the specific angular momentum $\ell$ decrease. Consequently the disk moves  towards  the attractor possibly   shrinking  from the  maximum elongation $\lambda_{\times}$, reached at  the early phases of the process   where accretion starts.
 Following the decrease of the $K$ parameter we assume  the accretion    could even  lead to a stabilization  of the inner ring with parameters   $\ell_i\in \mathbf{L1}$  and $K_i\in \mathbf{K0}$.  Consequently  a  $\mathbf{\cc^1_{\times}\rightarrow \cc_1}$ transition may rise as a {second} and {last} stage of 1-loop--see Table\il(\ref{Table:loops}).
 On the other hand,  as the magnitude of the specific angular momentum $\ell_i$   decreases as well,  the ring moves towards the attractor, moving away from the outer ring, while the outer edge of the disk moves inward,  finally  blocking the loop, preventing  any contact with the outer ring,  and  a  transition $\mathbf{\cc_i\rightarrow \cc^i_{\times} \rightarrow \cc_i}$, involving only the inner vertex, can happen.

\textbf{\textsl{2.} Loops after interaction between the rings }
In general,  an {interaction} of an inner torus   with the surrounding material or  {collision} with the outer torus can lead to increase of matter and $K$, implying destabilization towards a $\cc^1_{\times}$ or $\oo_{\times}^2$  phase. Possibly  also a split of the disk can occur,  increases of  the angular momentum  which could lead to an  $\mathbf{\cc_i\rightarrow \cc^i_{\times}}$ transition.
However, we note that if the magnitude of the specific momentum increases, the torus  moves outward, increasing therefore   the probability of collision with the outer ring and,  if the momentum magnitude   is sufficiently high, i.e. $\ell_i\in \{\mathbf{L2}, \mathbf{L3}\}$,  this prevents  emergence of a further stage of  accretion.
  If, on the other hand,   $K$ increases in $\mathbf{K0}$, and in  $]K_{\min}, K_{\max}[$ for $\ell\in\{\mathbf{L1}, \mathbf{L2}\}$,  the torus outer edge   moves outward and  its inner margin moves inward, increasing, from one side,  the probability of  collision with the outer ring,  and on the other  side, the torus instability finally leading, to  the  accretion (point \textbf{\textsl{1.}}), if $\ell_i\in\mathbf{L1}$.

The accretion  of the inner ring  generally  \emph{blocks} the loop:
if the inner ring is accreting  onto the source, preserving the separation from the outer ring, there is possibility to  evolve in a loop,
due to the stabilization of the  inner vertex that would return under appropriate conditions in the starting  equilibrium phase (assuming the ringed disk would not evolve towards a $\mathbf{C}_{\odot}^x{}^2$ phase,  the conditions for  this  behavior to occur are discussed in Sec.\il(\ref{Sec:non-rigid})).
 According to the evolutive graph  of Fig.\il(\ref{Fig:doub-grap-ll-cor}), the $\cc_{\times}^1$ disk   can return to an equilibrium  configuration  even for enhanced specific angular momentum, undergoing transition  from a  $\mathbf{L1}$ range to  $\mathbf{L2}$ or $\mathbf{L3}$, through an  evolutive line which   `` brings '' a new different decoration for the initial vertex. However,  such an increase in magnitude of specific angular momentum (due to some  specified process) would imply the outer ring to be far enough ($\ell_o/\ell_i>1$), or to be  small enough, to prevent collision.

On the other hand, if the outer disk  is not quiescent but  it collides with the inner one, then the ringed disk passes from a $\mathbf{C}^2$ phase to a $\mathbf{\cc}^2_{\odot}$ one. This may occur   due to decrease (increase) of specific angular momentum  $\ell_o$  ($\ell_i$) or growing of  $K_i$  or $K_o$ (or a combination of these  possibilities).
Under this  very simplified  scheme, the interaction between the two tori  would lead to a \emph{merge} in a single disk, destroying therefore the double system.  This will happen in \emph{competition} with a destabilization towards accretion which involves the inner torus only for these couples.

The ringed disk could  start, for example,  from   a $\mathbf{\cc}_{\times}^2$ phase,  where the inner ring is accreting into the source,  and it undergoes a $\mathbf{\cc}^{\times}_{\odot}{}^2$ phase, where the loss of specific angular momentum    $\ell_i$   and decreasing of $K_i$ parameter, should be compared with the support of matter from the outer configuration, which  could feed the accretion. As a consequence,  this  process  should  be considered as  case of collision-inducing-accretion  leading possibly to a stabilizing  evolutive loop. It may be also induced by  an increasing angular momentum, blocking the accretion from the inner torus, causing   a merging with the outer torus. Otherwise  collision may  be predominant, with consequent merging of the two rings in one accreting disk--Table\il(\ref{Table:loops}).

We can conclude that the double ring systems are less likely to form for corotating rings that would merge or collide  that for counterrotating rings:   corotating ones  are  less likely to form around  faster  attractors, while the formation of counterrotating double rings is   favored by the rotation of  the attractor.
 Therefore, we should search for a double system of separated counterrotating rings for fast attractor, while merging should characterize the corotating rings--see Fig.\il(\ref{Fig:Goatms}).
The \emph{collision}  inducing  merging can follow  the  increase of $K$ and the increase
of $|\ell_i|$, or the  increase of  $K$ and decrease of $|\ell_o|$. The increase of $K_i$ is compatible with the onset of the growth phase of $\cc^i_{\times}$ where in fact $\lambda_{\times}>\lambda$, but if the accretion is  associated with a decrease of both $K_i$ and $|\ell_i|$  (according to \textbf{1.})--this  favors the separation   and prevents   the disks collision.
To summarize; the processes described in \textbf{1.} and \textbf{2.} deal with  a competition between the loss of the angular momentum,  which causes the tori to shrink and to shift inwardly, and the  feeding of matter causing increasing  $K$ and  possibly $\ell$.  Thus,  assuming   the state $\cc^i_{\times}<\cc_o$ does not involve collision, the accretion of the inner  ring inevitably leads  to separate  the tori avoiding the   merging,    therefore we have to ensure the initial condition for    $\lambda_i=\lambda^i_{\times}$ guaranteing that the two tori are separated, and they  \emph{remain} separated until the outer configuration would change its morphology.

Finally,  the role of the radii $r_{\mathcal{M}}$ and $\bar{\mathfrak{r}}_{\mathcal{M}}$ may have impact on the loop production. These radii correspond to the maximum point  of the variation of the  tori fluid specific angular momenta magnitude  $\ell$ with respect to the orbital distance from the attractor, in other words they are solutions of $\partial_r\partial_r\ell=0$ \cite{open}.  The maximum $\ell_{\mathcal{M}}$ is associated to a torus centered in $r_{\mathcal{M}}$ and with critical point in  $\bar{\mathfrak{r}}_{\mathcal{M}}$. This implies that  the rings of  $\ell$corotating couple  with fixed  angular momentum magnitude difference $\ell_o-\ell_i=\epsilon$,  are increasingly closer as their angular momentum approaches $\ell_{\mathcal{M}}$ (say, for initial $r_{\min}>r_{\mathcal{M}}$, for  $\ell_i=\ell_{\mathcal{M}}+\kappa_i$ and $\ell_o=\ell_{\mathcal{M}}+\kappa_i+\epsilon$ with $\epsilon=$constant, and decreasing  $\kappa_i$)  approaching the black hole.  The location of momenta  $\ell_{\mathcal{M}}^{\pm}\in \mathbf{Li}$  depends on  rotation of the torus with respect to the attractor and the dimensionless spin of the attractors-- Fig.\il(\ref{Figs:Ptherepdiverg}) and  \cite{ringed,open}.

\textbf{The emergence of the Loops for $\ell$counterrotating couples $\cc^-<\cc^+$}
   %%%%%%%%%%%%%%%%%%%%%%%%%%%%%%%%%%%%%%%%%
	%%%%%%%%%%%%%%%%%%%%%%%%%%%%%%%
%%%%%%%%%%%%%%%%%%%%Loops  at %%%%%%%%%%%%%%%%%%%%%%%%%%%%%%%%%%%%
We focus on the $\ell$counterrotating couples $\cc^-<\cc^+$, analyzed in  Sec.\il(\ref{Sec:tex-dual}).
We consider    loops involving evolutive phases with accretion,
 with possible  interaction by collision, leading to a $\mathbf{C}^2_{\odot}$ ringed disk.
 First, we note that  if evolution towards accretion involves the outer torus only, then   collision of the first Roche lobe  on the inner  torus may  be unavoidable. The  angular  momentum magnitude may decrease, the torus would loose material, finally  going inwardly and then making inevitable collision of the first Roche lobe with the inner torus that  would acquire  momentum and mass. Therefore, this process would give rise  to the destruction of the couple, and then  there would be no loop.
  Processes  $\mathbf{(a)}$ and $\mathbf{(c)}$ of Eq.\il(\ref{Eq:a-bbh}) and  Eq.\il(\ref{Eq:non-sile-voic}), describe the evolutions of one vertex, however, although similar, their evolutions towards loop is very different. In fact the situation for  $\mathbf{(a)}$ is analogue to all the other cases  where the inner ring is in accretion
as all the $\ell$corotating couples
described  in Sec.\il(\ref{Sec:lc-or}) and the
 $\ell$counterrotating ones  in the static spacetimes or the
 $\cc_{\times}^+<\cc^-$ addressed in Sec.\il(\ref{Sec:coun-co}).d
Consequently,  a loop  may be attached to the unstable  vertex of  the $\cc_{\times}^-<\cc^+$  state  as  in Fig.\il(\ref{Table:loops}),
when the outer configuration is  inert (does not loose angular momentum, otherwise giving rise to a $\mathbf{(b)}$ path or a  possible collision
with formation of a $\mathbf{C}^2_{\odot}$ ringed disk), and no correlation between the two rings torus so that the evolution of the state proceeds through  independent evolution  of  the two vertices.  The collision, for loss of angular momentum  or thickening of $\cc^+$,  with
 formation of a  $\mathbf{C}_{\odot}^x$ system
would result in a loop (with correlation)
with an outer counterrotating torus which may  also be quiescent, i.e., non accreting.
 In the case of  the $\mathbf{(c)}$ process, we can trace the following   qualitative consideration regarding the possible generation of loop.  A stabilized loop  attached to the outer configuration should not be likely in this scenario  because  the outer accreting torus,  losing angular momentum and matter (decreasing  $ K$ and   $-\ell_+$), moves inward. The competition of these two processes, however, could be compensated by the fact that  material  thickens on the inner torus which then should stretch outwardly. In this  very simplified  model for the evolution of the system   after collision,  the double system would seem to be destined to merge. However, if the outer torus stabilizes,
then in this particular situation the torus may be attached  to a   loop
in which, for example, the  outer ring is stable,  while the inner one is in accretion, giving rise, for example,  to the  sequence of processes $\mathbf{(c)}\rightarrow \mathbf{(a)}$ with a loop.
Similarly, we can  draw some  general consideration on   the paths $\mathbf{(b)}$ and $\mathbf{(d)}$.
A loop in $\mathbf{(b)}$ could be extended  for example, by continuing the path $\mathbf{(b)}$ with  the  $\mathbf{(a)}$ evolution, for stabilization of the inner ring  and  consequently preceding  a
$\mathbf{(c)}$ process.
It is however necessary to discuss  the  $\cc^-_{\times}<\cc^+_{\times}$ segment,
  as final state of  $\mathbf{(b)}$  and  $\mathbf{(d)}$. Assuming that   this state will not be constrained by the  initial conditions (the different paths)  to distinguish the two cases in this way, then  the inner torus in accretion onto the black hole and the outer torus in accretion  on the inner torus  could lead to a possible evolution for stabilizing the outer torus or otherwise a  merging.
\section{Comments on Table 4%\il(\ref{Table:REDUCtIN})
}\label{App:notes-tables}
We provide some comments on  Table\il(\ref{Table:REDUCtIN}).
The analysis refers to Fig.\il(\ref{Fig:OWayveShowno}), where the geodesic complementary structure  $\bar{\mathbf{R}}_{\mathrm{N}}$ is represented.
\subsubsection{Couples: $\pp^+<\pp^-$}
We consider the
decoration of the equilibrium vertices  with angular momentum classes  for the states   $\pp^+<\pp^-$ of the graph in Fig.\il(\ref{Fig:CC-CONT}) and proof of the results of Table\il(\ref{Table:REDUCtIN}).
The following properties holds:
\begin{itemize}
\item[]
All the couples  $\pp^+_3<\pp^-$ are $\pp_3^+<\pp_3^-$.  In fact for   $\pp_3^+<\pp^-$,  it has to be  $\bar{\mathfrak{r}}_{\gamma}^+<r_{\min}^+<r_{\min}^-$,
 which is realized only for corotating tori  $\pp_3^-$. This means that if the   inner ring is counterrotating and sufficiently far from the attractor, then the second (outer) counterrotating ring must be $\pp_3^+$.
\\
\item[]
Viceversa, the couples $\pp^+<\pp_3^-$ with  $r_{\mso}^+<r_{\min}^+<r_{\min}^-$   and  $r_{\min}^->\bar{\mathfrak{r}}_{\gamma}^+$
have no  constraints on the angular momentum of the inner ring.
\\
\item[]
The couples  $\pp_2^+<\pp^-$ are characterized by the relations  $\bar{\mathfrak{r}}_{\mbo}^+<r_{\min}^+<r_{\min}^-$, which implies  that
for $a >a_{{}_u}$,  there are only couples $\pp_2^+<\pp_3^-$,
but if  $ a<a_{{}_u}$,  there are $\pp_2^+<\pp_3^-$ and $\pp_2^+<\pp_2^-$, but not all the configurations $\pp_2^-$ are eligible.
\\
\item[]
If  $\pp^+<\pp_2^-$, then $r_{\mso}^+<r_{\min}^+<r_{\min}^-<\bar{\mathfrak{r}}_{\gamma}^-$, and there are no
$\pp_3^+<\pp_2^-$ couples, while,  remarkably, there are no
$\pp^+<\pp_2^-$ couples around  attractors with dimensionless spins $a>\breve{a}_{\aleph}$.
For slower attractors, $a\in]a_{{}_u},\breve{a}_{\aleph}[$, there are only  $\pp_1^+<\pp_2^-$ couples.
Then there is a small class of  attractors with the spins  $]\tilde{a}_{\aleph},a_{{}_u}[$
where
$\bar{\mathfrak{r}}_{\mbo}^-<r_{\mso}^+<\bar{\mathfrak{r}}_{\mbo}^+<\bar{\mathfrak{r}}_{\gamma}^-$ and,
accordingly, there are   $\pp_2^+<\pp_2^-$  and $\pp_1^+<\pp_2^-$.
For even lower spins,   $a<\tilde{a}_{\aleph}$,  where $r_{\mso}^+<\bar{\mathfrak{r}}_{\mbo}^-<\bar{\mathfrak{r}}_{\mbo}^+<\bar{\mathfrak{r}}_{\gamma}^-$,
 there may be both couples $\pp_1^+<\pp_2^-$  and   $\pp_2^+<\pp_2^-$, but not all the  $\pp_2^+$ fulfill this property.
\\
\item[]
Finally, we  focus on the configurations $\pp_1$ with  $\ell\in\mathbf{L1}$. Analyzing the
couples  $\pp_1^+<\pp^-$, which satisfy the property  $r_{\mso}^+<r_{\min}^+<r_{\min}^-$  then  for attractors with    $a>\breve{a}_{\aleph}$, there are only couples $\pp_1^+<\pp_3^-$, while  for $a\in]\tilde{a}_{\aleph},\breve{a}_{\aleph}[$ there are   $\pp_1^+<\pp_3^-$  and part of $\pp_1^+<\pp_2^-$.
For slower attractors, $a<\tilde{a}_{\aleph}$, there are also the couples  $\pp_1^+<\pp_1^-$.
\\
\item[]
The couples, $\pp^+<\pp_1^-$, for which $r_{\mso}^+<r_{\min}^+<r_{\min}^-<\bar{\mathfrak{r}}_{\mbo}^-$, are possible only for  $a<\tilde{a}_{\aleph}$ as  $\pp_1^+<\pp_1^-$.
\end{itemize}
\subsubsection{Couples $\pp^-<\pp^+$}\label{Sec:app-submmp}
We deal with the decoration of the equilibrium vertices   with angular momentum ranges for the states   $\pp^-<\pp^+$--\textsl{scheme III} of Fig.\il(\ref{Table:Torc}) as in Table\il(\ref{Table:REDUCtIN}).
We make reference to Fig.\il(\ref{Fig:OWayveShowno}).
\begin{description}
\item[]
We start with the couple  $\pp^-=\pp_3^-$,
 where   $\bar{\mathfrak{r}}_{\gamma}^-<r_{\min}^-<r_{\min}^+$.
Then for  $a>\breve{a}_{\aleph}$, couples $\pp_3^-<\pp^+$ exist for any counterrotating topology.
Instead, if the attractor has dimensionless spin  $a\in]a_{{}_{u}},\breve{a}_{\aleph}[$ only part of the  $\pp_1^+$ configurations, $\pp_2^+$  and $\pp_3^+$ fulfills the condition.
For slower attractors,
$a<a_{{}_{u}}$, only part of  $\pp_2^+$ and all  $\pp_3^+$ configurations satisfy the condition.
\item[]
For  $\pp^+=\pp^+_3$, there is  $r_{\mso}^-<r_{\min}^-<r_{\min}^+$, then there could be
any corotating topology $\pp_i^-<\pp^+_3$.
\item[]
If $\pp^-=\pp_2^-$, there is   $\bar{\mathfrak{r}}_{\mbo}^-<r_{\min}^-<r_{\min}^+$
thus,
for  $a>\tilde{a}_{\aleph} $, any configuration $\pp^+$ can be in the couple $\pp_2^-<\pp^+$;
 for  $a<\tilde{a}_{\aleph}$ only part of  $\pp_1^+$ configurations and all $\pp_2$ and  $\pp_3$ are in $\pp_2^-<\pp^+$.
\item[]
 If
 $\pp^+=\pp_2^+$, there is
$r_{\mso}^-<r_{\min}^-<r_{\min}^+<\bar{\mathfrak{r}}_{\gamma}^+$,
   any counterrotating configurations may be in the couple.
\item[]
If  $\pp^-=\pp_1^-$, there is    $r_{\mso}^-<r_{\min}^-<r_{\min}^+$
and  $\pp^+$ can be in any class of angular momentum with some further restrictions on  $\pp_1^+$ for  $a<\tilde{a}_{\aleph}$.
If $\pp^+=\pp_1^+$, there is
$r_{\mso}^-<r_{\min}^-<r_{\min}^+<\bar{\mathfrak{r}}_{\mbo}^+$
 for  $a>a_{{}_u}$, $\pp^-$ can be in any angular momentum range, and for
 $a<a_{{}_{u}}$, $\pp^-$ can be  $\pp_2$ or $\pp_1$. Note that there is the radius  $r_{\mathcal{M}}^-$
 crossing $\bar{\mathfrak{r}}_{\mbo}^-$  for  $a\approx 0.35 M$.
\end{description}
\subsubsection{Comments on the constrained criticality}\label{App:comm-con-crit}
Here we provide some notes on the results of Table\il(\ref{Table:REDUCtIN}) on the criticality order, discussing also  the order of decorated state lines for the counterrotating couples of Fig.\il(\ref{Table:statescc}).
These constraints on the location of the instability points   are  consequences of the geodesic structure of spacetime, as represented in  Fig.\il(\ref{Figs:Ptherepdiverg}). However, we  considered also  restrictions  provided by \cite{open}, based on the geodesic structure  $\bar{\mathbf{R}}_{\mathrm{N}}$ and the relation between the angular momenta $\ell_{\mathrm{N}}$-- Fig.\il(\ref{Figs:Ptherepdiverg}).
For $a >a_{\iota_a}$ the geodesic structure implies  $\cc_{\times}^-\prec \cc_{\times}^+$
 where $a_{\iota_a}=0.372583M:\;r_{\mbo}^+ = r_{\mso}^-$,
 where the critical sequentiality is not fixed. However
 it is possible to prove that the relation   $\cc_{\times}^-\prec \cc_{\times}^+$  extends also for couples orbiting
 around  slower attractors. In fact, for   $a<a_{\iota_a}$, there is $r_{\mbo}^-<r_{\mbo}^+ < r_{\mso}^-<r_{\mso}^+$,
However,  $r_{\mbo}^+\in! \cc_{\times}^-$,
 and  $r_{\mso}^+\non{\in} \cc_{\times}^-$--
this means in particular that
$r_{\mbo}^-<y_{3}^-<r_{\mbo}^+ <y_3^+< r_{\mso}^-<r_{\mso}^+$.
therefore there is  $\cc_{\times}^-\prec \cc_{\times}^+$, which closes the proof.
Note that we used  the assessment of $ r_{\mbo}^+\in! \cc_{\times}^-$  obtained from the considerations in \cite{open}.
Moreover there is
$r_{\mso}^-\non{\in }\cc_{\times}^+$ if $\ell_1\in]\ell_{\mso}^+, \ell_1(r_{\mso}^-)[,\;
r_{\mso}^-{\in }!\cc_{\times}^+,\; \ell_1\in]\ell_1(r_{\mso}^-),\ell_{\mbo}^+[$.
A similar analysis could be implemented for other topologies.
For  $a>a_{\iota}$,  where
$a_{\iota}=0.313708M:\;r_{\gamma}^+=r_{\mbo}^-$,
we have $\oo_{\times}^-\prec \oo_{\times}^+$.
While there is
$\oo_{\times}^-\prec \cc_{\times}^+$ in any geometry.
Finally, for   $a >a_{\gamma_+}^-=0.638285M:\; r_{\mso}^-=r_{\gamma}^+$, there is
 $\cc_{\times}^-\prec \oo_{\times}^+$.
%

%
%%


\end{document}